\PassOptionsToPackage{pdftex,
pdfencoding=auto,
pdfnewwindow=true,
pdfusetitle=true,
bookmarks=true,
bookmarksnumbered=true,
bookmarksopen=true,
pdfpagemode=UseThumbs,
bookmarksopenlevel=1,
pdfpagelabels=true,
breaklinks=true,
colorlinks=true,
}{hyperref}
\documentclass[amsmath,amssymb,superscriptaddress,twocolumn,notitlepage,nofootinbib]{revtex4-2}

\usepackage[utf8]{inputenx}
\usepackage[english]{babel}
\usepackage{graphicx}
\usepackage[usenames,dvipsnames]{xcolor}
\definecolor{purple}{RGB}{128,0,128}
\definecolor{ultramarine}{RGB}{63, 0, 255}
\definecolor{medblue}{RGB}{0, 0, 100}
\definecolor{googleblue}{RGB}{34, 0, 204}
\definecolor{panblue}{RGB}{0,24,150}
\definecolor{carmine}{RGB}{150, 0, 24}
\definecolor{gray}{RGB}{150, 150, 150}
\definecolor{darkgreen}{RGB}{0, 80, 0}

\usepackage[]{hyperref}
\hypersetup{linkcolor=carmine,citecolor=darkgreen,urlcolor=googleblue,anchorcolor=OliveGreen}

\usepackage{microtype}
\microtypecontext{spacing=nonfrench}
\microtypesetup{
expansion={true,nocompatibility},
protrusion={true,nocompatibility},
activate={true,nocompatibility},
tracking=true,
kerning=true,
spacing={true}
}
\usepackage[all=normal,floats=tight,mathspacing=tight,wordspacing=tight,paragraphs=normal,tracking=tight,charwidths=tight]{savetrees}
\everypar=\expandafter{\the\everypar\loosness=-1 }
\usepackage[style=american,autopunct=true]{csquotes} 
\usepackage{mathtools}

\usepackage{soul}

\newcommand{\ket}[1]{\left|#1\right\rangle}

\newcommand{\dyad}[1]{\left|#1\right\rangle\langle#1|}

\newcommand\blfootnote[1]{%
  \begingroup
  \renewcommand\thefootnote{}\footnote{#1}%
  \addtocounter{footnote}{-1}%
  \endgroup
}

\usepackage{bbold}
\usepackage[flushleft, neveradjust]{paralist}

\usepackage{subfigure}
\usepackage{enumitem}
\usepackage{verbatim}
\usepackage{verbatimbox}

\graphicspath{{./images/}}

\begin{document}
\onecolumngrid
\twocolumngrid

\setlist[description]{font=\normalfont\textbullet\space}

\title{Experimental nonclassicality in a causal network without assuming freedom of choice}

\author{Emanuele Polino}
\affiliation{Dipartimento di Fisica - Sapienza Universit\`{a} di Roma, P.le Aldo Moro 5, I-00185 Roma, Italy}

\author{Davide Poderini}
\affiliation{Dipartimento di Fisica - Sapienza Universit\`{a} di Roma, P.le Aldo Moro 5, I-00185 Roma, Italy}
\affiliation{International Institute of Physics, Federal University of Rio Grande do Norte, 59078-970, P. O. Box 1613, Natal, Brazil}
\author{Giovanni Rodari}
\author{Iris Agresti}
\author{Alessia Suprano}
\affiliation{Dipartimento di Fisica - Sapienza Universit\`{a} di Roma, P.le Aldo Moro 5, I-00185 Roma, Italy}

\author{Gonzalo Carvacho}
\affiliation{Dipartimento di Fisica - Sapienza Universit\`{a} di Roma, P.le Aldo Moro 5, I-00185 Roma, Italy}

\author{Elie Wolfe}
\email{ewolfe@perimeterinstitute.ca}
\affiliation{Perimeter Institute for Theoretical Physics, 31 Caroline St. N, Waterloo, Ontario, N2L 2Y5, Canada}

\author{Askery Canabarro}
\affiliation{International Institute of Physics, Federal University of Rio Grande do Norte, 59078-970, P. O. Box 1613, Natal, Brazil}
\affiliation{Grupo de F\'isica da Mat\'eria Condensada, N\'ucleo de Ci\^encias Exatas - NCEx, Campus Arapiraca, Universidade Federal de Alagoas, 57309-005, Arapiraca, AL, Brazil}

\author{George Moreno}
\affiliation{International Institute of Physics, Federal University of Rio Grande do Norte, 59078-970, P. O. Box 1613, Natal, Brazil}
\affiliation{Departamento de Computação, Universidade Federal Rural de Pernambuco, 52171-900, Recife, Pernambuco, Brazil}

\author{Giorgio Milani}
\affiliation{Dipartimento di Fisica - Sapienza Universit\`{a} di Roma, P.le Aldo Moro 5, I-00185 Roma, Italy}

\author{Robert W. Spekkens}
\affiliation{Perimeter Institute for Theoretical Physics, 31 Caroline St. N, Waterloo, Ontario, N2L 2Y5, Canada}

\author{Rafael Chaves}
\email{rafael.chaves@ufrn.br}
\affiliation{International Institute of Physics, Federal University of Rio Grande do Norte, 59078-970, P. O. Box 1613, Natal, Brazil}
\affiliation{School of Science and Technology, Federal University of Rio Grande do Norte, Natal, Brazil}

\author{Fabio Sciarrino} 
\email{fabio.sciarrino@uniroma1.it}

\affiliation{Dipartimento di Fisica - Sapienza Universit\`{a} di Roma, P.le Aldo Moro 5, I-00185 Roma, Italy}

\begin{abstract}
In a Bell experiment, it is natural to seek a causal account of correlations wherein only a common cause acts on the outcomes. For this causal structure, Bell inequality violations can be explained only if causal dependencies are modelled as intrinsically quantum. There also exists a vast landscape of causal structures beyond Bell that can witness nonclassicality, in some cases without even requiring free external inputs.
Here, we undertake a photonic experiment realizing one such example: the triangle causal network, consisting of three measurement stations pairwise connected by common causes and no external inputs. To demonstrate the nonclassicality of the data, we adapt and improve three known techniques: (i) a machine-learning-based heuristic test, (ii) a data-seeded inflation technique generating polynomial Bell-type inequalities and (iii) entropic inequalities.  The demonstrated experimental and data analysis tools are broadly applicable paving the way for future networks of growing complexity.  

\end{abstract}
\blfootnote{Corresponding authors:}
\maketitle
\section{Introduction}

Bell's theorem~\cite{bell1964einstein}, more than any other result, elucidates the manner in which quantum theory necessitates a departure from a classical worldview~\cite{Brunner,scarani2019bell}. Recently, it has been realized that it can be understood as a no-go result for providing a satisfactory account of quantum correlations using a classical causal model~\cite{wood2015lesson,fritz2016beyond,wiseman2017causarum,schmid2020unscrambling}. Under this reframing, violating a Bell inequality can be understood as attesting to the necessity of using an intrinsically quantum notion of a causal model to achieve a causal account of the correlations \cite{fritz2016beyond,chaves2015information,cavalcanti2014modifications,costa2016quantum,allen2017quantum,barrett2019quantum,wiseman2017causarum,wolfe2021quantum}, 
and thus as witnessing nonclassicality.
Furthermore, it becomes clear that such an analysis can be generalized to causal structures that are distinct from the Bell scenario ~\cite{yurke1992einstein,fritz2016beyond,henson2014theory,branciard2012bilocal,branciard2010characterizing,wolfe2019inflation,renou2019genuine,pozas2019bounding,navascues2020genuine,aaberg2020semidefinite,gisin2019entanglement,chaves2018quantum,tavakoli2021bell,gebhart2021genuine}.

Such generalizations are highly relevant to the problem of developing quantum technologies.  In the context of the Bell scenario alone, the possibility of witnessing nonclassicality has applications ranging from quantum cryptography  \cite{pirandola2019advances} to self-testing  \cite{vsupic2020self} and communication complexity problems \cite{brukner2004bell}, as well as device-independent information processing \cite{acin2007device,acin2016certified}, where the processing can be accomplished while relaxing what needs to be known about the inner workings of the devices. Given that tasks such as these are also of interest in arbitrary quantum networks~\cite{wehner2018quantum,kimble2008quantum,briegel1998quantum}, which can have complex topologies, it is evident that there is a need for new data analysis tools appropriate for witnessing nonclassicality in generic causal structures (see review in Ref.\cite{tavakoli2021bell}).
 Moreover, so far, all the demonstrations of quantum nonlocality, in the Bell scenario (Fig.1(a)) or in complex networks, relied on the use of external inputs, variables whose values can be freely chosen by the experimenter and which serve to switch between different measurement settings \cite{scheidl2010violation,weihs1998violation,Shalm,Giustina,Hensen}.
The free choice of measurements lies at the basis of Bell's theorem \cite{hooft2007free} and in experimental demonstrations, this freedom has to be assumed, or at best made be as plausible as possible \cite{big2018challenging,rauch2018cosmic}. 
By contrast, quantum networks with several independent sources allow the demonstration of nonclassicality without the need for external freely chosen inputs, replacing the freedom of choice assumption with the assumption of independence of the sources \cite{fritz2016beyond,renou2019genuine,abiuso2022single,chaves2021causal,Boreiri2022towards}.

\begin{figure*}[t]
  \begin{center}
    \subfigure[\label{fig:Bell} \textbf{Bell Causal Structure}]
    {\centering
      \begin{minipage}[t]{0.30\textwidth}
      \centering\includegraphics[scale=0.8]{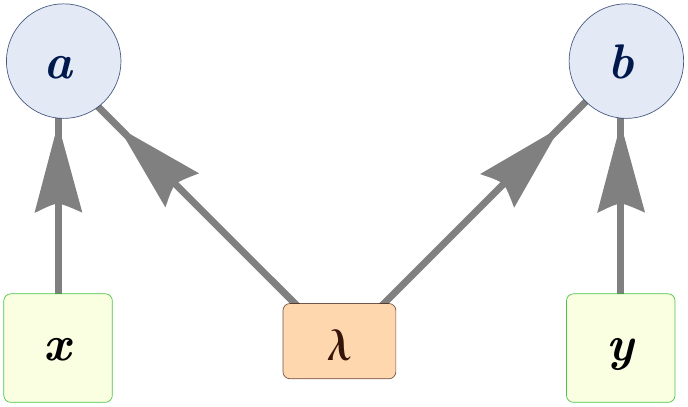}\end{minipage}
    }
   \hfill
       \subfigure[\label{fig:Tri} \textbf{Triangle Causal Structure}]
    {\centering
      \begin{minipage}[t]{0.3\textwidth}
      \centering\includegraphics[scale=0.8]{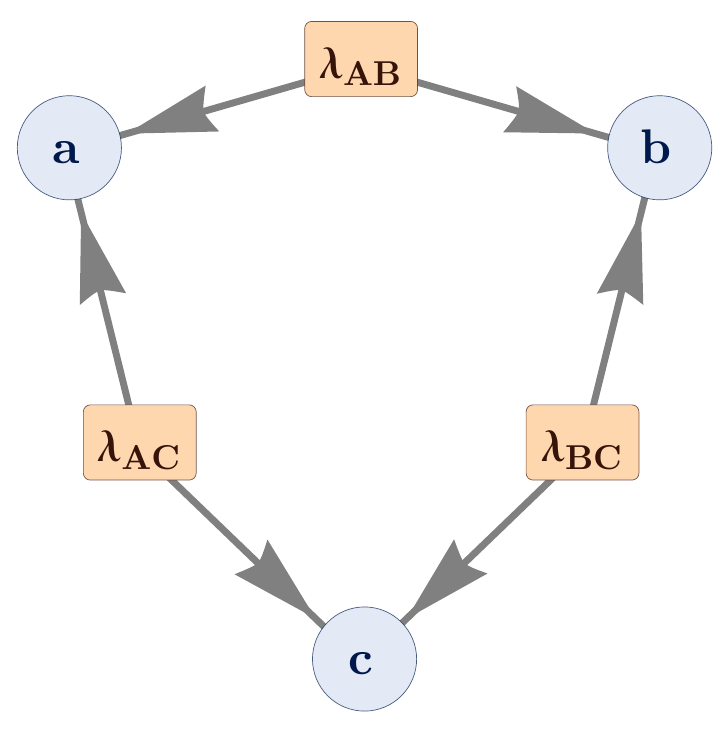}\end{minipage}
    }
    \hfill
    \subfigure[\label{fig:GHZ} \textbf{GHZ Causal Structure}]
    {\centering
      \begin{minipage}[t]{0.3\textwidth}
      \centering\includegraphics[scale=0.8]{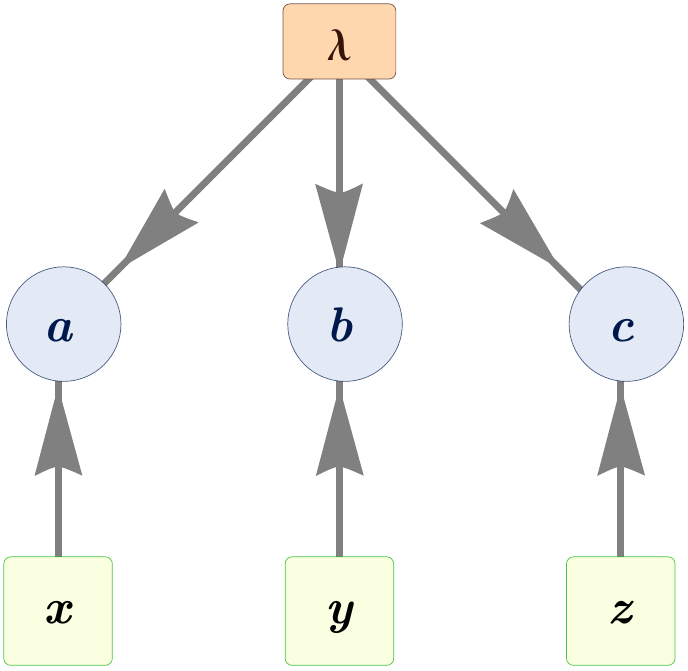}\end{minipage}
    }
  \end{center}
  \caption[]{\textbf{
  Directed acyclic graph (DAG) representation of different classical causal scenarios.}
  \subref{fig:Bell}~The Bell scenario is a causal structure in which a source $\lambda$ correlates the two parties having measurement outcomes $a$ and $b$ and choices $x$ and $y$, respectively.
  \subref{fig:Tri}~The triangle scenario involves three independent sources $\lambda_{AB}$,  $\lambda_{BC}$ and $\lambda_{AC}$ which establish correlations between pairwise stations $A$, $B$ and $C$. Note that measurements in the triangle scenario do not depend on external inputs. 
  \subref{fig:GHZ}~The tripartite Bell scenario is also known as the GHZ scenario, after the theorists who identified a nonlocal game for this scenario which quantum theory predicts can be won with 100\% probability.
  \label{fig:dags}
  }
\end{figure*}

In spite of its significance, this challenge remains largely unexplored, especially from the experimental perspective.
This work is a contribution to this effort. We undertake the experimental investigation of a causal structure that has attracted growing attention \cite{branciard2012bilocal,chaves2014causal,henson2014theory,steudel2015information,fritz2016beyond,wolfe2019inflation,fraser2018causal,renou2019genuine,aaberg2020semidefinite,gisin2019entanglement,pusey2019quantum,kraft2021quantum,vsupic2020quantum,krivachy2019neural,renou2019limits,baumer2021demonstrating,abiuso2022single,Sekatski2022partial}: the ``triangle scenario'', depicted in Fig.~\ref{fig:Tri}.  Here, three distant parties each receives a share from two out of three independent sources, and in stark contrast to the Bell scenario, each party implements a single measurement on the systems in its lab, rather than having the freedom to choose among a set of incompatible measurements.

Using a versatile photonic setup with three independent sources (one sharing entanglement and two sharing classical correlations) and the feedforward of classical information by means of fast optical switches, we provide the first experimental demonstration of classically unrealizable correlations in the triangle structure without the use of external inputs.
Importantly, witnessing nonclassicality in this new kind of causal structure goes beyond the standard Bell inequality violation and requires a radically different approach. In the course of doing so, we have enhanced some of the existing tools for testing nonclassicality in generic causal structures both from the experimental and the theoretical perspectives.  
These enhancements are in the service of making the tools applicable   to generic causal structures and arbitrary data, thus paving the way for future experiments in causal networks of growing size and complexity.

\section{Results}

\subsection{Beyond Bell's theorem}\label{BeyondBell}

Leveraging Bell's theorem, Fritz \cite{fritz2016beyond} showed the existence of a distribution in the triangle scenario that is realizable quantumly but not classically.
Fritz's result is best understood as a quantum no-go theorem akin to Bell's 1964 no-go theorem \cite{bell1964einstein} or the tripartite Greenberger-Horn-Zeilinger (GHZ) argument \cite{greenberger1989going}.  As with the distributions described in those works, Fritz's distribution has the feature that certain variables are  perfectly correlated, something that is predicted by quantum theory to be possible in principle, but which can never be realized in a real experiment given the unavoidable presence of noise.  

It was Clauser, Horne, Shimony, and Holt (CHSH) who first demonstrated how to turn Bell's argument into an experimental test, by deriving noise-robust inequalities~\cite{clauser1969proposed}.
Similarly, in the tripartite Bell scenario  (Fig.~\ref{fig:dags}\subref{fig:GHZ}), the step from the GHZ argument to the possibility of a noise-robust test  was achieved by Mermin's inequality \cite{mermin1990}.
In the case of the triangle scenario, classical causal compatibility inequalities have also been derived \cite{fraser2018causal} but these unfortunately require a degree of sensitivity higher than can reasonably be achieved in current experimental tests. Note that the inequalities derived in Ref.~\cite{vsupic2020quantum}, by contrast, are not noise-robust because they apply only to distributions exhibiting  perfect correlations between certain variables, analogously to Bell's 1964 inequality.  New techniques are therefore required to witness nonclassicality  in the triangle scenario for the sort of experimental data achievable at present.

Developing new data analysis techniques is also motivated by considerations of utility. If all one seeks to do is to demonstrate the existence of nonclassicality in a given causal structure, then it is clearly sufficient to implement a dedicated experiment that targets a specific distribution  and to test an inequality that is known to be able to witness nonclassicality  for the targeted distribution.
If, on the other hand, one seeks to use nonclassicality in a given causal structure as a resource for various information-processing tasks, then it is clearly of greater utility to have a test that is able to witness nonclassicality for  any distribution that is not classically realizable in the given causal structure.

In some cases, this higher bar 
can  be met by determining all of the classical causal compatibility inequalities associated to a given causal structure and testing for violations of any of these \cite{Brunner}. Unfortunately, however, such a complete characterization soon becomes out of reach, even for seemingly simple scenarios \cite{geiger1999quantifier,Brunner}. In order to be able to witness nonclassicality on arbitrary data, therefore, it is better to seek
a ``satisfiability'' algorithm, which takes as its input a concrete example of data, and answers the question of classical realizability for that data alone, and in the case of a negative answer,  identifies an inequality that is optimized for witnessing its nonclassicality.

We here propose a data-seeded algorithm of this sort that can be used for a generic causal structure. 
This is achieved by leveraging the fact that the inflation technique for causal inference \cite{wolfe2019inflation} can reduce the satisfiability problem to a linear program. We also pursue a second route to witnessing nonclassicality on generic data. In this approach, one foregoes  deriving inequalities altogether and one simply performs a statistical hypothesis test where the hypothesis is the compatibility of the data with a classical causal model for the given causal structure. Specifically, one implements a variation of the parameters of the model---some of which make explicit reference to the hidden (i.e., unobserved) variables---to try and find the best fit to the data, and one considers the hypothesis falsified at some level of confidence when no good fit can be found.
We here show that such hypothesis testing on experimental data can be made feasible for causal networks using the machine learning technique developed in \cite{krivachy2019neural} where the topology of the causal network is mapped to the topology of a neural network.
Finally, suitably mapping the triangle network to a generalization of Bell's scenario that incorporates the possibility of measurement dependence (i.e., that abandons the free choice assumption), we also witness the nonclassicality of the data by using an entropic approach, recently introduced in \cite{chaves2021causal}.  

Note that for the triangle scenario, our goal is to witness nonclassicality of the experimentally realized distribution assuming  only that the causal relations among the three measurement nodes and sources are those described by the triangle scenario. 
If one were to avail oneself of additional assumptions, in particular, assumptions regarding the causal relations among variables within a given laboratory, then one could witness nonclassicality of our experimental data using standard Bell inequalities.
Since such additional assumptions do not hold for all setups that can realize a distribution exhibiting a quantum-classical gap, an analysis which leveraged these additional causal assumptions would not achieve the goal of being applicable to arbitrary data.

\begin{figure*}[t!]
\begin{center}
\includegraphics[
      width=0.9\textwidth]{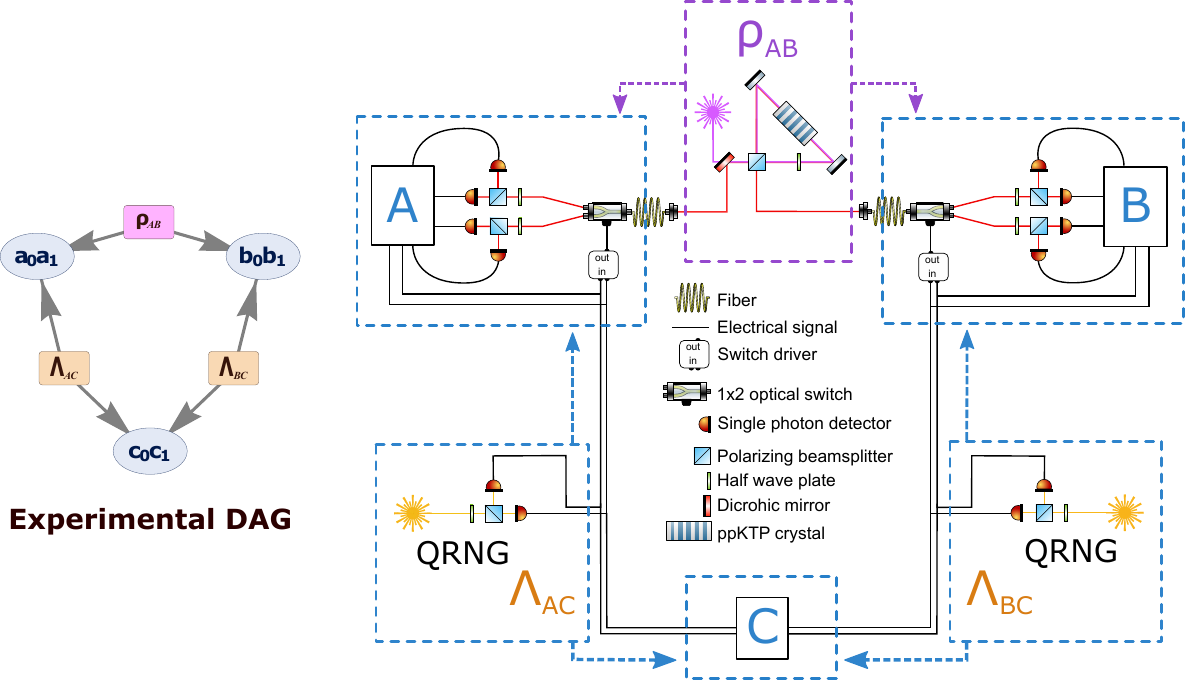}
\end{center}
\caption{\label{fig:appNEW}
\textbf{Experimental implementation of the triangle network.} The source $\rho_{AB}$ generates  polarization-entangled photon pairs in the singlet state $\ket{\Psi^-}$, by pumping with a continuous wave UV laser a periodically poled potassium titanyl phosphate (ppKTP) crystal. Conversely, $\Lambda_{AC}$ ($\Lambda_{BC}$) produces classically correlated states $(\dyad{00} + \dyad{11})/2$ obtained by splitting the output signal of two single-photon avalanche photodiodes subjected to environmental light noise. In nodes A and B, to implement the measurements needed to reconstruct the probability distribution $p(a,b,c)$, the photons from the source $\rho_{AB}$ are collected by the input single-mode fiber (SMF) of a 5ns rise-time optical switch. In Fritz-like distributions, the measurement result $a_0$ ($b_0$) on part of the source $\Lambda_{AC}$ ($\Lambda_{BC}$) determines the observable to be measured on the photon coming from $\rho_{AB}$, leading to outcomes $a_1$ ($b_1$). In our implementation, this is achieved by appropriately driving the optical switches through a specially designed electronic driver which receives signals coming from $\Lambda_{AC}$ ($\Lambda_{BC}$) and drives the output port of the optical switch based on the results $a_0$ ($b_0$). The bit $a_1$ ($b_1$) is obtained by performing a polarization measurement on the photons produced by the ppKTP source through a half-waveplate (HWP) and a polarizing beam splitter (PBS), implemented in fiber. In node $C$, $c_0$ and $c_1$ are measured independently by directly feeding the electrical signals produced by $\Lambda_{AC}$ and $\Lambda_{BC}$ into a time to digital converter (TDC).}
\end{figure*}

\subsection{The causal modeling perspective on Bell's theorem}
Bell's theorem can be seen as a particular instance of a causal inference problem where for a given hypothesis about the causal structure of the experiment, one inquires
whether a classical causal model is able to reproduce the observations \cite{wood2015lesson, wiseman2017causarum}.
In a Bell experiment, a source distributes physical systems between two distant observers --Alice and Bob--,
they choose the values of their setting variables, denoted by $x$ and $y$ respectively (these determine which of a set of incompatible measurements is implemented at each lab), and then they register the outcomes, denoted by $a$ and $b$ respectively.  For simplicity here, we represent the variables and their values with the same letter.
The natural causal structure to hypothesize in such an  experiment is the one depicted in Fig.~\ref{fig:Bell}, termed the ``Bell scenario''.

The assumption of a classical causal model implies that the observed distribution can be decomposed as
\begin{equation}
\label{eq:lhv}
p(a,b \vert x,y)= \sum_{\lambda} p(\lambda)p(a\vert x,\lambda)p(b\vert y,\lambda).
\end{equation}
This decomposition is familiar in discussions of Bell's theorem as what follows from assuming a hidden variable model satisfying local causality and certain other conditions \cite{hall2011relaxed,chaves2015unifying}, but it can also be understood as a simple consequence of the causal Markov condition \cite{pearl2009causality} under the assumption that the causal structure is that of the Bell scenario~\cite{wood2015lesson,wiseman2017causarum}.

In turn, for a quantum causal model, sources of correlations are not copies of a variable $\lambda$ that is probabilistically distributed but rather pairs of systems that are in a joint quantum state $\rho$ (potentially entangled).
Similarly, dependencies among nodes are not represented by conditional probabilities such as $p(a\vert x,\lambda)$ but by the quantum analogues thereof, completely positive and trace preserving (CPTP) maps, which, in the particular case of a measurement, correspond to a positive operator-valued measure (POVM). Operationally, the quantum description is given by Born's rule, implying that

\begin{equation}
p_{\mathrm{Q}}(a,b \vert x,y)= \mathrm{Tr} \left[ \left( M^A_{a|x} \otimes M^B_{b|y} \right)\rho_{AB} \right],
\end{equation}
where $\{ M^A_{a|x}\}_a$ and  $\{ M^B_{b|y}\}_b$ are POVMs on $A$ and $B$ respectively.

Bell's theorem \cite{bell1964einstein} asserts that the quantum description can lead to an observable distribution that fails to have a classical explanation in terms of the causal model \eqref{eq:lhv}.

\subsection{The triangle scenario}
Among the simplest quantum networks beyond the paradigmatic Bell causal structure is the triangle scenario of Fig.~\ref{fig:Tri}. It is distinguished from the tripartite Bell scenario (depicted in Fig.~\ref{fig:GHZ}) by the fact that the distant parties are not connected by a 3-way source, but by three 2-way sources. 

In the triangle scenario, the correlations that admit a classical realization, i.e., those that are compatible with a classical causal model with the structure of Fig.~\ref{fig:Tri}, can be written as:
\begin{align}\begin{split}
\label{eq:trianglec}
p(a,b,c)
&=\smashoperator[r]{\sum_{\lambda_{AB},\lambda_{BC},\lambda_{AC}}}\quad p(\lambda_{AB})\,p(\lambda_{BC})\,p(\lambda_{AC})\\
&p(a\vert \lambda_{AB},\lambda_{AC})\;p(b\vert \lambda_{AB},\lambda_{BC})\;p(c\vert \lambda_{AC},\lambda_{BC}).
\end{split}\end{align}

By contrast, the correlations which admit of a quantum realization in the triangle network are given by
\begin{align}\label{qexpress}
&p_{\mathrm{Q}}(a,b,c)=
\\\nonumber \mathrm{Tr}&\left({\rho_{A B}} \otimes {\rho_{A C}} \otimes {\rho_{B C}}\,\cdot \,  {M^{A}_a} \otimes {M^{B}_b} \otimes {M^{C}_c} \right)\;,
\end{align}
where $\rho_{A B}$ denotes the density operator of the state shared between the nodes $\{A,B\}$ (likewise for $\rho_{A C}$ and $\rho_{B C}$), while $\{M^{A}_a\}_a$ denotes a POVM on the subsystem in station $A$ (similarly for $\{M^{B}_b\}_b$ and $\{M^{C}_c\}_c$). 

Recently, it has been theoretically and experimentally demonstrated that a quantum triangle network with a setting variable at each station can give rise to nonclassical correlations \cite{suprano2022experimental}.
This result, however, employs measurement choices for each of the observers. Here, we go a significant step beyond, showing that nonclassical correlations can emerge even without any freedom of choice.

\subsection{The Fritz distribution}

In Fritz's example~\cite{fritz2012beyond} of a distribution $p_{\mathrm{Q}}(a,b,c)$ that is not classically realizable, $a$, $b$ and $c$ are 4-valued variables, each of which is conceptualized as a pair of binary variables, $a=(a_0,a_1)$, $b=(b_0,b_1)$ and $c=(c_0,c_1)$. Moreover, one can  decompose the quantum system $A$ as $A=(A_0,A_1)$, where $A_0$ is the subsystem  appearing in $\rho_{AC}$ and $A_1$ is the subsystem  appearing in $\rho_{AB}$; analogously for $B=(B_0,B_1)$ and $C=(C_0,C_1)$.
The example is realized by taking the three POVMs in Eq.~\eqref{qexpress} to have the following form:
\begin{align}\begin{split}\label{eq:FritzPOVMs}
& M^{C_0 C_1}_{(c_0,c_1)}=M^{C_0}_{c_0}\otimes M^{C_1}_{c_1}, \\
&
M^{A_0 A_1}_{(a_0,a_1)}=M^{A_0}_{a_0}\otimes M^{A_1}_{a_1|a_0},
\\
&
M^{B_0 B_1}_{(b_0,b_1)}=M^{B_0}_{b_0}\otimes M^{B_1}_{b_1|b_0},
\end{split}
\end{align}
where $\{ M^{C_0}_{c_0}\}_{c_0},$ $\{M^{C_1}_{c_1}\}_{c_1},$ $\{M^{A_0}_{a_0}\}_{a_0},$ $\{M^{B_0}_{b_0}\}_{b_0}$ are all 
measurements 
of the $\sigma_z$ Pauli observable, $ \{ M^{A_1}_{a_1|a_0}\}_{a_1}$ corresponds to one of the two Pauli observables among $\{\sigma_x, \sigma_z\}$ depending on the value of $a_0$, and $\{ M^{B_1}_{b_1|b_0} \}_{b_1}$ corresponds to one of the two observables among  $\{{(\sigma_x+\sigma_z)/\sqrt{2}}, {(\sigma_x-\sigma_z)/\sqrt{2}}\}$  depending on the value of $b_0$. In Fritz's description of a genuinely quantum distribution in the triangle scenario, the state $\rho_{AB}$
is taken to be, for example, a singlet state $\ket{\Psi^-} = (\ket{01} - \ket{10})/\sqrt{2}$; while $\rho_{AC}$ and $\rho_{BC}$ are
maximally entangled states $(\ket{00} + \ket{11})/\sqrt{2}$. 
However, since all the measurements on $\rho_{AC}$ and $\rho_{BC}$ are of $\sigma_z$, it is sufficient to take these to be a classically correlated state, namely:
\begin{equation}
    \Lambda_{AC} = \Lambda_{BC} =
    (\dyad{00} + \dyad{11})/2.
    \label{eqn:Cstate}
\end{equation}

As noted in Ref.~\cite{fritz2016beyond}, to see that Fritz's distribution is not classically realizable, it suffices to make a connection to a Bell scenario between Alice and Bob.  Note that the variables $a_0$ and $b_0$ determine the measurements that are implemented on $A_1$ and $B_1$.  In this respect, they are akin to measurement settings $x$ and $y$ in the usual scenario.  However, because $a_0$ and $b_0$ are outputs in the triangle scenario, they could in principle depend on the common source between Alice and Bob. In the usual Bell scenario, of course, if the setting variable $x$ (or $y$) is correlated with $\lambda_{AB}$, one cannot derive the Bell inequalities.  
The assumption that $x$ and $y$ are not correlated with $\lambda_{AB}$ is termed measurement independence (or freedom of choice) and is a consequence of the hypothesis that the causal structure for the usual Bell scenario is that of Fig.~\ref{fig:Bell}.

For the Fritz distribution in the triangle scenario, one can still infer that $a_0$ and $\lambda_{AB}$ are uncorrelated, but now this follows from the fact that $a_0$ is perfectly correlated with the outcome $c_0$, which is causally disconnected from $\lambda_{AB}$.  Similarly, the lack of correlation between $b_0$ and $\lambda_{AB}$ is inferred from the perfect correlation between $b_0$ and $c_1$ and the fact that $c_1$ is causally disconnected from $\lambda_{AB}$.  
If one considers the conditional distribution $p(a_1,b_1 \vert a_0,b_0)$ that is obtained by making the appropriate Bayesian inversion on a distribution $p(a,b,c)$ that is classically realizable in the triangle scenario, then  given the independence of $a_0$ (and $b_0$) from $\lambda_{AB}$, this conditional distribution should satisfy the standard Bell inequalities.  The fact that the measurements in Fritz's example have been chosen to ensure that the conditional $p_{Q}(a_1,b_1 \vert a_0,b_0)$   violates a standard Bell inequality implies that the distribution $p_{Q}(a ,b,c )$ is not classically realizable in the triangle scenario.

Any experiment that aims to realize the Fritz distribution in the triangle scenario has the goal of realizing the ideal states and measurements specified above, but due to the inevitability of noise, the states and measurements that are actually implemented are necessarily noisy versions of these. This implies that the correlations between $a_0$ and $c_0$ and between $b_0$ and $c_1$ will  not be perfect, which in turn blocks the inference from the classical realizability of $p(a,b,c)$ in the triangle scenario to the classical realizability of $p(a_1,b_1|a_0,b_0)$ in the standard Bell scenario. 
As such, to witness nonclassicality in such an experiment, one must go beyond the techniques that witness nonclassicality in a standard Bell experiment. 

It is worth reiterating here a point made in the beginning of subsection "Beyond Bell's theorem", that our goal is to witness nonclassicality using a data analysis technique that assumes  only the causal structure of the triangle scenario.  If we associate a laboratory with each of the nodes in the causal structure, then even though our particular experiment involves specific causal relations between systems  within the laboratories, the data analysis cannot make use of this extra structure.  In other words, we seek a data analysis technique that can witness nonclassicality without assuming any such extra structure. 
This is the sort of assumption that is appropriate for the device-independent paradigm, wherein the experimental devices are presumed to be supplied by an adversary. All that is presumed to be guaranteed is that the causal relations among the laboratories are the ones specified by the triangle scenario.  If one could avail oneself of the extra structure that is present in the experiment but not part of the description of the triangle scenario, then standard Bell inequalities would be sufficient to witness nonclassicality. For instance, if one could assume that Alice's output $a_0$ was a faithful copy of the classical randomness she shares with Charlie and that Bob's output $b_0$ was a faithful copy of the classical randomness he shares with Charlie, then one could infer that neither $a_0$ nor $b_0$ could depend on $\Lambda_{AB}$ and consequently having $p(a_1, b_1|a_0,b_0)$ violate a Bell inequality would be sufficient to witness nonclassicality.  As a second example, if one could assume that the pair of variables $c_0$ and $c_1$ that are outputs of Charlie's laboratory are such that $c_0$ depends only on the source shared with Alice and $c_1$ depends only on the source shared with Bob, then the causal structure being assumed is equivalent to a 4-party line-like structure rather than a triangle scenario.  In this case, the full set of Bell inequalities for the conditional distribution $p(a,b|c_0,c_1)$ (where $a=(a_0,a_1)$ and $b=(b_0,b_1)$) are the necessary and sufficient conditions for classicality~\cite{evans2016margins}.

In order to be able to witness the nonclassicality of our data assuming only the triangle causal structure, therefore, we cannot rely on standard Bell inequalities.
This is why we must have recourse to new data-analysis techniques, such as those presented in subsections "Bounding measurement dependence and violating an entropic inequality for the triangle network", "Violation of a causal compatibility inequality" and "Bounding measurement dependence and violating an entropic inequality for the triangle network".

\begin{figure*}
  \begin{center}
    \subfigure[\label{fig:fritz_theo} \textbf{Ideal Fritz distribution} ]
    {\centering
      \centering\includegraphics[width=0.495\textwidth]{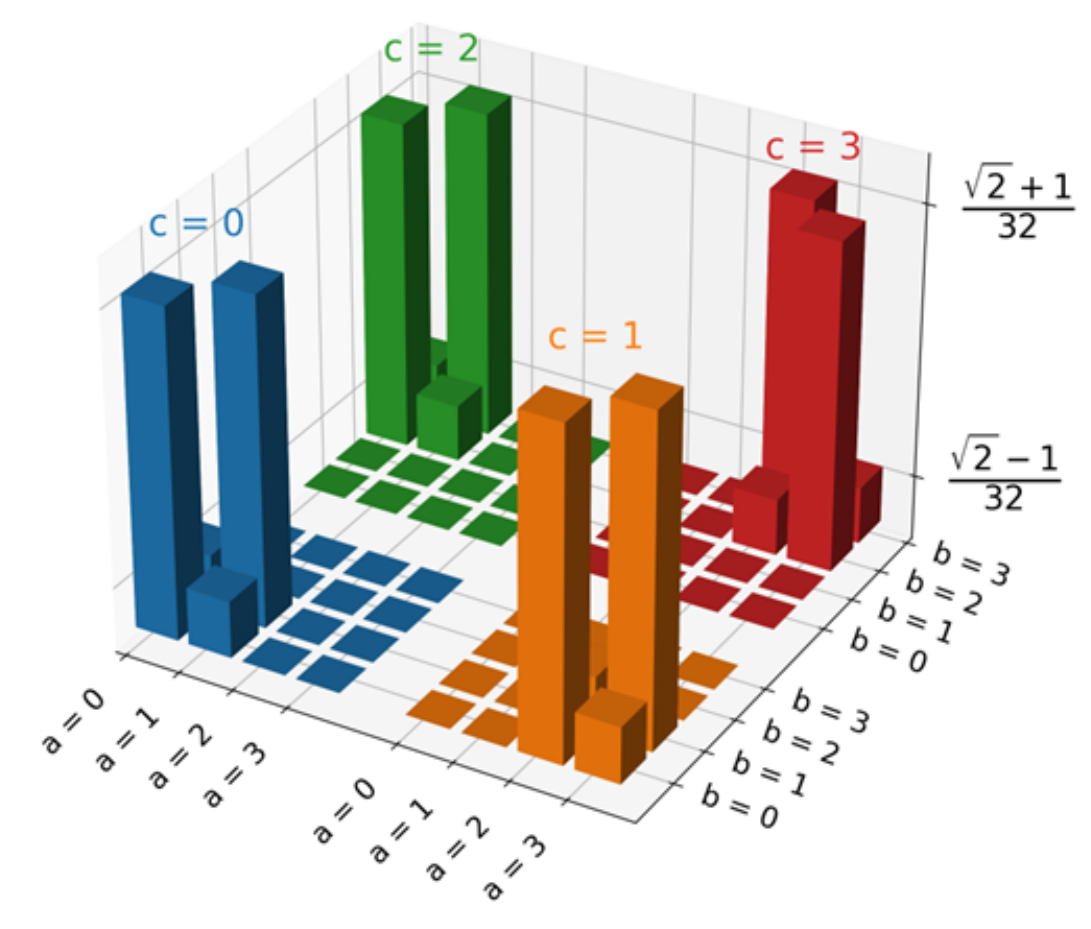}
    }
    \hfill
    \subfigure[\label{fig:fritz_expdata} \textbf{Experimental Fritz distribution}]
    {\centering\includegraphics[width=0.495\textwidth]{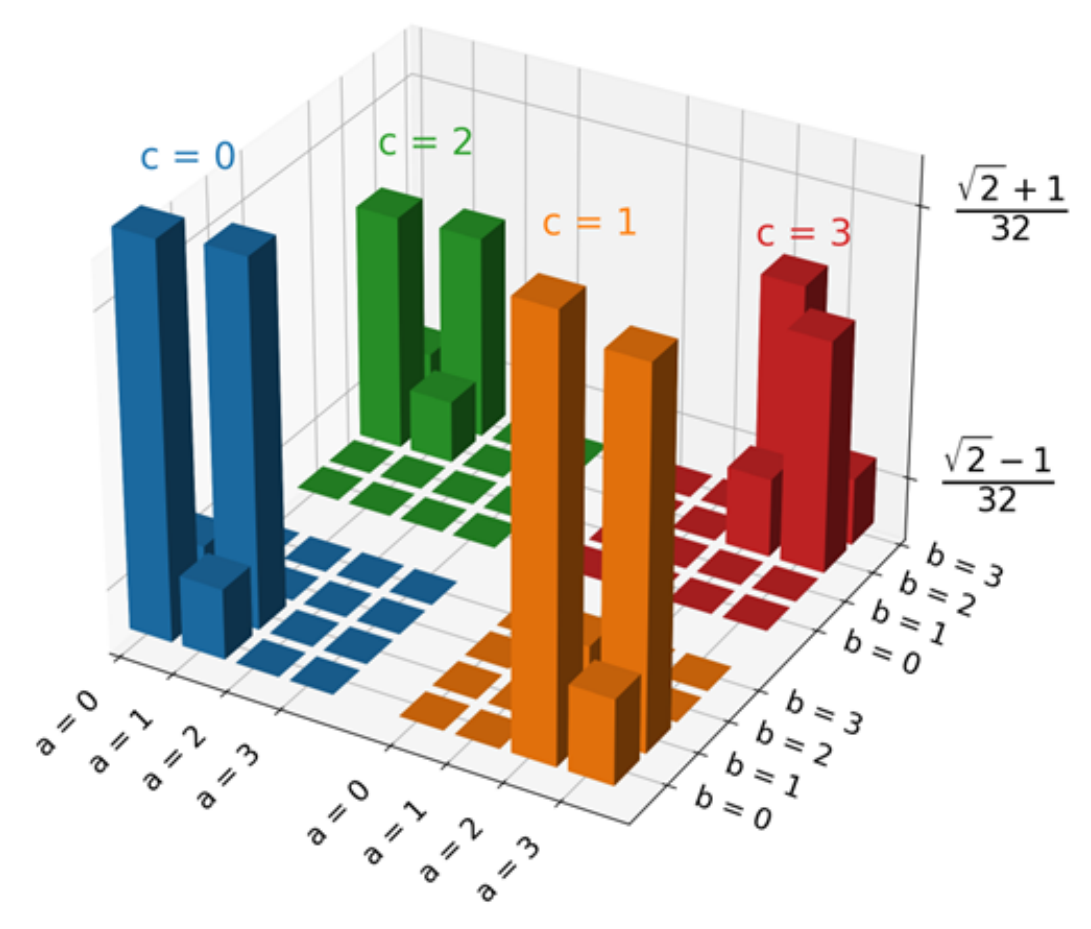}
    }
  \end{center}
\centering
\caption[]{\textbf{The Fritz distribution, theoretical versus experimental.}
\subref{fig:fritz_theo} Ideal Fritz distribution  computed by choosing $\rho_{AB} = (\ket{HV} - \ket{VH})/\sqrt 2$ (a noiseless singlet state); $\Lambda_{AC} = \Lambda_{BC}$ as classically, perfectly correlated mixed states; and the ideal measurement operators described in Eq.~\eqref{eq:FritzPOVMs}.
\subref{fig:fritz_expdata} Experimental distribution measured in an experimental run. The error bars are calculated using Poissonian statistics and are not visible in the plot. The three indexes $a$, $b$, and $c$ indicate the measurement results, ranging from 0 to 3, corresponding to the three nodes $A$, $B$ and $C$, respectively. The chart bars representing the terms of the probability distribution have different colors based on the value of the outcome $c$.
}
\label{fig:fritz_}
\end{figure*}

\subsection{Experimental Setup}

In our experimental implementation, we used the polarization degrees of freedom of a pair of photons as the two qubits distributed by the source shared between $A$ and $B$, with the $\sigma_z$ eigenstates corresponding to the $\{ \ket{H},\ket{V}\}$ basis of linear polarization. We investigated quantum correlations arising in the triangle network 
where we aim to have the source between $A$ and $B$ prepare the singlet state.
Meanwhile, for the source shared by $A$ and $C$ and the source shared by $B$ and $C$, we aim to have these prepare the classically correlated state of Eq.~\eqref{eqn:Cstate}.

Recent years have seen the first experimental implementations of causal structures with a number of independent sources~\cite{sun2019experimental,poderini2020experimental,suprano2022experimental,carvacho2022quantum}.
In our implementation, the pair of photons associated to the source between $A$ and $B$ are at a wavelength of $810$nm, and are generated through spontaneous parametric down-conversion in a ppKTP nonlinear crystal pumped with a $405$nm UV CW-laser, placed inside a Sagnac interferometric geometry \cite{kim2006phase,fedrizzi2007wavelength}, depicted in the box labelled $\rho_{AB}$ in Fig.~\ref{fig:appNEW}.
To implement the classically correlated sources $\Lambda_{AC}$ and $\Lambda_{BC}$, electrical pulses randomly generated by the shot-noise of distant pairs of single-photon detectors are locally split (boxes labelled $\Lambda_{AC}$ and $\Lambda_{BC}$ in Fig.~\ref{fig:appNEW}); then they are sent to the stations $A$, $C$  and $B$, $C$, respectively,
by means of $20$m-long electrical cables. Detection of such signals gives values for the bits $a_0$, $b_0$, $c_0$, $c_1$.  

Note that this electrical signal sets up classical correlations (i.e., shared randomness) between Charlie and Alice (Bob), and this is a faithful implementation of the state in Eq.(6).

Due to the probabilistic nature of photon generation and random shot-noise events from detectors, justifying the independence of different sources turns out to be very demanding.  This is the reason why the first experimental realization of quantum networks \cite{carvacho2017experimental,saunders2017experimental,big2018challenging} actually involved a single laser source, thereby requiring a device-dependent justification for the supposed independence of the generated quantum states that relies on the knowledge of the inner process of photon generation. Using spatially separated non-synchronized sources, of different natures, enforces the independence of the sources, also having direct applications in quantum communication protocols. Note, however, that the independence of the sources still remains an assumption, considering that this assumption can always be violated by superdeterministic models\cite{hossenfelder2020rethinking}.

To experimentally achieve the implementation of the separable measurement operators as in Eq.~\eqref{eq:FritzPOVMs}, the electrical signals arriving at $A$ and $B$ determine the state of ultra-fast optical switches (Nano Speed Ultra-Fast 1x2 by company Photonwares with a switching time equal to $\sim$ 8ns) that affect the measurements on the photons coming from $\rho_{AB}$. More specifically, based on which one of the two signals arrives in $A$ ($B$) from $\Lambda_{AC}$ ($\Lambda_{BC}$), the switch  will send  the photon from $\rho_{AB}$ to two fibers connected to the measurement setups implementing the different polarization measurements. The measurement of the photons is performed by polarization controllers defining the measurement basis followed by in-fiber polarizing beam splitters (PBS) and single photon detectors.
Finally, the four detectors in $A$ ($B$) are electronically connected to a time-to-digital converter, located in the measurement station. The signal from the photon counting, together with the signal from source $\Lambda_{AC}$ ($\Lambda_{BC}$) generate the 
4-valued outcome $a$ ($b$).  Conversely, in station $C$ the 
4-valued outcome $c$ is given by the two classical signals  from $\Lambda_{AC}$ and $\Lambda_{BC}$.
Note that the electronic signals generated by the detectors are sent to three separated time-to-digital converters, one for each measurement station $A$, $B$, $C$, and the recorded events are sent for data processing to a computer located outside the laboratory.

We record experimental events by first choosing a small window $w_1 \sim 4.1$ns, to filter in the  signals produced simultaneously from the same source $\Lambda_i$. This allows us to account mostly for 2-fold events which are due to the same entangled pair, or the same split signal, thus filtering out most of the experimental noise due to the detectors' dark counts and residual environmental light.
The 6-fold coincidence events are finally computed by employing a time window equal to $w_2 \sim 20 \mu$s inside which an event is defined by the arrival of three two-fold coincidences (see Supplementary Note $1$ for more details on data analysis). Such a choice of value for the 6-fold coincidence window represents a compromise between two different requirements. On one side, we want to make such a window as narrow as possible to approximately achieve simultaneity, with respect to both the generation and the measurements, which in principle could lead to an implementation directly addressing the locality loophole. On the other, a broader window is necessary to detect a large enough number of 6-fold coincidences, enhancing the events' rate and thus leading to sufficiently small errors on the measured probabilities in smaller measurement times.

In this demonstration, we do not attempt to achieve space-like separation between the registration of the outcomes $a$, $b$ and $c$.  Achieving such a separation would provide the strongest possible justification for the lack of causal influences between the outcomes $a$, $b$ and $c$.  It is important to note, however, that it would still  not justify the lack of a 3-way common cause. 

Furthermore, due to the low efficiencies of the single photon detectors ($\eta \sim 0.5$) and the fact that the threshold values required for closing the detector loophole in the triangle scenario are not yet known, we rely on the fair-sampling assumption. 
On this point, we note
that even for the much simpler case of the Bell scenario, closing the detector loophole required decades of effort.

\begin{figure*}[t]
  \begin{center}
    \subfigure[\label{fig:DLA} \textbf{Deep learning illustration.}]
    {\centering\includegraphics[scale = 0.88]{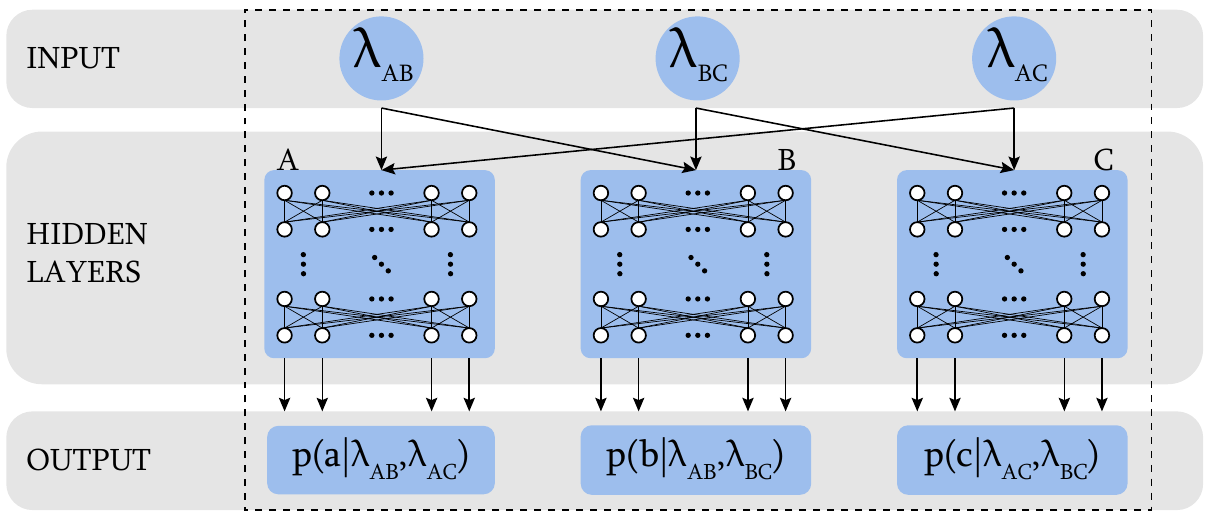}    }
    \subfigure[\label{fig:DLR} \textbf{ML nonclassicality detection.}]
    {\centering\includegraphics[scale = 0.4625]{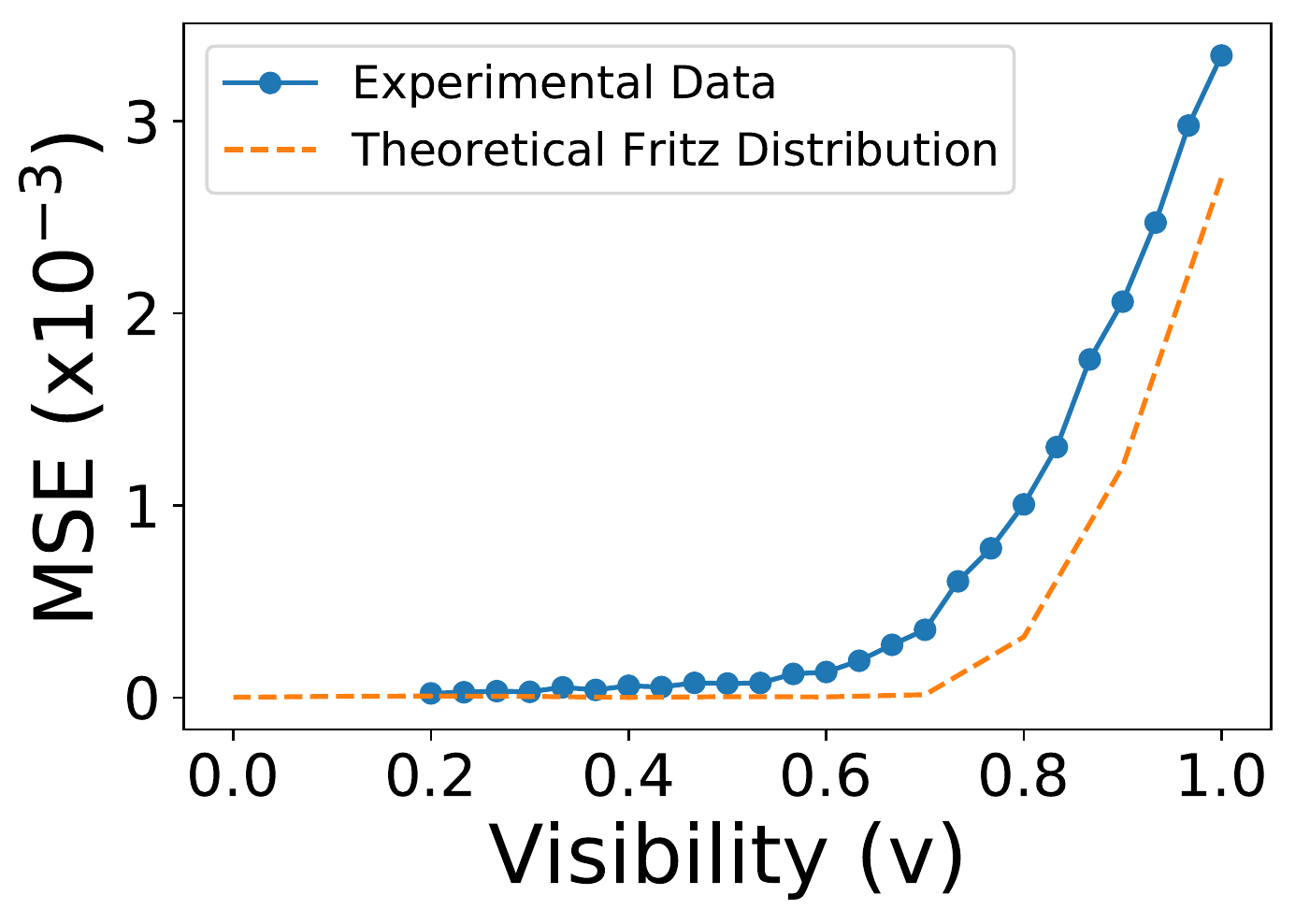}    }
  \end{center}
\caption{ \textbf{Neural network for the triangle network.} \subref{fig:DLA}~Neural Network capable of reproducing distributions compatible with the triangle configuration, where number of layers varies from $3$ to $6$ and number of neurons is either $16$ or $32$, yielding $8$ distinct architectures.  The three sources $\lambda_{AB}$, $\lambda_{BC}$ and $\lambda_{AC}$ send information to three parties, Alice, Bob, and Charlie, each receiving, respectively, the pairs $ \{ \lambda_{AB}, \lambda_{AC} \}$, $ \{ \lambda_{AB}, \lambda_{BC} \}$ and $ \{ \lambda_{AC}, \lambda_{BC} \}$. \subref{fig:DLR}~The minimum mean square error (\textsf{MSE}) distance achieved by the machine as function of the visibility for the experimental data (solid line) and the comparison with the same distance for theoretical Fritz distribution (dashed line). For distinct visibility values, a different ML architecture is the optimum one, strengthening the advantage of using an assembly of oracles. See \hyperref[{sec:Methods}]{Methods}
and Supplementary Note $3$ for specific details.   
}
\label{fig:DL}
\end{figure*}

\begin{figure*}
  \begin{center}
    \subfigure[\label{fig:inflation} \textbf{Triangle network second-order inflation} ]
    {\centering
      \centering\includegraphics[width=0.40\textwidth]{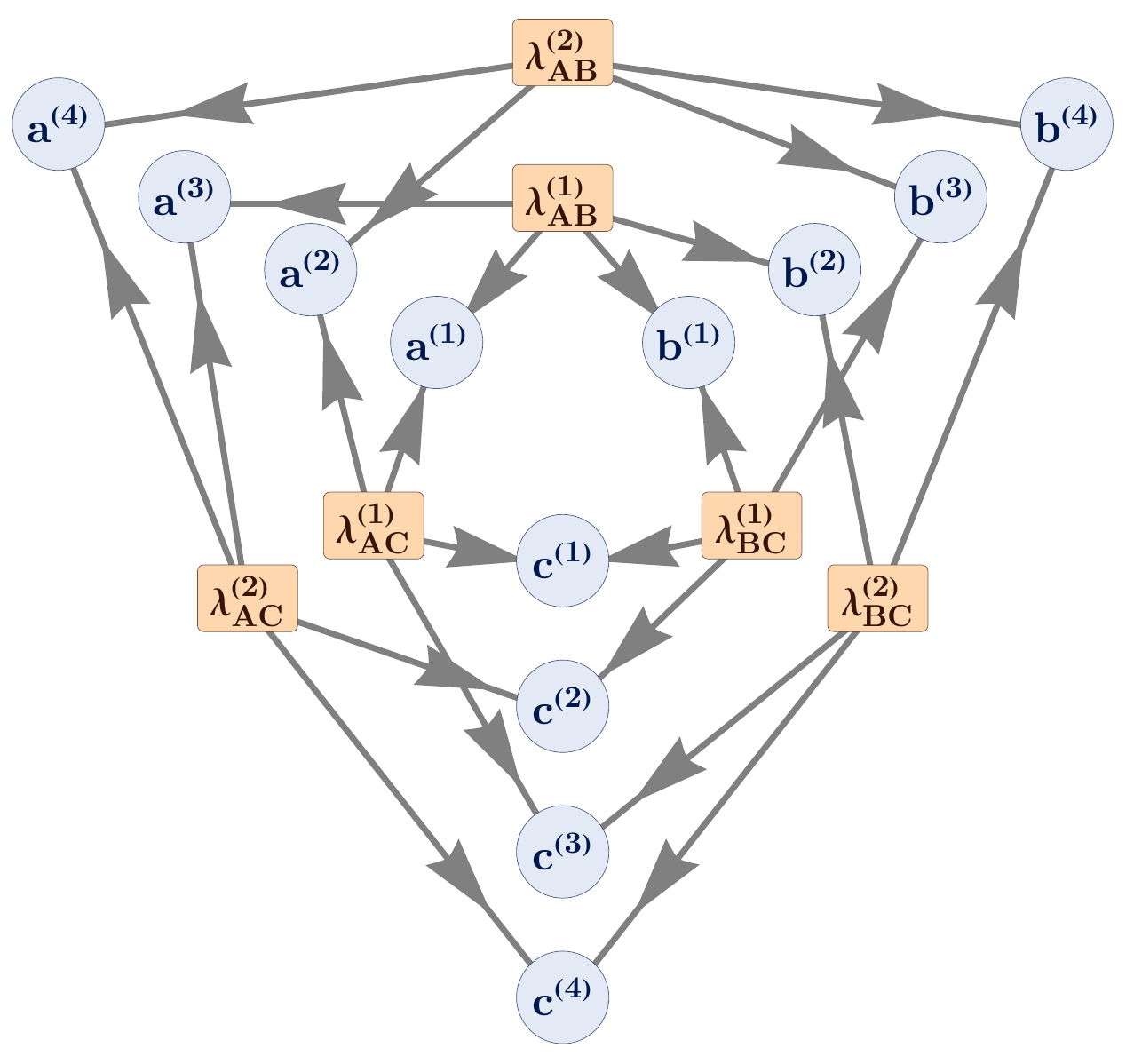}
    }
    \hfill
    \subfigure[\label{fig:ineq} \textbf{The coefficients of a quadratic inequality}]
    {\centering\includegraphics[width=0.59\textwidth]{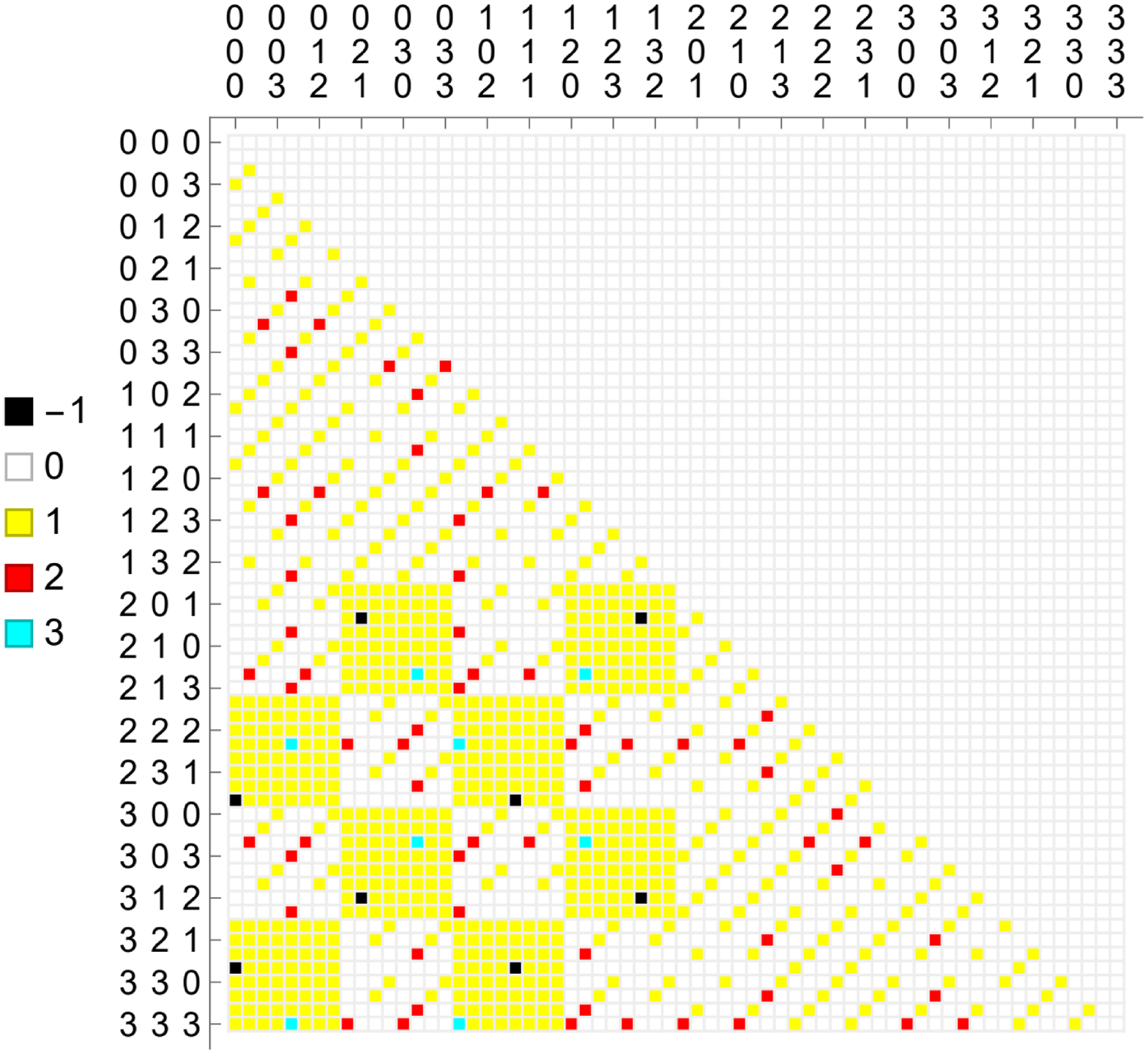}
    }
  \end{center}
\centering
\caption[]{ \textbf{Inflation technique for the triangle network.}
\subref{fig:inflation}
The second order inflation graph of the triangle network. Such an inflation doubles the number of latent variables relative to the triangle scenario, having six latent variables  $\{\lambda_{AB}^{(1)}, \lambda_{AB}^{(2)},\lambda_{BC}^{(1)}, \lambda_{BC}^{(2)}, \lambda_{AC}^{(1)}, \lambda_{AC}^{(2)}\}$. The inflation quadruples the number of observable random variables of the triangle scenario, having twelve observable random variables $\{a^{(1)}, b^{(1)}, c^{(1)},a^{(2)}, b^{(2)}, c^{(2)},a^{(3)},b^{(3)},c^{(3)},a^{(4)},b^{(4)},c^{(4)}\}$. Distributions compatible with this inflated structure satisfy symmetry properties, and have marginals corresponding to products of triangle-compatible distribution. This can be exploited to derive suitable causal compatibility inequalities that are violated by the experimental data.
\subref{fig:ineq} This plot depicts the $64\times 64$ coefficients ${y}_{a_1 b_1 c_1 a_2 b_2 c_3}$ for a quadratic inequality of the form of Eq.~\eqref{eq:webinfl} such that the left-hand side is nonnegative on all distributions compatible with the classical triangle scenario, but which  evaluates to the negative number $V_{exp} = -0.02436 \pm 0.00016$ on our experimental data. The $x$-axis ranges over the values of $(a_1,b_1,c_1)$ while the $y$-axis ranges over the values of $(a_2,b_2,c_2)$, and the color at a given point denotes the value of ${y}_{a_1 b_1 c_1 a_2 b_2 c_3}$ according to the mapping set out in the legend.
}
\end{figure*}

\subsection{Experimental Results}

As stated above, in order to realize the Fritz distribution, it is sufficient to share entanglement only between Alice and Bob's measurement stations, since Alice and Charlie as well as Bob and Charlie can merely share classical correlations.  Moreover, using such classical sources (in our case, a doubled electronic signal)
makes it possible to experimentally achieve correlations between Alice and Charlie and between Bob and Charlie that can be almost perfect for the duration of the experiment. Recall that perfect correlation is required for the logic of Fritz's argument to go through, but demonstrating perfect correlations can never be done  in an experiment and, importantly, demonstrating nonclassicality in the triangle network in the manner described by Fritz would boil down to violating a standard Bell inequality (sometimes referred to as disguised network nonlocality \cite{tavakoli2014nonlocal}). So, we did not use this approach here, as it is the goal of our work to introduce and validate data analysis techniques that would be applicable for  any example of a quantum-classical gap in the triangle scenario, including gaps based on distributions that, unlike
Fritz's,  could be noisy.
Fig.~\ref{fig:fritz_} provides a comparison between the theoretical Fritz distribution reported in panel ~\ref{fig:fritz_}a, obtainable with noiseless states and measurement operators, and the experimentally achieved one reported in panel ~\ref{fig:fritz_}b. The latter one was reconstructed from $\sim 1.4\cdot 10^6$ events collected in $\sim 10$ hours of data taking, achieving a 6-fold coincidence rate of $\sim 38.7$Hz (see Supplementary Note $2$ for the complete distribution).

Even with our approach, employing ultra-fast optical switches and classical correlations shared between $A$ and $C$ and between $B$ and $C$,
the measurement outcomes on the state $\Lambda_{AC}$ are not perfectly correlated, nor those on  $\Lambda_{BC}$, contrary to the ideal Fritz distribution: specifically, the probability of anti-correlation in each case is found to be $p_{\rm anticorr} = 3 \cdot 10^{-5}$. As argued, it is the practical impossibility of achieving perfect correlations which necessitates implementing a hypothesis test for compatibility or a test of causal compatibility inequalities.
In what follows, we will focus on three possible avenues: 
machine learning techniques \cite{bharti2020machine,krivachy2019neural,canabarro2019machine}, the inflation method \cite{wolfe2019inflation, fraser2018causal, navascues2020inflation, wolfe2021quantum} and finally, recently derived entropic inequalities~\cite{chaves2021causal}.

\subsection{
Excluding the hypothesis of classicality with machine learning}\label{sec:excluding}

We follow the approach in \cite{krivachy2019neural}, the central idea of which is to encode the structure of the causal network under test in the topology of a neural network.
Consider the triangle network with quaternary outputs as depicted in Fig.~\ref{fig:Tri}, where three sources $\lambda_{AB}$, $\lambda_{BC}$ and $\lambda_{AC}$ send information to three parties, Alice, Bob, and Charlie, each receiving, respectively, the pairs $ (\lambda_{AB}, \lambda_{AC} )$, $(\lambda_{AB}, \lambda_{BC})$ and $(\lambda_{AC}, \lambda_{BC})$, as schematically shown in Fig.~\ref{fig:DL} (a). After locally processing the inputs, they flag a number $a,b,c \in \{0,1,2,3\}$, by sampling the probability distributions $p(a|\lambda_{AB}, \lambda_{AC})$, $p(b|\lambda_{AB}, \lambda_{BC})$ and $p(c|\lambda_{BC}, \lambda_{AC})$ respectively. 
In the machine learning algorithm, the input layers to the multilayer perceptrons (MLPs) are composed of the independent uniformly distributed random numbers in the unit interval, i.e. $\lambda_{AB},\lambda_{BC}, \lambda_{AC} \in [0,1]$, with the restriction in the flow of information mirroring the causal structure of the triangle network: The $A$-block of the hidden layer receives random numbers ($\lambda_{AB}, \lambda_{AC}$), the $B$-block receives ($\lambda_{AB}, \lambda_{BC}$) and the $C$-block receives ($\lambda_{BC}, \lambda_{AC}$). 
Therefore, individual inputs belong to $\mathbb{R}^2$ (i.e. they have length 2). For the training, we provide batches of ($N_{\text{batch}},2$) dimension for the corresponding MLP for each of the three blocks.

If a certain probability distribution $p(a,b,c)$ is compatible with a classical causal model on the triangle causal structure, then a set of three independent neural networks mimicking the topology of the triangle should be able to reproduce the distribution. By numerically sampling over different values of 
the random numbers 
$\lambda_{AB}$, $\lambda_{BC}$ and $\lambda_{AC}$ one can construct the approximation $\tilde{p}(a,b,c)$ by averaging the Cartesian
product of the output conditional probabilities corresponding to each party.  See \hyperref[{sec:Methods}]{Methods} for more details. 

In turn, if the distribution under test is nonclassical, the neural network will be unable to mimic the distribution perfectly, producing considerable errors.
To quantify how much the machine model can approximate the target/experimental distribution,  we employ the element-wise mean square error (\textsf{MSE}), also termed as L2-norm error, between $p(a,b,c)$ and $\tilde{p}(a,b,c)$.  This is given by ${\mathsf{MSE}= \frac{1}{64}  \vert p(a,b,c)-\tilde{p}(a,b,c) \vert_2}$ and can be understood as a measure of nonclassicality \cite{canabarro2019machine}.
By repeated iterations, the neural network can be optimized in order to minimize this distance, since it should be close to zero  if the target distribution has a classical
model that the machine manages to approximate. Clearly, however, even if the distribution is compatible with the triangle network, due to numerical precision and the finite size of the neural network, the distance will never be exactly zero. To address this issue, we mix our experimental probability $p(a,b,c)$ with the flat distribution $p_I(a,b,c)=1/64$, which is compatible with the triangle structure, so that the machine is asked to retrieve the best possible model for the mixed distribution $\tilde{p}=v \; p +(1-v) \; p_I$.  
If $p$ has no classical explanation, then we expect that, as one increases the weight $v$ of $p$ in the mixed distribution $\tilde{p}$, there is a range of values wherein a classical model of $\tilde{p}$ remains possible and \textsf{MSE} is very small, but that there exists a threshold value beyond which \textsf{MSE} begins to increase, 
and the machine cannot make an almost perfect approximation anymore.

As shown in Fig.~\ref{fig:DL} (b), only
 below a certain threshold value around $v_{\rm{crit}}=1/\sqrt{2}$ \cite{krivachy2019neural}, can the machine learn $\tilde{p}$ 
 while it fails to do so for higher values of $v$.
This analysis gives a strong indication of the nonclassicality of $p$, but given that there is no guarantee that the machine finds the optimal parameters, it does not guarantee it.
To overcome this limitation, in the following we present two alternative techniques.

\subsection{
Violation of a causal compatibility inequality}\label{sec:violation}
In order to demonstrate the nonclassicality of the experimental data relative to the triangle causal network,
we seek to identify some causal inequalities which must be satisfied by all distributions compatible with the \emph{classical} triangle network but which are violated by our experimental statistics. To this end, we turn to the inflation technique for causal inference introduced in Ref.~\cite{wolfe2019inflation}.

As detailed in the \hyperref[{sec:Methods}]{Methods}, the inflation technique relates \emph{compatibility with a given causal structure} $\mathcal{G}$ 
to \emph{feasibility of a linear program (LP)}. If the LP related to an inflation of $\mathcal{G}$ (see Fig.~\ref{fig:inflation}) is found to be infeasible, then evidently $p$ is incompatible with $\mathcal{G}$. In our case,  $\mathcal{G}$ is taken to be the classical triangle scenario causal structure depicted in Fig.~\ref{fig:Tri}.

In the case of infeasibility, the algorithm returns an infeasibility witness, in the form of an inequality.  In this way, we can 
find a causal compatibility inequality tailored to the specific experimental data we obtained.
Using the second order inflation of the triangle network shown in Fig.~\ref{fig:inflation}, one can derive causal compatibility inequalities (satisfied by all triangle-compatible 
$p(a,b,c)$) of the form
\begin{align}
V \equiv
\hspace{-12ex}\smashoperator[r]{\sum_{\hspace{10ex}\substack{a_1 b_1 c_1 \\a_2 b_2 c_2 
}\in \{0,1,2,3\}^{\times 6}
}}
\;{y}_{a_1 b_1 c_1 a_2 b_2 c_2} 
p(a_1, b_1, c_1)p(a_2, b_2, c_2)
\geq 0\,,
\label{eq:webinfl}
\end{align}
where the $y$ are real coefficients.

As further detailed in the \hyperref[{sec:Methods}]{Methods}, the LP of the inflation technique may be specially adapted to yield \emph{elegant looking} causal compatibility inequalities; namely, where \emph{sets} of monomials are each associated to a single (i.e., uniform) coefficient. Working with such an \emph{adapted} LP can be orders of magnitude less computationally demanding as compared to the unadapted LP. However, it may be the case that despite a given distribution leading to infeasibility in the \emph{unadapted} primal LP, there may not exist any inequality \emph{with restricted coefficients} capable of witnessing that fact. As such, one is motivated to carefully select a coefficient restriction which \emph{matches} the specifically-targeted distribution: one should only impose that a pair of monomials should share a uniform coefficient in the inequality if the given distribution would lead to both monomials being evaluated to the same numerical value (within a small tolerance). One cannot impose arbitrary coefficient uniformity restrictions. The \hyperref[{sec:Methods}]{Methods} contains an explanation for why certain special coefficient restrictions may be justifiable. We employed the \emph{ideal} theoretical Fritz distribution as our guide when selecting our LP adaptation, rendering moot the selection of a numerical tolerance. We stress, however, that a theoretical guide is \emph{not} a prerequisite for optimally adapting the inflation technique LP to witness the nonclassicality of experimental data: it is perfectly possible to isolate the near-symmetries in the experimental data without the educated guess provided by a theoretical model.

\begin{figure}[t!]
\begin{center}
\includegraphics[width=0.9\columnwidth]{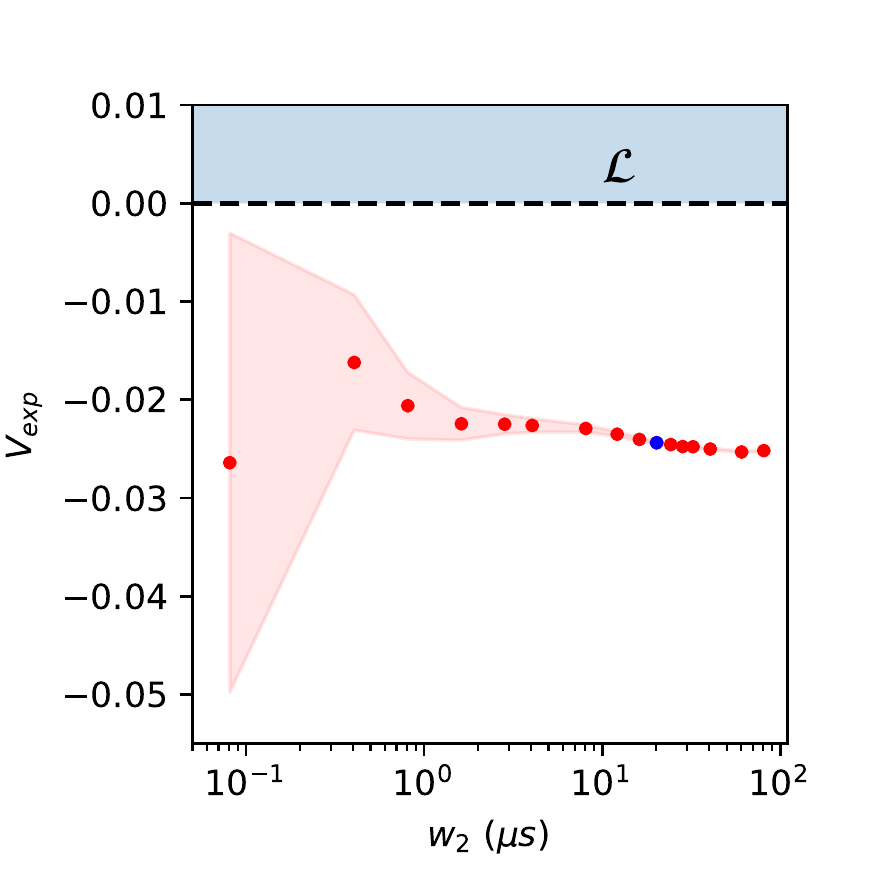}
\end{center}
\caption{\textbf{Inflation inequality violation vs 6-fold coincidence window.} Values of violation of the causal compatibility inequality, which has been optimized over the experimental data corresponding to the blue point (a window $w_2 \sim 20 \mu s$), as a function of the 6-fold coincidence window $w_2$. The red shadowed area represents the statistical error on the computed value of $V_{exp}$, estimated employing Monte Carlo methods. The blue shadowed region $\mathcal{L}$ indicates the values obtainable by a classical causal model.}
\label{fig:bigwin}
\end{figure}

The infeasibility witness obtained by the program for our data yields
an inequality of the form of Eq.~\eqref{eq:webinfl} which  is violated by the experimental data by
several standard deviations: in this way, we unambiguously demonstrate the emergence of nonclassicality in the triangle network, without relying on Bell's theorem.
We depict the particular coefficients ${y}_{a_1 b_1 c_1 a_2 b_2 c_2}$ defining the inequality that we obtained from the adapted LP in Fig.~\ref{fig:ineq}.
Denoting the value that the data gives for the left-hand-side of this inequality by $V_{\rm exp}$,
we obtain $V_{exp} = -0.02436 \pm 0.00016$ (using a 6-fold coincidence window $w_2 \sim 20 \mu s$),
corresponding to a violation  of the inequality by $152$ standard deviations. 
In Fig.~\ref{fig:bigwin}, we plot $V_{\rm exp}$
as a function of the choice of the 6-fold coincidence window $w_2$. As expected, by increasing $w_2$, we increase the detection rate of 6-fold events, in turn decreasing the statistical error on the computed value of $V_{exp}$, shown in the figure with the red shadowed area.

\subsection{Bounding measurement dependence and violating an entropic inequality for the triangle network}
\label{sec:entropic}

Another approach that can be used to robustly demonstrate the nonclassicality of the generated data is to map the triangle network into a modification of the Bell scenario, in a similar way to Fritz's original proof of nonclassicality in the triangle scenario.
In this modification, any amount of measurement dependence is in principle allowed between the hidden variable and the measurement settings. Consequently, even though the scenario is related to Bell's, the nonclassicality exhibited necessarily goes beyond that which one finds in Bell's scenario because in the latter measurement dependence allows for a classical account of any correlations. Indeed, in a Bell scenario where causal influences between the source and the measurement settings  are allowed, some amount of measurement independence has to be \emph{assumed} in order to witness nonclassicality from the data \cite{barrett2011much,hall2011relaxed,putz2014arbitrarily}, otherwise, any violation of a Bell inequality can be explained by classical local models \cite{brans1988bell}. 

\begin{figure}
  \begin{center}
    \subfigure[\label{fig:freechoice_dag_bell} \textbf{Bell scenario without freedom of choice.} ]
    {\centering
      \centering\includegraphics[width=0.85\columnwidth]{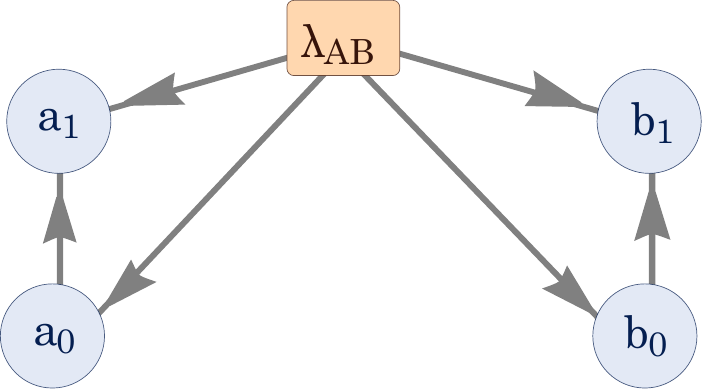}
    }
    \hfill
    \subfigure[\label{fig:freechoice_dag_all} \textbf{Extended Bell scenario as a triangle network.}]
    {\centering\includegraphics[width=0.85\columnwidth]{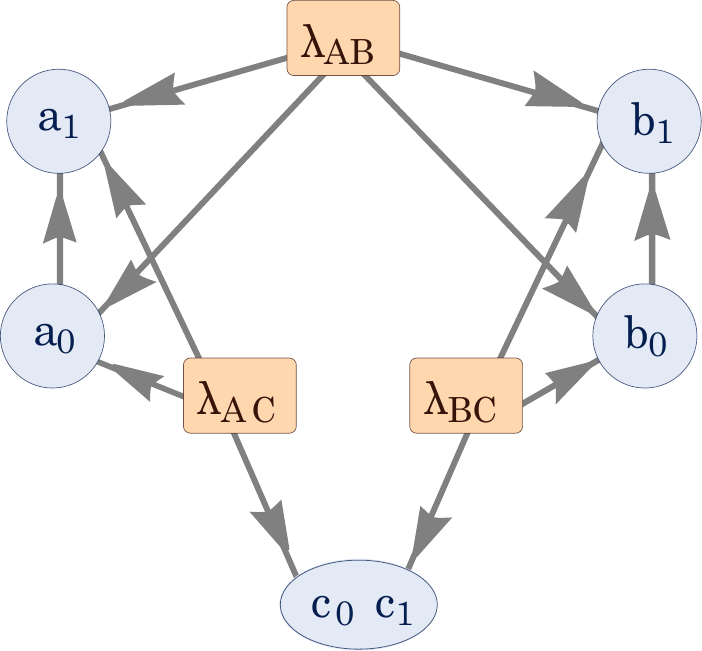}
    }
  \end{center}
\centering
\caption[]{\textbf{Triangle scenario from extended Bell scenario.}
\subref{fig:freechoice_dag_bell} Extended Bell scenario with measurement dependence. Relative to the standard Bell scenario, the source $\lambda_{AB}$ is presumed to influence not only the outcomes 
$a_1$ and $b_1$, but the setting variables $a_0$ and $b_0$ as well. This allows for measurement dependence and can describe superdeterministic models \cite{hossenfelder2020rethinking}.
\subref{fig:freechoice_dag_all} Extended Bell causal structure with measurement dependence mapped into the triangle scenario.  
Relative to the standard Bell scenario, one posits an additional laboratory, associated to Charlie, an additional source $\lambda_{AC}$ between Alice and Charlie and an additional source $\lambda_{BC}$ between Bob and Charlie. 
Correlations between $a_0$ and $c_0,c_1$ imply an upper bound on the potential dependence of $a_0$ on $\lambda_{AB}$, described by the entropic inequality in Eq.~\eqref{eq:ineqCHSH2}.  Similarly, correlations between $b_0$ and $c_0,c_1$ imply an upper bound on the potential dependence of $b_0$ on $\lambda_{AB}$.
}
\label{fig:freechoicedag}
\end{figure}

In the modified scenario, one can use entropic inequalities to put an upper bound on the amount of measurement dependence, as demonstrated in  \cite{chaves2021causal}.
The modification can be understood as a two-step departure from Bell's scenario.  In the first step, depicted in Fig.~\ref{fig:freechoicedag}(a), one allows there to be a common cause not only on the pair of outcome variables, but on  all four of the observed variables, meaning that a given outcome variable shares a common cause with the setting variable at the opposite wing; this is a relaxation of the assumption of freedom of choice~\cite{hall2010local,hall2011relaxed,chaves2015unifying,big2018challenging}. In the second step, one introduces additional observed variables $c_0 c_1$ and a variable $\lambda_{AC}$ that is a common cause to Alice's setting and outcome ($a_0$, $a_1$) and $c_0 c_1$, as well as a variable $\lambda_{BC}$ that is a common cause to Bob's setting and outcome ($b_0$, $b_1$) and $c_0 c_1$ (see Fig.\ref{fig:freechoicedag}(b)).

Referring to the DAG of the triangle network
shown in Fig.~\ref{fig:appNEW}, we map the measurement settings of the two stations $A$ and $B$ of the Bell scenario to the variables $a_0$ and $b_0$, and the measurement outcomes are mapped to the  variables $a_1$ and $b_1$.  It is clear, therefore, that if one lumps $a_0$ and $a_1$ together, and similarly for $b_0$ and $b_!$, the modified Bell scenario can be seen to have the form of the triangle network.

In this modified Bell scenario, shown in Fig.~\ref{fig:freechoicedag}(b),
one can lower bound the measurement dependence, quantified via the mutual information $I(\lambda_{AB}:a_0,b_0)$ between the source $\lambda_{AB}$ and the measurement settings $a_0$ and $b_0$, relating it with the violation of the CHSH inequality \cite{chaves2015unifying,hall2020measurement}. Further, employing the entropic approach \cite{chaves2014inferring,fritz2012entropic,chaves2014causal,chaves2015information}, this mutual information can also be upper bounded by an entropic function that involves only observable variables and so can be extracted directly from the experimental data. Combining both the upper and lower bounds on $I(\lambda_{AB}:a_0,b_0)$, one arrives at a Bell inequality blending probabilities and entropies, the violation of which witnesses the nonclassicality of the data, irrespectively of any potential measurement dependence $I(\lambda_{AB}:a_0,b_0)$ present in the experiment. This inequality is given by (see Ref.\cite{chaves2021causal} for the further details)
\begin{equation}
\label{eq:ineqCHSH2}
\mathcal{E}\equiv 2- S^{CHSH}+ \sqrt{\frac{16\;\Theta(a_0,b_0,C)}{\log_2 e}} \geq 0\,,
\end{equation}
where $S^{CHSH}$ is the standard CHSH quantity evaluated on $P(a_1 b_1|a_0 b_0))$~\cite{clauser1969proposed}, and
\begin{align}
\Theta(&a_0,b_0,C)\coloneqq 
\\&\min \begin{cases} \begin{smallmatrix} H(a_0,b_0 \vert C)\,,\end{smallmatrix} \\ 
\begin{smallmatrix} H(a_0,b_0) -I(a_0:b_0:C)
   -I(a_0:C) -I(b_0:C)\,,
\end{smallmatrix}\\
\begin{smallmatrix} 
H(a_0,b_0)+H(C) -2I(a_0:b_0:C)-2I(a_0:C)  -2I(b_0:C)\,,
\end{smallmatrix}
\end{cases}\nonumber
\end{align} 
with $I(a_0:b_0:C)\coloneqq H(a_0,b_0,C)-H(a_0,b_0)-H(a_0,C)-H(b_0,C)+H(a_0)+H(b_0)+H(C)$ the tripartite mutual information and $H(X)=-\sum_{x}p(x)\log{p(x)}$ the Shannon entropy relative to the variable $X$.

Using the experimental data in Fig.~\ref{fig:fritz_}, we obtain a value $\mathcal{E}_{exp}=-0.340\pm0.001$, violating the bound of Eq.~\eqref{eq:ineqCHSH2} by $340$ standard deviations and thereby demonstrating nonclassicality.

\section{Discussion}

The triangle scenario has particular novelty as a means of witnessing nonclassicality insofar as there is no known way to obtain classical causal compatibility inequalities for it from standard Bell inequalities.  This is in contrast to the two other causal structures distinct from the Bell scenario that have been experimentally investigated previously, namely, the instrumental scenario  \cite{chaves2018quantum} and the bilocality scenario \cite{saunders2017experimental,carvacho2017experimental,sun2019experimental,li2022testing,wu2022experimental,carvacho2022quantum}.  In the case of the instrumental scenario, it suffices to process the Bell inequalities by forcing equality between the value of the setting variable at one wing and the value of the outcome variable at the opposite wing~\cite{van2019quantum}.  In the case of the bilocality scenario, by post-selecting on the outcome of the measurement that accesses both sources, the other two measurements can be proven to satisfy the Bell inequalities in a classical causal model (via an analogue of entanglement-swapping) \cite{branciard2010characterizing,branciard2012bilocal}.  Such short-cuts to deriving noise-robust causal compatibility inequalities, however, are not available in the triangle scenario.

Another peculiar aspect of the triangle scenario is the possibility to show new forms of nonclassicality that do not require the use of external inputs freely chosen by the experimenter, but instead rely on the assumption of independence of the sources, as shown by Fritz 
\cite{fritz2016beyond}. In this work, we realized for the first time a triangle network without external inputs, proving the emergence of nonclassicality
in this new regime, up to detection  and locality loopholes. This has been possible by employing fast feed-forward of measurement in an optical setup comprising an entangled photon source and two sources of classical correlations.
In order to demonstrate the nonclassicality of the experimental data, we had to extend pre-existing data analysis techniques, making them suitable to detect nonclassicality in noisy distributions.

The data analysis techniques we have presented here are also distinguished insofar as they have  the capacity to witness nonclassicality for  any distribution that might arise in an experiment, whereas previous experiments witnessing nonclassicality in causal structures beyond Bell have used tools that can only witness the nonclassicality of limited classes of target distributions. This approach thus extends data-seeded techniques previously limited to the standard bipartite Bell scenario
\cite{Brunner,scarani2019bell,elliott2009linear,zhang2011asymptotically} to the realm of more complex causal networks.

\color{black}

The employed data analysis techniques and aspects of our photonic setup 
provide a scalable platform in which nonclassicality can be witnessed in networks of growing size and of arbitrary topology.
In particular, the implemented measurements are based on local wirings, i.e., separable measurements with classical feedbacks, making the approach scalable.
Furthermore, it is widely speculated that the triangle scenario may admit  distributions which
imply a no-go result whose logic is entirely independent of that of Bell's theorem~\cite{renou2019genuine,gisin2020constraints}. These are likely to require entangled measurements as well as three sources of entanglement, and consequently, integrating such measurements and sources into our set-up may open the way to experimentally targeting distributions which are thought to exhibit these new types of nonclassicality.

Finally, this work can also pave the way for future applications in quantum communications involving several sources and measurement stations.

\section{Methods}\label{sec:Methods}

\subsection{Details on the machine learning implementation}\label{sec:MethodsML}

The number of samples we sum over, i.e., the batch size,  is ${N_{\text{batch}} = 10000}$. We decided to vary the architecture of the neural network using different number of layers ${(n_{\text{layers}} = [3,4,5,6])}$ and number of neurons ${(n_{\text{neurons}} = [16, 32])}$, accounting for an assembly of $8$ neural networks independently trained, in order to obtain better approximations by taking the minimum or the average of the predictions. The ensemble of networks also reduces the probability of being trapped in optimization local minima and enhances the relative expressive power of the method in comparison to a single architecture; see Fig.~\ref{fig:DL} (a). As pointed out in Ref.~\cite{krivachy2019neural}, ideal values for parameters and hyper-parameters vary for distinct triangle scenarios, therefore the strength of the ensemble approach also varies. The reader is referred to the Supplementary Note $3$ for more specific details.

\subsection{Details on the Inflation Technique}\label{sec:inflationmethods}

At its core, the inflation technique at $n^{th}$ order shows that 
\begin{description}
\item[IF] A distribution $p$ is compatible with a given classical causal structure $\mathcal{G}$ 
\item[THEN] For the $n^{th}$ order inflation graph  $\mathcal{G}'$ induced by $\mathcal{G}$ there must exist some larger distribution $p'$ pertaining to the observable nodes in $\mathcal{G}'$ such that
\begin{enumerate}
\item $p'$ possesses certain symmetry properties related to automorphisms of  $\mathcal{G}'$, and
\item the distribution $p^{\otimes n}$ --- defined as $n$ identical but independently distributed (I.I.D.) copies of $p$  --- arises as a marginal distribution of $p'$.
\end{enumerate}
\end{description}
These conditions implicitly define a linear program (LP). In the Supplementary Note $4$, we elaborate on the required marginal symmetry properties which must be satisfied by distributions compatible with the second order inflation graph depicted in Fig.~\ref{fig:inflation}. 

Farkas' duality lemma tells us how to extract a \emph{certificate of infeasibility} whenever a LP  is infeasible~\cite{Andersen2001}. Note that Farkas' lemma applies to convex optimization in general~\cite{Dinh2014}; linear programming is just a special case. For the primal LP defined by second order inflation, the certificate of infeasibility is a dual vector ${\boldsymbol{y}}$ such that ${{\boldsymbol{y}}\cdot p^{\otimes 2}\geq 0}$ holds for all instances of $p^{\otimes 2}$ which make the primal LP feasible. Given such a dual vector ${\boldsymbol{y}}$, one certifies the infeasibility of $p_{\text{nonclassical}}^{\otimes 2}$ --- i.e., one certifies the \emph{incompatibility} of $p_{\text{nonclassical}}$ with a classical causal model with the structure $\mathcal{G}$  --- whenever one finds  that ${{\boldsymbol{y}}\cdot p_{\text{nonclassical}}^{\otimes 2} < 0}$. 
Hence, the certificate ${\boldsymbol{y}}$ yields a quadratic polynomial inequality satisfied by all distributions $p$ which are compatible with $\mathcal{G}$.

We employed the \enquote{hierarchy} version of inflation defined in Ref.~\cite{navascues2020inflation} due to its computationally efficient and data-agnostic implementation.

\sloppypar The second order inflation graph of the classical triangle network is depicted in Fig.~\ref{fig:inflation}, and the $p'$ which is posited to exist would pertain to the twelve observable random variables depicted in Fig.~\ref{fig:inflation}, namely 
$\{a^{(1)}$, $b^{(1)}$, $c^{(1)}$, $a^{(2)}$, $b^{(2)}$, $c^{(2)}$, $a^{(3)}$, $b^{(3)}$, $c^{(3)}$, $a^{(4)}$, $b^{(4)}$, $c^{(4)}\}$.

The LP implied by inflation is as follows. The condition for the existence of $p'$ can be understood as a collection of very many inequality constraints (every probability which makes up $p'$ must be nonnegative) along with one equality constraint (the sum of all probabilities comprising $p'$ totals unity). The symmetry requirements of $p'$ can be understood as equality constraints relating the various probabilities comprising $p'$. Finally, the requirement that $p^{\otimes 2}$ is a marginal of $p'$ can be understood as equating 
$p^{\otimes 2}$ 
evaluated at a particular set of values for its arguments
to a sum over all those probabilities of $p'$ which agree on these values.
In other words, if $p$ is compatible with $\mathcal{G}$, then some collection of equality and inequality constraints are simultaneously satisfiable;
i.e., some LP should be feasible. 

The Farkas infeasibility certificate of the LP defined by inflation constitutes quadratic inequalities which are satisfied by all triangle-compatible distributions but violated by the nonclassical distribution whose triangle-incompatibility is witnessed by inflation. See Supplementary Note $4$ for an explicit walk-through of the inflation technique in full detail.

\subsection{Adapting polytope membership LPs to yield symmetric inequalities}
It can be insightful to compare the LP defined by inflation to the more familiar LP associated with Bell nonlocality. In Bell nonlocality, a family of conditional probability distributions (a.k.a. a \enquote{correlation}) is said to admit a local hidden variable model (LHVM) if and only if corresponding \emph{vector of all conditional probabilities} lies within the local polytope. When a correlation does not admit a LHVM explanation, then we can always find a separating hyperplane (typically a facet of the local polytope) such that the vector of conditional probabilities associated with the given correlation lies strictly to one side of the hyperplane whereas all LHVM-explainable correlations correspond to vectors of conditional probabilities in or on the other side of the hyperplane. Thus, hyperplanes which distinguish all LHVM-explainable vectors from some other are equivalent to Bell inequalities; these hyperplanes which correspond to facets of the local polytope are equivalent to \emph{facet-defining} Bell inequalities.

The picture is quite similar when thinking about the LP associated with inflation. Instead of vectors of conditional probabilities, however, we are considering vectors whose elements are products of unconditional probabilities, i.e., vectors of probability monomials. The LP of inflation similarly defines a polytope: a vector of monomials is in the polytope iff the primal LP is feasible; the objective of the dual LP is to return a separating hyperplane such that 
\begin{compactenum}
\item the given vector of monomials is as far from the hyperplane as possible, and 
\item such that all vectors which would make the primal LP feasible lie on or on the other side of the hyperplane.
\end{compactenum}

Without loss of generality, a polytope may be defined in terms of its extremal points. Let $M^{d,n}$ be a $d\times n$ matrix whose $n$ columns correspond to the extremal points of the polytope, each of which is a vector in dimension $d$, and where we have introduced a notation of marking an object's dimension in superscript for pedagogical clarity in what follows.
A vector ${v}^{d}$ lies withing the polytope (technically, the LP formulations here apply to both bounded polytopes and unbounded polycones) if and only if
\begin{align}\label{eq:primalsat}\begin{split}
\text{there exists some}&\quad {{x}^{n}}\\
\text{such that}&\quad M^{d,n}\cdot {x}^{n} = {v}^{d}\,,\\
\text{where}&\quad{x}^{n}\geq {\mathbb{0}}^{n}\,.\\
\end{split}\end{align}
We can relax the \emph{satisfiability} LP of Eq.~\eqref{eq:primalsat} into an \emph{optimization} problem which measures the \emph{degree of primal infeasibility}. One natural measure of the infeasibility of Eq.~\eqref{eq:primalsat} is defined by the following optimization problem:
\begin{align}\label{eq:primalopt}\begin{split}
\max_{{x}^{n},\; {s}^{n}}&\quad  -{\mathbb{1}}^{n}\cdot{s}^{n} \\
\text{such that}&\quad M^{d,n}\cdot \left({x}^{n}-{s}^{n}\right) = {v}^{d}\,,\\
\text{where}&\quad {x}^{n}\geq {\mathbb{0}}^{n}\quad\text{and}\quad
{s}^{n}\geq {\mathbb{0}}^{n}\,.
\end{split}\end{align}
Note that if the LP of Eq.~\eqref{eq:primalsat} can be satisfied, then the objective in Eq.~\eqref{eq:primalopt} can be reach up to 0; conversely, if the objective in Eq.~\eqref{eq:primalopt} is strictly negative over all variables which satisfy that LP's conditions, then the LP in Eq.~\eqref{eq:primalsat} is evidently infeasible. 
The formal dual to the above LP can then be used to extract optimal separating hyperplanes. The astute reader may notice that even the reformulated LP as given in Eq.~\eqref{eq:primalopt} may not always be feasible; it can only be satisfied if ${v}^{d}$ is wholly in the \emph{linear span} of the columns of $M^{d,n}$. If ${v}^{d}$ has some component orthogonal to that linear span, then the primal formulation in Eq.~\eqref{eq:primalopt} is infeasible and the dual formulation in Eq.~\eqref{eq:dualopt} is unbounded. See Appendix B of Ref.~\cite{cao2022experimental} for alternative relaxations of an LP satisfiability problem into an optimization problem, and the connection therein to distance measures such as robustness and nonlocal fraction. Namely,
\begin{align}\label{eq:dualopt}\begin{split}
\min_{{y}^{d}}&\quad  {y}^{d}\cdot{v}^{d} \\
\text{such that}&\quad{\mathbb{0}}^{n} \leq {y}^{d}\cdot M^{d,n} \leq {\mathbb{1}}^{n}.
\end{split}\end{align}
Indeed, the \emph{weak duality theorem} in linear programming ensures that regardless of the feasibility of Eq.~\eqref{eq:primalsat}, it holds that for \emph{every} $y^d$ satisfying the condition of Eq.~\eqref{eq:dualopt} and \emph{every} $x^n$, $s^n$ satisfying the conditions of Eq.~\eqref{eq:primalopt}, it is always the case that ${y}^{d}\cdot{v}^{d} \geq -{\mathbb{1}}^{n}\cdot{s}^{n}$. So, if \emph{any} $y^d$ can be found satisfying the condition of Eq.~\eqref{eq:dualopt} such that ${y}^{d}\cdot{v}^{d}\leq 0$, this serves as a certificate of the infeasibility of Eq.~\eqref{eq:primalsat}.

Now, the matrix $M^{d,n}$ which defines the polytope may exhibit \emph{inherent symmetries}. An inherent symmetry of a matrix is a pair of permutation operations $\pi^{d,d}_{\text{row}}$ and $\pi^{n,n}_{\text{col}}$, acting respectively on the row space and column space of the matrix, such that if \emph{both} the row permutation and the column permutation are performed the matrix is invariant. That is,
\begin{align}
M^{d,n} = \pi^{d,d}_{\text{row}} \cdot M^{d,n} \cdot \pi^{n,n}_{\text{col}}.
\end{align}
Whenever such an inherent symmetry can be identified, it can be used to transform feasible solutions of both the primal and dual formulations into new solutions: Suppose we have a collection of vectors $v^d$, $y^d$, $x^n$, $s^n$ 
such that all of the conditions of both Eq.~\eqref{eq:primalopt} and Eq.~\eqref{eq:dualopt} are satisfied. Then, acting on all the vectors with the inherent symmetry leads to a new solution pair to both the primal and dual LP formulations, with the same duality gap (if any). Accordingly, we have that the \emph{symmetrized} inequality ${y^\prime}^d \geq 0$ where ${y^\prime}^d \coloneqq \frac{y^d + \pi^{d,d}_{\text{row}}\cdot y^d }{2}$ is also a valid inequality. When ${y}^d$ is an \emph{optimal} solution to the dual LP in Eq.~\eqref{eq:dualopt}, then symmetrized inequality ${y^\prime}^d$ is also optimal if $v^d$ is \emph{invariant} under the inherent symmetry operation $\pi^{d,d}_{\text{row}}$.

This is what allows us to restrict the coefficients of the separating hyperplanes. Suppose we find a bunch of different inherent symmetries of the matrix which defines the polytope; these can be used to construct a \emph{group} with well-defined actions on both the row and column spaces. We can then \emph{twirl} the matrix with respect to this group: 
We collect columns which map to each other under the group action, and replace each \emph{orbit} of columns with a single new column given by the mean of the orbit. We do the same to the rows. This twirling operation thus yields a substantially smaller matrix, say, ${M^\prime}^{d^\prime, n^\prime}$. 
Given a vector $v^d$ in the row space of the matrix, we can apply the same twirling to obtain ${v^\prime}^{d^\prime}$, essentially projecting the vector to the symmetric subspace of the group. We now can obtain a separating hyperplane ${y^\prime}^{d^\prime}$ by applying the dual formulation of the LP in this symmetric subspace. To convert this hyperplane in the symmetric subspace to a hyperplane in the full row space we de-twirl: namely, each row in a given orbit is uniformly associated with the coefficient of that orbit in the symmetric subspace.

There is no loss of generality whatsoever in using this symmetry-adapted version of the LP if the target vector $v^d$ is also invariant under the group.
So, in general, the most efficient way to exploit inherent symmetries in linear programming is to identify the largest symmetry group (acting on both row and column spaces) which leaves both $M^{d,n}$ and $v^d$ invariant.

For more information regarding exploiting symmetry in linear programming see Refs.~\cite{BancalSymmetricBell,
bremner2009polyhedral, PANDA, SDPSymmetry}. 

\color{black}

\subsection{Robustness to noise added by varying 2-fold coincidence window}

We study the behavior of the nonlocality tests over the addition of noise due to the enlargement of the two-fold coincidence window $w_1$. Increasing such a window causes the increase of accidental counts, affecting both the events from the entangled source  and those relative to classically correlated signals. From a practical point of view, such noise acts substantially as a white noise on the correlations, that is event pairs which are uniformly and randomly distributed. Considering such effects, we do expect that at some point, increasing the noise, our witnesses will not be able to detect a nonclassical behaviour anymore. This is, in fact, the case. We show the curve of the violation of the inequality, Eq.~\ref{eq:webinfl} and Fig.~\ref{fig:ineq} inferred by means of the inflation technique, as a function of the 2-fold window $w_1$ in Fig.~8. The same  study is performed with the value of the violation of the entropic inequality in Eq.~\eqref{eq:ineqCHSH2} as shown in Fig.~9.

\begin{figure}[t!]
\begin{center}
\includegraphics[width=0.9\columnwidth]{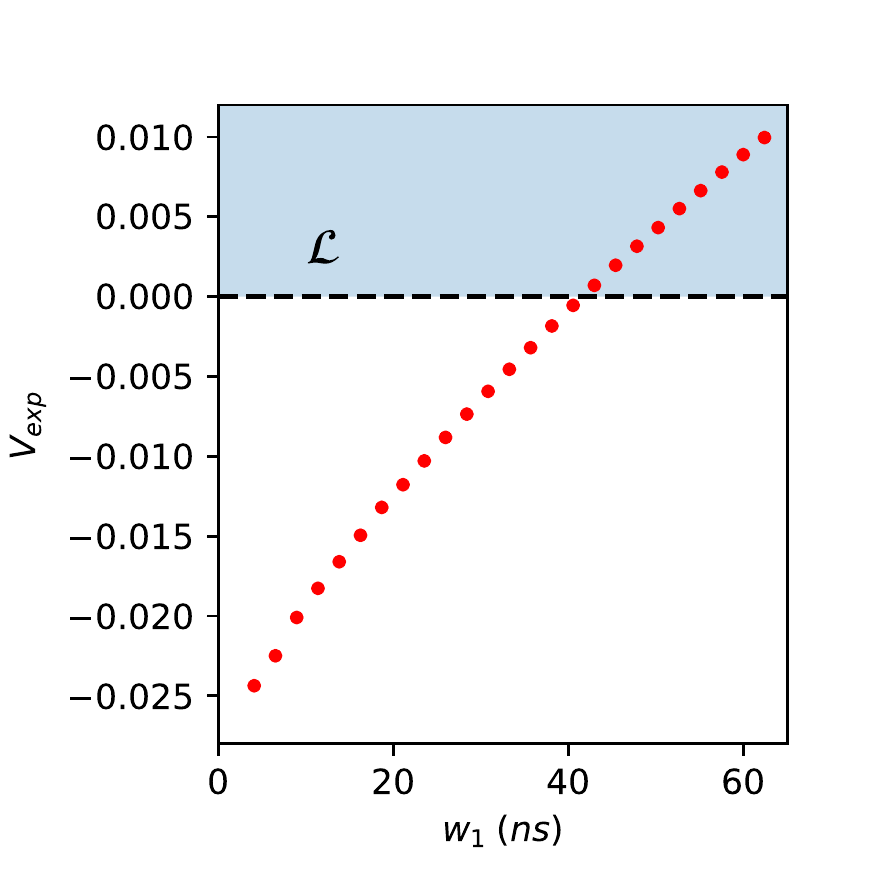}
\end{center}
\label{fig:8}
\caption{\textbf{Robustness of the inflation inequality to experimental noise.} In this plot we show the achieved value for the inequality from inflation technique, as function of the two-fold coincidence window $w_1$. As expected, since this increases uncorrelated 2-fold events, the measured correlations become "local" for large enough windows. The leftmost point corresponds to the result reported in subsection "Bounding measurement dependence and violating
an entropic inequality for the triangle network", i.e. a window $w_1 \sim 4.1 ns$, the plotted error bars are calculated through Monte Carlo technique assuming Poissonian statistics and are smaller than the size of the points. The blue shadowed region $\mathcal{L}$ indicates the values obtainable by a classical causal model.}
\end{figure}

\begin{figure}[t!]
\begin{center}
\includegraphics[width=0.9\columnwidth]{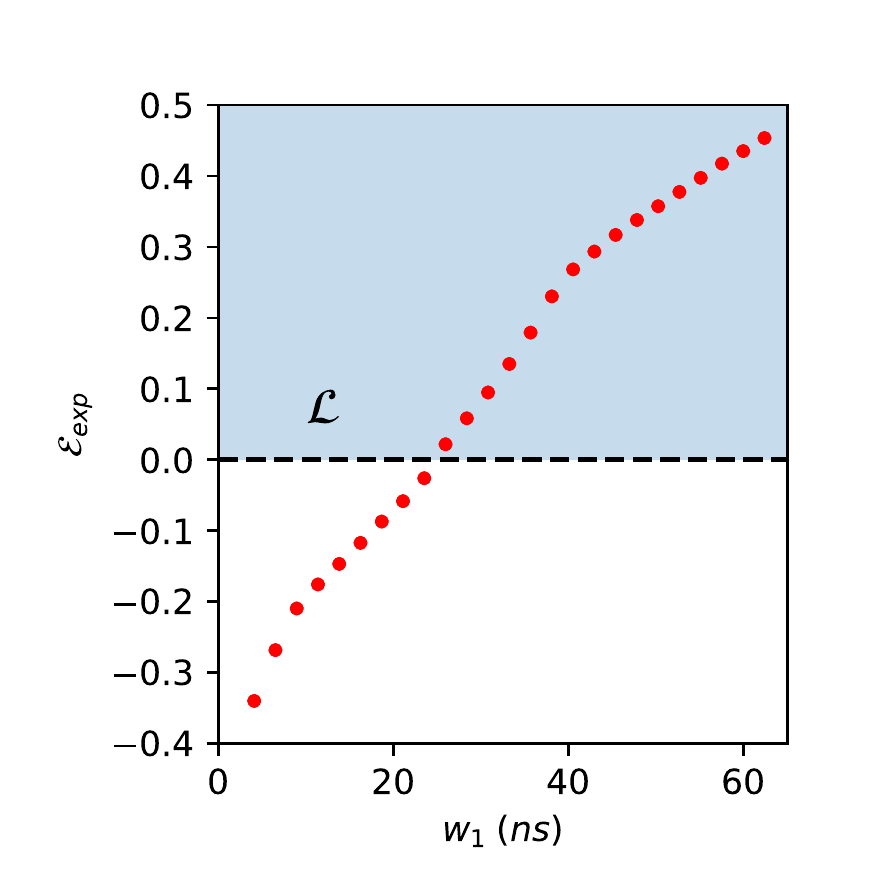}
\end{center}
\label{fig:9}
\caption{\textbf{Robustness of the entropic technique to experimental noise.} Values of the quantity $\mathcal{E}_{exp}$, as defined in Eq.~\eqref{eq:ineqCHSH2}, as function of the two-fold coincidence window $w_1$. If $w_1 > 25ns$, due to the noise, the entropic witness is not able to detect nonclassicality, resulting in a strictly positive value of the quantity $\mathcal{E}_{exp}$. 
The leftmost point corresponds to the result reported in subsection "Bounding measurement dependence and violating
an entropic inequality for the triangle network", i.e. a window $w_1 \sim 4.1 ns$, the plotted error bars are calculated through Monte Carlo technique assuming Poissonian statistics and are smaller than the size of the points.}
\end{figure}

\section*{Data availability}
 The data that support the findings of this study are available in the Supplementary Information and from the corresponding author upon request.

\section*{Code availability}
 All the custom code developed for this study is available from the corresponding author upon request.


\begin{thebibliography}{97}%
\makeatletter
\providecommand \@ifxundefined [1]{%
 \@ifx{#1\undefined}
}%
\providecommand \@ifnum [1]{%
 \ifnum #1\expandafter \@firstoftwo
 \else \expandafter \@secondoftwo
 \fi
}%
\providecommand \@ifx [1]{%
 \ifx #1\expandafter \@firstoftwo
 \else \expandafter \@secondoftwo
 \fi
}%
\providecommand \natexlab [1]{#1}%
\providecommand \enquote  [1]{``#1''}%
\providecommand \bibnamefont  [1]{#1}%
\providecommand \bibfnamefont [1]{#1}%
\providecommand \citenamefont [1]{#1}%
\providecommand \href@noop [0]{\@secondoftwo}%
\providecommand \href [0]{\begingroup \@sanitize@url \@href}%
\providecommand \@href[1]{\@@startlink{#1}\@@href}%
\providecommand \@@href[1]{\endgroup#1\@@endlink}%
\providecommand \@sanitize@url [0]{\catcode `\\12\catcode `\$12\catcode
  `\&12\catcode `\#12\catcode `\^12\catcode `\_12\catcode `\%12\relax}%
\providecommand \@@startlink[1]{}%
\providecommand \@@endlink[0]{}%
\providecommand \url  [0]{\begingroup\@sanitize@url \@url }%
\providecommand \@url [1]{\endgroup\@href {#1}{\urlprefix }}%
\providecommand \urlprefix  [0]{URL }%
\providecommand \Eprint [0]{\href }%
\providecommand \doibase [0]{https://doi.org/}%
\providecommand \selectlanguage [0]{\@gobble}%
\providecommand \bibinfo  [0]{\@secondoftwo}%
\providecommand \bibfield  [0]{\@secondoftwo}%
\providecommand \translation [1]{[#1]}%
\providecommand \BibitemOpen [0]{}%
\providecommand \bibitemStop [0]{}%
\providecommand \bibitemNoStop [0]{.\EOS\space}%
\providecommand \EOS [0]{\spacefactor3000\relax}%
\providecommand \BibitemShut  [1]{\csname bibitem#1\endcsname}%
\let\auto@bib@innerbib\@empty
\bibitem [{\citenamefont {Bell}(1964)}]{bell1964einstein}%
  \BibitemOpen
  \bibfield  {author} {\bibinfo {author} {\bibfnamefont {J.~S.}\ \bibnamefont
  {Bell}},\ }\bibfield  {title} {\enquote {\bibinfo {title} {{On the Einstein
  Podolsky Rosen paradox}},}\ }\href
  {https://doi.org/10.1103/PhysicsPhysiqueFizika.1.195} {\bibfield  {journal}
  {\bibinfo  {journal} {Physics Physique Fizika}\ }\textbf {\bibinfo {volume}
  {1}},\ \bibinfo {pages} {195} (\bibinfo {year} {1964})}\BibitemShut {NoStop}%
\bibitem [{\citenamefont {Brunner}\ \emph {et~al.}(2014)\citenamefont
  {Brunner}, \citenamefont {Cavalcanti}, \citenamefont {Pironio}, \citenamefont
  {Scarani},\ and\ \citenamefont {Wehner}}]{Brunner}%
  \BibitemOpen
  \bibfield  {author} {\bibinfo {author} {\bibfnamefont {N.}~\bibnamefont
  {Brunner}}, \bibinfo {author} {\bibfnamefont {D.}~\bibnamefont {Cavalcanti}},
  \bibinfo {author} {\bibfnamefont {S.}~\bibnamefont {Pironio}}, \bibinfo
  {author} {\bibfnamefont {V.}~\bibnamefont {Scarani}},\ and\ \bibinfo {author}
  {\bibfnamefont {S.}~\bibnamefont {Wehner}},\ }\bibfield  {title} {\enquote
  {\bibinfo {title} {{Bell nonlocality}},}\ }\href
  {https://doi.org/10.1103/RevModPhys.86.419} {\bibfield  {journal} {\bibinfo
  {journal} {Rev. Mod. Phys.}\ }\textbf {\bibinfo {volume} {86}},\ \bibinfo
  {pages} {419} (\bibinfo {year} {2014})}\BibitemShut {NoStop}%
\bibitem [{\citenamefont {Scarani}(2019)}]{scarani2019bell}%
  \BibitemOpen
  \bibfield  {author} {\bibinfo {author} {\bibfnamefont {V.}~\bibnamefont
  {Scarani}},\ }\href {https://doi.org/10.1093/oso/9780198788416.001.0001}
  {\emph {\bibinfo {title} {{Bell nonlocality}}}}\ (\bibinfo  {publisher}
  {Oxford University Press},\ \bibinfo {year} {2019})\BibitemShut {NoStop}%
\bibitem [{\citenamefont {Wood}\ and\ \citenamefont
  {Spekkens}(2015)}]{wood2015lesson}%
  \BibitemOpen
  \bibfield  {author} {\bibinfo {author} {\bibfnamefont {C.~J.}\ \bibnamefont
  {Wood}}\ and\ \bibinfo {author} {\bibfnamefont {R.~W.}\ \bibnamefont
  {Spekkens}},\ }\bibfield  {title} {\enquote {\bibinfo {title} {{The lesson of
  causal discovery algorithms for quantum correlations Causal explanations of
  Bell-inequality violations require fine-tuning}},}\ }\href
  {https://doi.org/10.1088/1367-2630/17/3/033002} {\bibfield  {journal}
  {\bibinfo  {journal} {New J. Phys.}\ }\textbf {\bibinfo {volume} {17}},\
  \bibinfo {pages} {033002} (\bibinfo {year} {2015})}\BibitemShut {NoStop}%
\bibitem [{\citenamefont {Fritz}(2016)}]{fritz2016beyond}%
  \BibitemOpen
  \bibfield  {author} {\bibinfo {author} {\bibfnamefont {T.}~\bibnamefont
  {Fritz}},\ }\bibfield  {title} {\enquote {\bibinfo {title} {{Beyond Bell's
  theorem II: Scenarios with arbitrary causal structure}},}\ }\href
  {https://doi.org/10.1007/s00220-015-2495-5} {\bibfield  {journal} {\bibinfo
  {journal} {Comm. Math. Phys.}\ }\textbf {\bibinfo {volume} {341}},\ \bibinfo
  {pages} {391} (\bibinfo {year} {2016})}\BibitemShut {NoStop}%
\bibitem [{\citenamefont {Wiseman}\ and\ \citenamefont
  {Cavalcanti}(2017)}]{wiseman2017causarum}%
  \BibitemOpen
  \bibfield  {author} {\bibinfo {author} {\bibfnamefont {H.~M.}\ \bibnamefont
  {Wiseman}}\ and\ \bibinfo {author} {\bibfnamefont {E.~G.}\ \bibnamefont
  {Cavalcanti}},\ }\bibfield  {title} {\enquote {\bibinfo {title} {{Causarum
  Investigatio and the two Bell's theorems of John Bell}},}\ }in\ \href
  {https://doi.org/10.1007/978-3-319-38987-5\_6} {\emph {\bibinfo {booktitle}
  {Quantum [Un] Speakables II}}}\ (\bibinfo  {publisher} {Springer},\ \bibinfo
  {year} {2017})\ pp.\ \bibinfo {pages} {119--142}\BibitemShut {NoStop}%
\bibitem [{\citenamefont {Schmid}\ \emph {et~al.}(2020)\citenamefont {Schmid},
  \citenamefont {Selby},\ and\ \citenamefont
  {Spekkens}}]{schmid2020unscrambling}%
  \BibitemOpen
  \bibfield  {author} {\bibinfo {author} {\bibfnamefont {D.}~\bibnamefont
  {Schmid}}, \bibinfo {author} {\bibfnamefont {J.~H.}\ \bibnamefont {Selby}},\
  and\ \bibinfo {author} {\bibfnamefont {R.~W.}\ \bibnamefont {Spekkens}},\
  }\bibfield  {title} {\enquote {\bibinfo {title} {{Unscrambling the omelette
  of causation and inference: The framework of causal-inferential theories}},}\
  }\href {https://arxiv.org/abs/2009.03297} {\bibfield  {journal} {\bibinfo
  {journal} {arXiv:2009.03297}\ } (\bibinfo {year} {2020})}\BibitemShut
  {NoStop}%
\bibitem [{\citenamefont {Chaves}\ \emph
  {et~al.}(2015{\natexlab{a}})\citenamefont {Chaves}, \citenamefont {Majenz},\
  and\ \citenamefont {Gross}}]{chaves2015information}%
  \BibitemOpen
  \bibfield  {author} {\bibinfo {author} {\bibfnamefont {R.}~\bibnamefont
  {Chaves}}, \bibinfo {author} {\bibfnamefont {C.}~\bibnamefont {Majenz}},\
  and\ \bibinfo {author} {\bibfnamefont {D.}~\bibnamefont {Gross}},\ }\bibfield
   {title} {\enquote {\bibinfo {title} {{Information--theoretic implications of
  quantum causal structures}},}\ }\href {https://doi.org/10.1038/ncomms6766}
  {\bibfield  {journal} {\bibinfo  {journal} {Nature comm.}\ }\textbf {\bibinfo
  {volume} {6}},\ \bibinfo {pages} {1} (\bibinfo {year}
  {2015}{\natexlab{a}})}\BibitemShut {NoStop}%
\bibitem [{\citenamefont {Cavalcanti}\ and\ \citenamefont
  {Lal}(2014)}]{cavalcanti2014modifications}%
  \BibitemOpen
  \bibfield  {author} {\bibinfo {author} {\bibfnamefont {E.~G.}\ \bibnamefont
  {Cavalcanti}}\ and\ \bibinfo {author} {\bibfnamefont {R.}~\bibnamefont
  {Lal}},\ }\bibfield  {title} {\enquote {\bibinfo {title} {{On modifications
  of Reichenbach's principle of common cause in light of Bell's theorem}},}\
  }\href@noop {} {\bibfield  {journal} {\bibinfo  {journal} {J. Phys. A.}\
  }\textbf {\bibinfo {volume} {47}},\ \bibinfo {pages} {424018} (\bibinfo
  {year} {2014})}\BibitemShut {NoStop}%
\bibitem [{\citenamefont {Costa}\ and\ \citenamefont
  {Shrapnel}(2016)}]{costa2016quantum}%
  \BibitemOpen
  \bibfield  {author} {\bibinfo {author} {\bibfnamefont {F.}~\bibnamefont
  {Costa}}\ and\ \bibinfo {author} {\bibfnamefont {S.}~\bibnamefont
  {Shrapnel}},\ }\bibfield  {title} {\enquote {\bibinfo {title} {{Quantum
  causal modelling}},}\ }\href {https://doi.org/10.1088/1367-2630/18/6/063032}
  {\bibfield  {journal} {\bibinfo  {journal} {New J. Phys.}\ }\textbf {\bibinfo
  {volume} {18}},\ \bibinfo {pages} {063032} (\bibinfo {year}
  {2016})}\BibitemShut {NoStop}%
\bibitem [{\citenamefont {Allen}\ \emph {et~al.}(2017)\citenamefont {Allen},
  \citenamefont {Barrett}, \citenamefont {Horsman}, \citenamefont {Lee},\ and\
  \citenamefont {Spekkens}}]{allen2017quantum}%
  \BibitemOpen
  \bibfield  {author} {\bibinfo {author} {\bibfnamefont {J.-M.~A.}\
  \bibnamefont {Allen}}, \bibinfo {author} {\bibfnamefont {J.}~\bibnamefont
  {Barrett}}, \bibinfo {author} {\bibfnamefont {D.~C.}\ \bibnamefont
  {Horsman}}, \bibinfo {author} {\bibfnamefont {C.~M.}\ \bibnamefont {Lee}},\
  and\ \bibinfo {author} {\bibfnamefont {R.~W.}\ \bibnamefont {Spekkens}},\
  }\bibfield  {title} {\enquote {\bibinfo {title} {{Quantum common causes and
  quantum causal models}},}\ }\href {https://doi.org/10.1103/PhysRevX.7.031021}
  {\bibfield  {journal} {\bibinfo  {journal} {Phys. Rev. X}\ }\textbf {\bibinfo
  {volume} {7}},\ \bibinfo {pages} {031021} (\bibinfo {year}
  {2017})}\BibitemShut {NoStop}%
\bibitem [{\citenamefont {Barrett}\ \emph {et~al.}(2019)\citenamefont
  {Barrett}, \citenamefont {Lorenz},\ and\ \citenamefont
  {Oreshkov}}]{barrett2019quantum}%
  \BibitemOpen
  \bibfield  {author} {\bibinfo {author} {\bibfnamefont {J.}~\bibnamefont
  {Barrett}}, \bibinfo {author} {\bibfnamefont {R.}~\bibnamefont {Lorenz}},\
  and\ \bibinfo {author} {\bibfnamefont {O.}~\bibnamefont {Oreshkov}},\
  }\href@noop {} {\enquote {\bibinfo {title} {{Quantum Causal Models}},}\ }
  (\bibinfo {year} {2019}),\ \Eprint {https://arxiv.org/abs/1906.10726}
  {arXiv:1906.10726 [quant-ph]} \BibitemShut {NoStop}%
\bibitem [{\citenamefont {Wolfe}\ \emph {et~al.}(2021)\citenamefont {Wolfe},
  \citenamefont {Pozas-Kerstjens}, \citenamefont {Grinberg}, \citenamefont
  {Rosset}, \citenamefont {Ac{\'i}n},\ and\ \citenamefont
  {Navascu{\'e}s}}]{wolfe2021quantum}%
  \BibitemOpen
  \bibfield  {author} {\bibinfo {author} {\bibfnamefont {E.}~\bibnamefont
  {Wolfe}}, \bibinfo {author} {\bibfnamefont {A.}~\bibnamefont
  {Pozas-Kerstjens}}, \bibinfo {author} {\bibfnamefont {M.}~\bibnamefont
  {Grinberg}}, \bibinfo {author} {\bibfnamefont {D.}~\bibnamefont {Rosset}},
  \bibinfo {author} {\bibfnamefont {A.}~\bibnamefont {Ac{\'i}n}},\ and\
  \bibinfo {author} {\bibfnamefont {M.}~\bibnamefont {Navascu{\'e}s}},\
  }\bibfield  {title} {\enquote {\bibinfo {title} {{Quantum inflation: A
  general approach to quantum causal compatibility}},}\ }\href
  {https://link.aps.org/doi/10.1103/PhysRevX.11.021043} {\bibfield  {journal}
  {\bibinfo  {journal} {Phys. Rev. X}\ }\textbf {\bibinfo {volume} {11}},\
  \bibinfo {pages} {021043} (\bibinfo {year} {2021})}\BibitemShut {NoStop}%
\bibitem [{\citenamefont {Yurke}\ and\ \citenamefont
  {Stoler}(1992)}]{yurke1992einstein}%
  \BibitemOpen
  \bibfield  {author} {\bibinfo {author} {\bibfnamefont {B.}~\bibnamefont
  {Yurke}}\ and\ \bibinfo {author} {\bibfnamefont {D.}~\bibnamefont {Stoler}},\
  }\bibfield  {title} {\enquote {\bibinfo {title} {{Einstein-Podolsky-Rosen
  effects from independent particle sources}},}\ }\href
  {https://link.aps.org/doi/10.1103/PhysRevLett.68.1251} {\bibfield  {journal}
  {\bibinfo  {journal} {Phys. Rev. Lett.}\ }\textbf {\bibinfo {volume} {68}},\
  \bibinfo {pages} {1251} (\bibinfo {year} {1992})}\BibitemShut {NoStop}%
\bibitem [{\citenamefont {Henson}\ \emph {et~al.}(2014)\citenamefont {Henson},
  \citenamefont {Lal},\ and\ \citenamefont {Pusey}}]{henson2014theory}%
  \BibitemOpen
  \bibfield  {author} {\bibinfo {author} {\bibfnamefont {J.}~\bibnamefont
  {Henson}}, \bibinfo {author} {\bibfnamefont {R.}~\bibnamefont {Lal}},\ and\
  \bibinfo {author} {\bibfnamefont {M.~F.}\ \bibnamefont {Pusey}},\ }\bibfield
  {title} {\enquote {\bibinfo {title} {{Theory-independent limits on
  correlations from generalized Bayesian networks}},}\ }\href
  {https://doi.org/10.1088/1367-2630/16/11/113043} {\bibfield  {journal}
  {\bibinfo  {journal} {New J. Phys.}\ }\textbf {\bibinfo {volume} {16}},\
  \bibinfo {pages} {113043} (\bibinfo {year} {2014})}\BibitemShut {NoStop}%
\bibitem [{\citenamefont {Branciard}\ \emph {et~al.}(2012)\citenamefont
  {Branciard}, \citenamefont {Rosset}, \citenamefont {Gisin},\ and\
  \citenamefont {Pironio}}]{branciard2012bilocal}%
  \BibitemOpen
  \bibfield  {author} {\bibinfo {author} {\bibfnamefont {C.}~\bibnamefont
  {Branciard}}, \bibinfo {author} {\bibfnamefont {D.}~\bibnamefont {Rosset}},
  \bibinfo {author} {\bibfnamefont {N.}~\bibnamefont {Gisin}},\ and\ \bibinfo
  {author} {\bibfnamefont {S.}~\bibnamefont {Pironio}},\ }\bibfield  {title}
  {\enquote {\bibinfo {title} {{Bilocal versus nonbilocal correlations in
  entanglement-swapping experiments}},}\ }\href
  {https://doi.org/10.1103/PhysRevA.85.032119} {\bibfield  {journal} {\bibinfo
  {journal} {Phys. Rev. A}\ }\textbf {\bibinfo {volume} {85}},\ \bibinfo
  {pages} {032119} (\bibinfo {year} {2012})}\BibitemShut {NoStop}%
\bibitem [{\citenamefont {Branciard}\ \emph {et~al.}(2010)\citenamefont
  {Branciard}, \citenamefont {Gisin},\ and\ \citenamefont
  {Pironio}}]{branciard2010characterizing}%
  \BibitemOpen
  \bibfield  {author} {\bibinfo {author} {\bibfnamefont {C.}~\bibnamefont
  {Branciard}}, \bibinfo {author} {\bibfnamefont {N.}~\bibnamefont {Gisin}},\
  and\ \bibinfo {author} {\bibfnamefont {S.}~\bibnamefont {Pironio}},\
  }\bibfield  {title} {\enquote {\bibinfo {title} {{Characterizing the nonlocal
  correlations created via entanglement swapping}},}\ }\href
  {https://doi.org/10.1103/PhysRevLett.104.170401} {\bibfield  {journal}
  {\bibinfo  {journal} {Phys. Rev. Lett.}\ }\textbf {\bibinfo {volume} {104}},\
  \bibinfo {pages} {170401} (\bibinfo {year} {2010})}\BibitemShut {NoStop}%
\bibitem [{\citenamefont {Wolfe}\ \emph {et~al.}(2019)\citenamefont {Wolfe},
  \citenamefont {Spekkens},\ and\ \citenamefont {Fritz}}]{wolfe2019inflation}%
  \BibitemOpen
  \bibfield  {author} {\bibinfo {author} {\bibfnamefont {E.}~\bibnamefont
  {Wolfe}}, \bibinfo {author} {\bibfnamefont {R.~W.}\ \bibnamefont
  {Spekkens}},\ and\ \bibinfo {author} {\bibfnamefont {T.}~\bibnamefont
  {Fritz}},\ }\bibfield  {title} {\enquote {\bibinfo {title} {{The Inflation
  Technique for Causal Inference with Latent Variables}},}\ }\href
  {https://dx.doi.org/10.1515/jci-2017-0020} {\bibfield  {journal} {\bibinfo
  {journal} {J. Causal Inference}\ }\textbf {\bibinfo {volume} {7}} (\bibinfo
  {year} {2019})}\BibitemShut {NoStop}%
\bibitem [{\citenamefont {Renou}\ \emph
  {et~al.}(2019{\natexlab{a}})\citenamefont {Renou}, \citenamefont {B{a}umer},
  \citenamefont {Boreiri}, \citenamefont {Brunner}, \citenamefont {Gisin},\
  and\ \citenamefont {Beigi}}]{renou2019genuine}%
  \BibitemOpen
  \bibfield  {author} {\bibinfo {author} {\bibfnamefont {M.-O.}\ \bibnamefont
  {Renou}}, \bibinfo {author} {\bibfnamefont {E.}~\bibnamefont {B{a}umer}},
  \bibinfo {author} {\bibfnamefont {S.}~\bibnamefont {Boreiri}}, \bibinfo
  {author} {\bibfnamefont {N.}~\bibnamefont {Brunner}}, \bibinfo {author}
  {\bibfnamefont {N.}~\bibnamefont {Gisin}},\ and\ \bibinfo {author}
  {\bibfnamefont {S.}~\bibnamefont {Beigi}},\ }\bibfield  {title} {\enquote
  {\bibinfo {title} {{Genuine quantum nonlocality in the triangle network}},}\
  }\href {https://doi.org/10.1103/PhysRevLett.123.140401} {\bibfield  {journal}
  {\bibinfo  {journal} {Phys. Rev. Lett.}\ }\textbf {\bibinfo {volume} {123}},\
  \bibinfo {pages} {140401} (\bibinfo {year} {2019}{\natexlab{a}})}\BibitemShut
  {NoStop}%
\bibitem [{\citenamefont {Pozas-Kerstjens}\ \emph {et~al.}(2019)\citenamefont
  {Pozas-Kerstjens}, \citenamefont {Rabelo}, \citenamefont {Rudnicki},
  \citenamefont {Chaves}, \citenamefont {Cavalcanti}, \citenamefont
  {Navascu\'{e}s},\ and\ \citenamefont {Ac{\'i}n}}]{pozas2019bounding}%
  \BibitemOpen
  \bibfield  {author} {\bibinfo {author} {\bibfnamefont {A.}~\bibnamefont
  {Pozas-Kerstjens}}, \bibinfo {author} {\bibfnamefont {R.}~\bibnamefont
  {Rabelo}}, \bibinfo {author} {\bibfnamefont {L.}~\bibnamefont {Rudnicki}},
  \bibinfo {author} {\bibfnamefont {R.}~\bibnamefont {Chaves}}, \bibinfo
  {author} {\bibfnamefont {D.}~\bibnamefont {Cavalcanti}}, \bibinfo {author}
  {\bibfnamefont {M.}~\bibnamefont {Navascu\'{e}s}},\ and\ \bibinfo {author}
  {\bibfnamefont {A.}~\bibnamefont {Ac{\'i}n}},\ }\bibfield  {title} {\enquote
  {\bibinfo {title} {{Bounding the sets of classical and quantum correlations
  in networks}},}\ }\href {https://doi.org/10.1103/PhysRevLett.123.140503}
  {\bibfield  {journal} {\bibinfo  {journal} {Phys. Rev. Lett.}\ }\textbf
  {\bibinfo {volume} {123}},\ \bibinfo {pages} {140503} (\bibinfo {year}
  {2019})}\BibitemShut {NoStop}%
\bibitem [{\citenamefont {Navascues}\ \emph {et~al.}(2020)\citenamefont
  {Navascues}, \citenamefont {Wolfe}, \citenamefont {Rosset},\ and\
  \citenamefont {Pozas-Kerstjens}}]{navascues2020genuine}%
  \BibitemOpen
  \bibfield  {author} {\bibinfo {author} {\bibfnamefont {M.}~\bibnamefont
  {Navascues}}, \bibinfo {author} {\bibfnamefont {E.}~\bibnamefont {Wolfe}},
  \bibinfo {author} {\bibfnamefont {D.}~\bibnamefont {Rosset}},\ and\ \bibinfo
  {author} {\bibfnamefont {A.}~\bibnamefont {Pozas-Kerstjens}},\ }\bibfield
  {title} {\enquote {\bibinfo {title} {{Genuine network multipartite
  entanglement}},}\ }\href {https://doi.org/10.1103/PhysRevLett.125.240505}
  {\bibfield  {journal} {\bibinfo  {journal} {Phys. Rev. Lett.}\ }\textbf
  {\bibinfo {volume} {125}},\ \bibinfo {pages} {240505} (\bibinfo {year}
  {2020})}\BibitemShut {NoStop}%
\bibitem [{\citenamefont {{Kela}}\ \emph {et~al.}(2020)\citenamefont {{Kela}},
  \citenamefont {{Von Prillwitz}}, \citenamefont {{\AA{}berg}}, \citenamefont
  {{Chaves}},\ and\ \citenamefont {{Gross}}}]{aaberg2020semidefinite}%
  \BibitemOpen
  \bibfield  {author} {\bibinfo {author} {\bibfnamefont {A.}~\bibnamefont
  {{Kela}}}, \bibinfo {author} {\bibfnamefont {K.}~\bibnamefont {{Von
  Prillwitz}}}, \bibinfo {author} {\bibfnamefont {J.}~\bibnamefont
  {{\AA{}berg}}}, \bibinfo {author} {\bibfnamefont {R.}~\bibnamefont
  {{Chaves}}},\ and\ \bibinfo {author} {\bibfnamefont {D.}~\bibnamefont
  {{Gross}}},\ }\bibfield  {title} {\enquote {\bibinfo {title} {{Semidefinite
  Tests for Latent Causal Structures}},}\ }\href
  {https://doi.org/10.1109TIT.2019.2935755} {\bibfield  {journal} {\bibinfo
  {journal} {IEEE Trans. Info. Theo.}\ }\textbf {\bibinfo {volume} {66}},\
  \bibinfo {pages} {339} (\bibinfo {year} {2020})}\BibitemShut {NoStop}%
\bibitem [{\citenamefont {Gisin}(2019)}]{gisin2019entanglement}%
  \BibitemOpen
  \bibfield  {author} {\bibinfo {author} {\bibfnamefont {N.}~\bibnamefont
  {Gisin}},\ }\bibfield  {title} {\enquote {\bibinfo {title} {{Entanglement 25
  years after quantum teleportation testing joint measurements in quantum
  networks}},}\ }\href {https://doi.org/10.3390/e21030325} {\bibfield
  {journal} {\bibinfo  {journal} {Entropy}\ }\textbf {\bibinfo {volume} {21}},\
  \bibinfo {pages} {325} (\bibinfo {year} {2019})}\BibitemShut {NoStop}%
\bibitem [{\citenamefont {Chaves}\ \emph {et~al.}(2018)\citenamefont {Chaves},
  \citenamefont {Carvacho}, \citenamefont {Agresti}, \citenamefont {Di~Giulio},
  \citenamefont {Aolita}, \citenamefont {Giacomini},\ and\ \citenamefont
  {Sciarrino}}]{chaves2018quantum}%
  \BibitemOpen
  \bibfield  {author} {\bibinfo {author} {\bibfnamefont {R.}~\bibnamefont
  {Chaves}}, \bibinfo {author} {\bibfnamefont {G.}~\bibnamefont {Carvacho}},
  \bibinfo {author} {\bibfnamefont {I.}~\bibnamefont {Agresti}}, \bibinfo
  {author} {\bibfnamefont {V.}~\bibnamefont {Di~Giulio}}, \bibinfo {author}
  {\bibfnamefont {L.}~\bibnamefont {Aolita}}, \bibinfo {author} {\bibfnamefont
  {S.}~\bibnamefont {Giacomini}},\ and\ \bibinfo {author} {\bibfnamefont
  {F.}~\bibnamefont {Sciarrino}},\ }\bibfield  {title} {\enquote {\bibinfo
  {title} {{Quantum violation of an instrumental test}},}\ }\href
  {https://doi.org/10.1038/s41567-017-0008-5} {\bibfield  {journal} {\bibinfo
  {journal} {Nature Phys.}\ }\textbf {\bibinfo {volume} {14}},\ \bibinfo
  {pages} {291} (\bibinfo {year} {2018})}\BibitemShut {NoStop}%
\bibitem [{\citenamefont {Tavakoli}\ \emph {et~al.}(2021)\citenamefont
  {Tavakoli}, \citenamefont {Pozas-Kerstjens}, \citenamefont {Renou} \emph
  {et~al.}}]{tavakoli2021bell}%
  \BibitemOpen
  \bibfield  {author} {\bibinfo {author} {\bibfnamefont {A.}~\bibnamefont
  {Tavakoli}}, \bibinfo {author} {\bibfnamefont {A.}~\bibnamefont
  {Pozas-Kerstjens}}, \bibinfo {author} {\bibfnamefont {M.-O.}\ \bibnamefont
  {Renou}}, \emph {et~al.},\ }\bibfield  {title} {\enquote {\bibinfo {title}
  {Bell nonlocality in networks},}\ }\href@noop {} {\bibfield  {journal}
  {\bibinfo  {journal} {Reports on Progress in Physics}\ } (\bibinfo {year}
  {2021})}\BibitemShut {NoStop}%
\bibitem [{\citenamefont {Gebhart}\ \emph {et~al.}(2021)\citenamefont
  {Gebhart}, \citenamefont {Pezz{\`e}},\ and\ \citenamefont
  {Smerzi}}]{gebhart2021genuine}%
  \BibitemOpen
  \bibfield  {author} {\bibinfo {author} {\bibfnamefont {V.}~\bibnamefont
  {Gebhart}}, \bibinfo {author} {\bibfnamefont {L.}~\bibnamefont {Pezz{\`e}}},\
  and\ \bibinfo {author} {\bibfnamefont {A.}~\bibnamefont {Smerzi}},\
  }\bibfield  {title} {\enquote {\bibinfo {title} {{Genuine multipartite
  nonlocality with causal-diagram postselection}},}\ }\href@noop {} {\bibfield
  {journal} {\bibinfo  {journal} {Phys. Rev. Lett.}\ }\textbf {\bibinfo
  {volume} {127}},\ \bibinfo {pages} {140401} (\bibinfo {year}
  {2021})}\BibitemShut {NoStop}%
\bibitem [{\citenamefont {Pirandola}\ \emph {et~al.}(2020)\citenamefont
  {Pirandola}, \citenamefont {Andersen}, \citenamefont {Banchi}, \citenamefont
  {Berta}, \citenamefont {Bunandar}, \citenamefont {Colbeck}, \citenamefont
  {Englund}, \citenamefont {Gehring}, \citenamefont {Lupo}, \citenamefont
  {Ottaviani} \emph {et~al.}}]{pirandola2019advances}%
  \BibitemOpen
  \bibfield  {author} {\bibinfo {author} {\bibfnamefont {S.}~\bibnamefont
  {Pirandola}}, \bibinfo {author} {\bibfnamefont {U.~L.}\ \bibnamefont
  {Andersen}}, \bibinfo {author} {\bibfnamefont {L.}~\bibnamefont {Banchi}},
  \bibinfo {author} {\bibfnamefont {M.}~\bibnamefont {Berta}}, \bibinfo
  {author} {\bibfnamefont {D.}~\bibnamefont {Bunandar}}, \bibinfo {author}
  {\bibfnamefont {R.}~\bibnamefont {Colbeck}}, \bibinfo {author} {\bibfnamefont
  {D.}~\bibnamefont {Englund}}, \bibinfo {author} {\bibfnamefont
  {T.}~\bibnamefont {Gehring}}, \bibinfo {author} {\bibfnamefont
  {C.}~\bibnamefont {Lupo}}, \bibinfo {author} {\bibfnamefont {C.}~\bibnamefont
  {Ottaviani}}, \emph {et~al.},\ }\bibfield  {title} {\enquote {\bibinfo
  {title} {{Advances in quantum cryptography}},}\ }\href
  {https://doi.org/10.1364/AOP.361502} {\bibfield  {journal} {\bibinfo
  {journal} {Adv. Opt. Photon.}\ ,\ \bibinfo {pages} {1012}} (\bibinfo {year}
  {2020})}\BibitemShut {NoStop}%
\bibitem [{\citenamefont {{\v{S}}upi{\'c}}\ and\ \citenamefont
  {Bowles}(2020)}]{vsupic2020self}%
  \BibitemOpen
  \bibfield  {author} {\bibinfo {author} {\bibfnamefont {I.}~\bibnamefont
  {{\v{S}}upi{\'c}}}\ and\ \bibinfo {author} {\bibfnamefont {J.}~\bibnamefont
  {Bowles}},\ }\bibfield  {title} {\enquote {\bibinfo {title} {{Self-testing of
  quantum systems: a review}},}\ }\href
  {https://doi.org/10.22331/q-2020-09-30-337} {\bibfield  {journal} {\bibinfo
  {journal} {Quantum}\ }\textbf {\bibinfo {volume} {4}},\ \bibinfo {pages}
  {337} (\bibinfo {year} {2020})}\BibitemShut {NoStop}%
\bibitem [{\citenamefont {Brukner}\ \emph {et~al.}(2004)\citenamefont
  {Brukner}, \citenamefont {{\.Z}ukowski}, \citenamefont {Pan},\ and\
  \citenamefont {Zeilinger}}]{brukner2004bell}%
  \BibitemOpen
  \bibfield  {author} {\bibinfo {author} {\bibfnamefont {{\v{C}}.}~\bibnamefont
  {Brukner}}, \bibinfo {author} {\bibfnamefont {M.}~\bibnamefont
  {{\.Z}ukowski}}, \bibinfo {author} {\bibfnamefont {J.-W.}\ \bibnamefont
  {Pan}},\ and\ \bibinfo {author} {\bibfnamefont {A.}~\bibnamefont
  {Zeilinger}},\ }\bibfield  {title} {\enquote {\bibinfo {title} {{Bell's
  Inequalities and Quantum Communication Complexity}},}\ }\href
  {https://doi.org/10.1103/PhysRevLett.92.127901} {\bibfield  {journal}
  {\bibinfo  {journal} {Phys. Rev. Lett.}\ }\textbf {\bibinfo {volume} {92}},\
  \bibinfo {pages} {127901} (\bibinfo {year} {2004})}\BibitemShut {NoStop}%
\bibitem [{\citenamefont {Ac{\'i}n}\ \emph {et~al.}(2007)\citenamefont
  {Ac{\'i}n}, \citenamefont {Brunner}, \citenamefont {Gisin}, \citenamefont
  {Massar}, \citenamefont {Pironio},\ and\ \citenamefont
  {Scarani}}]{acin2007device}%
  \BibitemOpen
  \bibfield  {author} {\bibinfo {author} {\bibfnamefont {A.}~\bibnamefont
  {Ac{\'i}n}}, \bibinfo {author} {\bibfnamefont {N.}~\bibnamefont {Brunner}},
  \bibinfo {author} {\bibfnamefont {N.}~\bibnamefont {Gisin}}, \bibinfo
  {author} {\bibfnamefont {S.}~\bibnamefont {Massar}}, \bibinfo {author}
  {\bibfnamefont {S.}~\bibnamefont {Pironio}},\ and\ \bibinfo {author}
  {\bibfnamefont {V.}~\bibnamefont {Scarani}},\ }\bibfield  {title} {\enquote
  {\bibinfo {title} {{Device-independent security of quantum cryptography
  against collective attacks}},}\ }\href
  {https://doi.org/10.1103/PhysRevLett.98.230501} {\bibfield  {journal}
  {\bibinfo  {journal} {Phys. Rev. Lett.}\ }\textbf {\bibinfo {volume} {98}},\
  \bibinfo {pages} {230501} (\bibinfo {year} {2007})}\BibitemShut {NoStop}%
\bibitem [{\citenamefont {Ac{\'i}n}\ and\ \citenamefont
  {Masanes}(2016)}]{acin2016certified}%
  \BibitemOpen
  \bibfield  {author} {\bibinfo {author} {\bibfnamefont {A.}~\bibnamefont
  {Ac{\'i}n}}\ and\ \bibinfo {author} {\bibfnamefont {L.}~\bibnamefont
  {Masanes}},\ }\bibfield  {title} {\enquote {\bibinfo {title} {{Certified
  randomness in quantum physics}},}\ }\href
  {https://doi.org/10.1038/nature20119} {\bibfield  {journal} {\bibinfo
  {journal} {Nature}\ }\textbf {\bibinfo {volume} {540}},\ \bibinfo {pages}
  {213} (\bibinfo {year} {2016})}\BibitemShut {NoStop}%
\bibitem [{\citenamefont {Wehner}\ \emph {et~al.}(2018)\citenamefont {Wehner},
  \citenamefont {Elkouss},\ and\ \citenamefont {Hanson}}]{wehner2018quantum}%
  \BibitemOpen
  \bibfield  {author} {\bibinfo {author} {\bibfnamefont {S.}~\bibnamefont
  {Wehner}}, \bibinfo {author} {\bibfnamefont {D.}~\bibnamefont {Elkouss}},\
  and\ \bibinfo {author} {\bibfnamefont {R.}~\bibnamefont {Hanson}},\
  }\bibfield  {title} {\enquote {\bibinfo {title} {{Quantum internet: A vision
  for the road ahead}},}\ }\href {https://doi.org/10.1126/science.aam9288}
  {\bibfield  {journal} {\bibinfo  {journal} {Science}\ }\textbf {\bibinfo
  {volume} {362}},\ \bibinfo {pages} {eaam9288} (\bibinfo {year}
  {2018})}\BibitemShut {NoStop}%
\bibitem [{\citenamefont {Kimble}(2008)}]{kimble2008quantum}%
  \BibitemOpen
  \bibfield  {author} {\bibinfo {author} {\bibfnamefont {H.~J.}\ \bibnamefont
  {Kimble}},\ }\bibfield  {title} {\enquote {\bibinfo {title} {{The quantum
  internet}},}\ }\href {https://doi.org/10.1038/nature07127} {\bibfield
  {journal} {\bibinfo  {journal} {Nature}\ }\textbf {\bibinfo {volume} {453}},\
  \bibinfo {pages} {1023} (\bibinfo {year} {2008})}\BibitemShut {NoStop}%
\bibitem [{\citenamefont {Briegel}\ \emph {et~al.}(1998)\citenamefont
  {Briegel}, \citenamefont {D{\"u}r}, \citenamefont {Cirac},\ and\
  \citenamefont {Zoller}}]{briegel1998quantum}%
  \BibitemOpen
  \bibfield  {author} {\bibinfo {author} {\bibfnamefont {H.-J.}\ \bibnamefont
  {Briegel}}, \bibinfo {author} {\bibfnamefont {W.}~\bibnamefont {D{\"u}r}},
  \bibinfo {author} {\bibfnamefont {J.~I.}\ \bibnamefont {Cirac}},\ and\
  \bibinfo {author} {\bibfnamefont {P.}~\bibnamefont {Zoller}},\ }\bibfield
  {title} {\enquote {\bibinfo {title} {{Quantum repeaters: the role of
  imperfect local operations in quantum communication}},}\ }\href
  {https://doi.org/10.1103/PhysRevLett.81.5932} {\bibfield  {journal} {\bibinfo
   {journal} {Phys. Rev. Lett.}\ }\textbf {\bibinfo {volume} {81}},\ \bibinfo
  {pages} {5932} (\bibinfo {year} {1998})}\BibitemShut {NoStop}%
\bibitem [{\citenamefont {Scheidl}\ \emph {et~al.}(2010)\citenamefont
  {Scheidl}, \citenamefont {Ursin}, \citenamefont {Kofler}, \citenamefont
  {Ramelow}, \citenamefont {Ma}, \citenamefont {Herbst}, \citenamefont
  {Ratschbacher}, \citenamefont {Fedrizzi}, \citenamefont {Langford},
  \citenamefont {Jennewein} \emph {et~al.}}]{scheidl2010violation}%
  \BibitemOpen
  \bibfield  {author} {\bibinfo {author} {\bibfnamefont {T.}~\bibnamefont
  {Scheidl}}, \bibinfo {author} {\bibfnamefont {R.}~\bibnamefont {Ursin}},
  \bibinfo {author} {\bibfnamefont {J.}~\bibnamefont {Kofler}}, \bibinfo
  {author} {\bibfnamefont {S.}~\bibnamefont {Ramelow}}, \bibinfo {author}
  {\bibfnamefont {X.-S.}\ \bibnamefont {Ma}}, \bibinfo {author} {\bibfnamefont
  {T.}~\bibnamefont {Herbst}}, \bibinfo {author} {\bibfnamefont
  {L.}~\bibnamefont {Ratschbacher}}, \bibinfo {author} {\bibfnamefont
  {A.}~\bibnamefont {Fedrizzi}}, \bibinfo {author} {\bibfnamefont {N.~K.}\
  \bibnamefont {Langford}}, \bibinfo {author} {\bibfnamefont {T.}~\bibnamefont
  {Jennewein}}, \emph {et~al.},\ }\bibfield  {title} {\enquote {\bibinfo
  {title} {{Violation of local realism with freedom of choice}},}\ }\href
  {https://doi.org/10.1073/pnas.1002780107} {\bibfield  {journal} {\bibinfo
  {journal} {Proc. Nat. Acad. Sci.}\ }\textbf {\bibinfo {volume} {107}},\
  \bibinfo {pages} {19708} (\bibinfo {year} {2010})}\BibitemShut {NoStop}%
\bibitem [{\citenamefont {Weihs}\ \emph {et~al.}(1998)\citenamefont {Weihs},
  \citenamefont {Jennewein}, \citenamefont {Simon}, \citenamefont
  {Weinfurter},\ and\ \citenamefont {Zeilinger}}]{weihs1998violation}%
  \BibitemOpen
  \bibfield  {author} {\bibinfo {author} {\bibfnamefont {G.}~\bibnamefont
  {Weihs}}, \bibinfo {author} {\bibfnamefont {T.}~\bibnamefont {Jennewein}},
  \bibinfo {author} {\bibfnamefont {C.}~\bibnamefont {Simon}}, \bibinfo
  {author} {\bibfnamefont {H.}~\bibnamefont {Weinfurter}},\ and\ \bibinfo
  {author} {\bibfnamefont {A.}~\bibnamefont {Zeilinger}},\ }\bibfield  {title}
  {\enquote {\bibinfo {title} {{Violation of Bell's inequality under strict
  Einstein locality conditions}},}\ }\href
  {https://link.aps.org/doi/10.1103/PhysRevLett.81.5039} {\bibfield  {journal}
  {\bibinfo  {journal} {Phys. Rev. Lett.}\ }\textbf {\bibinfo {volume} {81}},\
  \bibinfo {pages} {5039} (\bibinfo {year} {1998})}\BibitemShut {NoStop}%
\bibitem [{\citenamefont {Shalm}\ \emph {et~al.}(2015)\citenamefont {Shalm},
  \citenamefont {Meyer-Scott}, \citenamefont {Christensen}, \citenamefont
  {Bierhorst}, \citenamefont {Wayne}, \citenamefont {Stevens}, \citenamefont
  {Gerrits}, \citenamefont {Glancy}, \citenamefont {Hamel}, \citenamefont
  {Allman} \emph {et~al.}}]{Shalm}%
  \BibitemOpen
  \bibfield  {author} {\bibinfo {author} {\bibfnamefont {L.~K.}\ \bibnamefont
  {Shalm}}, \bibinfo {author} {\bibfnamefont {E.}~\bibnamefont {Meyer-Scott}},
  \bibinfo {author} {\bibfnamefont {B.~G.}\ \bibnamefont {Christensen}},
  \bibinfo {author} {\bibfnamefont {P.}~\bibnamefont {Bierhorst}}, \bibinfo
  {author} {\bibfnamefont {M.~A.}\ \bibnamefont {Wayne}}, \bibinfo {author}
  {\bibfnamefont {M.~J.}\ \bibnamefont {Stevens}}, \bibinfo {author}
  {\bibfnamefont {T.}~\bibnamefont {Gerrits}}, \bibinfo {author} {\bibfnamefont
  {S.}~\bibnamefont {Glancy}}, \bibinfo {author} {\bibfnamefont {D.~R.}\
  \bibnamefont {Hamel}}, \bibinfo {author} {\bibfnamefont {M.~S.}\ \bibnamefont
  {Allman}}, \emph {et~al.},\ }\bibfield  {title} {\enquote {\bibinfo {title}
  {{Strong loophole-free test of local realism}},}\ }\href
  {https://doi.org/10.1103/PhysRevLett.115.250402} {\bibfield  {journal}
  {\bibinfo  {journal} {Phys. Rev. Lett.}\ }\textbf {\bibinfo {volume} {115}},\
  \bibinfo {pages} {250402} (\bibinfo {year} {2015})}\BibitemShut {NoStop}%
\bibitem [{\citenamefont {Giustina}\ \emph {et~al.}(2015)\citenamefont
  {Giustina}, \citenamefont {Versteegh}, \citenamefont {Wengerowsky},
  \citenamefont {Handsteiner}, \citenamefont {Hochrainer}, \citenamefont
  {Phelan}, \citenamefont {Steinlechner}, \citenamefont {Kofler}, \citenamefont
  {Larsson}, \citenamefont {Abell\'{a}n} \emph {et~al.}}]{Giustina}%
  \BibitemOpen
  \bibfield  {author} {\bibinfo {author} {\bibfnamefont {M.}~\bibnamefont
  {Giustina}}, \bibinfo {author} {\bibfnamefont {M.~A.}\ \bibnamefont
  {Versteegh}}, \bibinfo {author} {\bibfnamefont {S.}~\bibnamefont
  {Wengerowsky}}, \bibinfo {author} {\bibfnamefont {J.}~\bibnamefont
  {Handsteiner}}, \bibinfo {author} {\bibfnamefont {A.}~\bibnamefont
  {Hochrainer}}, \bibinfo {author} {\bibfnamefont {K.}~\bibnamefont {Phelan}},
  \bibinfo {author} {\bibfnamefont {F.}~\bibnamefont {Steinlechner}}, \bibinfo
  {author} {\bibfnamefont {J.}~\bibnamefont {Kofler}}, \bibinfo {author}
  {\bibfnamefont {J.-A.}\ \bibnamefont {Larsson}}, \bibinfo {author}
  {\bibfnamefont {C.}~\bibnamefont {Abell\'{a}n}}, \emph {et~al.},\ }\bibfield
  {title} {\enquote {\bibinfo {title} {{Significant-loophole-free test of
  Bell's theorem with entangled photons}},}\ }\href
  {https://doi.org/10.1103/PhysRevLett.115.250401} {\bibfield  {journal}
  {\bibinfo  {journal} {Phys. Rev. Lett.}\ }\textbf {\bibinfo {volume} {115}},\
  \bibinfo {pages} {250401} (\bibinfo {year} {2015})}\BibitemShut {NoStop}%
\bibitem [{\citenamefont {Hensen}\ \emph {et~al.}(2015)\citenamefont {Hensen},
  \citenamefont {Bernien}, \citenamefont {Dr\'{e}au}, \citenamefont {Reiserer},
  \citenamefont {Kalb}, \citenamefont {Blok}, \citenamefont {Ruitenberg},
  \citenamefont {Vermeulen}, \citenamefont {Schouten}, \citenamefont
  {Abell\'{a}n}, \citenamefont {Amaya}, \citenamefont {Pruneri}, \citenamefont
  {Mitchell}, \citenamefont {Markham}, \citenamefont {Twitchen}, \citenamefont
  {Elkouss}, \citenamefont {Wehner}, \citenamefont {Taminiau},\ and\
  \citenamefont {Hanson}}]{Hensen}%
  \BibitemOpen
  \bibfield  {author} {\bibinfo {author} {\bibfnamefont {B.}~\bibnamefont
  {Hensen}}, \bibinfo {author} {\bibfnamefont {H.}~\bibnamefont {Bernien}},
  \bibinfo {author} {\bibfnamefont {A.~E.}\ \bibnamefont {Dr\'{e}au}}, \bibinfo
  {author} {\bibfnamefont {A.}~\bibnamefont {Reiserer}}, \bibinfo {author}
  {\bibfnamefont {N.}~\bibnamefont {Kalb}}, \bibinfo {author} {\bibfnamefont
  {M.~S.}\ \bibnamefont {Blok}}, \bibinfo {author} {\bibfnamefont
  {J.}~\bibnamefont {Ruitenberg}}, \bibinfo {author} {\bibfnamefont {R.~F.}\
  \bibnamefont {Vermeulen}}, \bibinfo {author} {\bibfnamefont {R.~N.}\
  \bibnamefont {Schouten}}, \bibinfo {author} {\bibfnamefont {C.}~\bibnamefont
  {Abell\'{a}n}}, \bibinfo {author} {\bibfnamefont {W.}~\bibnamefont {Amaya}},
  \bibinfo {author} {\bibfnamefont {V.}~\bibnamefont {Pruneri}}, \bibinfo
  {author} {\bibfnamefont {M.~W.}\ \bibnamefont {Mitchell}}, \bibinfo {author}
  {\bibfnamefont {M.}~\bibnamefont {Markham}}, \bibinfo {author} {\bibfnamefont
  {D.~J.}\ \bibnamefont {Twitchen}}, \bibinfo {author} {\bibfnamefont
  {D.}~\bibnamefont {Elkouss}}, \bibinfo {author} {\bibfnamefont
  {S.}~\bibnamefont {Wehner}}, \bibinfo {author} {\bibfnamefont {T.~H.}\
  \bibnamefont {Taminiau}},\ and\ \bibinfo {author} {\bibfnamefont
  {R.}~\bibnamefont {Hanson}},\ }\bibfield  {title} {\enquote {\bibinfo {title}
  {{Loophole-free Bell inequality violation using electron spins separated by
  1.3 kilometres}},}\ }\href {https://doi.org/10.1038/nature15759} {\bibfield
  {journal} {\bibinfo  {journal} {Nature}\ }\textbf {\bibinfo {volume} {526}},\
  \bibinfo {pages} {682} (\bibinfo {year} {2015})}\BibitemShut {NoStop}%
\bibitem [{\citenamefont {Hooft}(2007)}]{hooft2007free}%
  \BibitemOpen
  \bibfield  {author} {\bibinfo {author} {\bibfnamefont {G.}~\bibnamefont
  {Hooft}},\ }\bibfield  {title} {\enquote {\bibinfo {title} {{The free-will
  postulate in quantum mechanics}},}\ }\href
  {https://arxiv.org/abs/quant-ph/0701097} {\bibfield  {journal} {\bibinfo
  {journal} {arXiv:quant-ph/0701097}\ } (\bibinfo {year} {2007})}\BibitemShut
  {NoStop}%
\bibitem [{\citenamefont {{BIG Bell Test Collaboration and
  others}}(2018)}]{big2018challenging}%
  \BibitemOpen
  \bibfield  {author} {\bibinfo {author} {\bibnamefont {{BIG Bell Test
  Collaboration and others}}},\ }\bibfield  {title} {\enquote {\bibinfo {title}
  {{Challenging local realism with human choices}},}\ }\href
  {https://doi.org/10.1038/s41586-018-0085-3} {\bibfield  {journal} {\bibinfo
  {journal} {Nature}\ }\textbf {\bibinfo {volume} {557}},\ \bibinfo {pages}
  {212} (\bibinfo {year} {2018})}\BibitemShut {NoStop}%
\bibitem [{\citenamefont {Rauch}\ \emph {et~al.}(2018)\citenamefont {Rauch},
  \citenamefont {Handsteiner}, \citenamefont {Hochrainer}, \citenamefont
  {Gallicchio}, \citenamefont {Friedman}, \citenamefont {Leung}, \citenamefont
  {Liu}, \citenamefont {Bulla}, \citenamefont {Ecker}, \citenamefont
  {Steinlechner} \emph {et~al.}}]{rauch2018cosmic}%
  \BibitemOpen
  \bibfield  {author} {\bibinfo {author} {\bibfnamefont {D.}~\bibnamefont
  {Rauch}}, \bibinfo {author} {\bibfnamefont {J.}~\bibnamefont {Handsteiner}},
  \bibinfo {author} {\bibfnamefont {A.}~\bibnamefont {Hochrainer}}, \bibinfo
  {author} {\bibfnamefont {J.}~\bibnamefont {Gallicchio}}, \bibinfo {author}
  {\bibfnamefont {A.~S.}\ \bibnamefont {Friedman}}, \bibinfo {author}
  {\bibfnamefont {C.}~\bibnamefont {Leung}}, \bibinfo {author} {\bibfnamefont
  {B.}~\bibnamefont {Liu}}, \bibinfo {author} {\bibfnamefont {L.}~\bibnamefont
  {Bulla}}, \bibinfo {author} {\bibfnamefont {S.}~\bibnamefont {Ecker}},
  \bibinfo {author} {\bibfnamefont {F.}~\bibnamefont {Steinlechner}}, \emph
  {et~al.},\ }\bibfield  {title} {\enquote {\bibinfo {title} {{Cosmic Bell test
  using random measurement settings from high-redshift quasars}},}\ }\href
  {https://doi.org/10.1103/PhysRevLett.121.080403} {\bibfield  {journal}
  {\bibinfo  {journal} {Phys. Rev. Lett.}\ }\textbf {\bibinfo {volume} {121}},\
  \bibinfo {pages} {080403} (\bibinfo {year} {2018})}\BibitemShut {NoStop}%
\bibitem [{\citenamefont {Abiuso}\ \emph {et~al.}(2022)\citenamefont {Abiuso},
  \citenamefont {Kriv{\'a}chy}, \citenamefont {Boghiu}, \citenamefont {Renou},
  \citenamefont {Pozas-Kerstjens},\ and\ \citenamefont
  {Ac{\'i}n}}]{abiuso2022single}%
  \BibitemOpen
  \bibfield  {author} {\bibinfo {author} {\bibfnamefont {P.}~\bibnamefont
  {Abiuso}}, \bibinfo {author} {\bibfnamefont {T.}~\bibnamefont
  {Kriv{\'a}chy}}, \bibinfo {author} {\bibfnamefont {E.-C.}\ \bibnamefont
  {Boghiu}}, \bibinfo {author} {\bibfnamefont {M.-O.}\ \bibnamefont {Renou}},
  \bibinfo {author} {\bibfnamefont {A.}~\bibnamefont {Pozas-Kerstjens}},\ and\
  \bibinfo {author} {\bibfnamefont {A.}~\bibnamefont {Ac{\'i}n}},\ }\bibfield
  {title} {\enquote {\bibinfo {title} {{Single-photon nonlocality in quantum
  networks}},}\ }\href {https://doi.org/10.1103/PhysRevResearch.4.L012041}
  {\bibfield  {journal} {\bibinfo  {journal} {Phys. Rev. Res.}\ }\textbf
  {\bibinfo {volume} {4}},\ \bibinfo {pages} {L012041} (\bibinfo {year}
  {2022})}\BibitemShut {NoStop}%
\bibitem [{\citenamefont {Chaves}\ \emph {et~al.}(2021)\citenamefont {Chaves},
  \citenamefont {Moreno}, \citenamefont {Polino}, \citenamefont {Poderini},
  \citenamefont {Agresti}, \citenamefont {Suprano}, \citenamefont {Barros},
  \citenamefont {Carvacho}, \citenamefont {Wolfe}, \citenamefont {Canabarro}
  \emph {et~al.}}]{chaves2021causal}%
  \BibitemOpen
  \bibfield  {author} {\bibinfo {author} {\bibfnamefont {R.}~\bibnamefont
  {Chaves}}, \bibinfo {author} {\bibfnamefont {G.}~\bibnamefont {Moreno}},
  \bibinfo {author} {\bibfnamefont {E.}~\bibnamefont {Polino}}, \bibinfo
  {author} {\bibfnamefont {D.}~\bibnamefont {Poderini}}, \bibinfo {author}
  {\bibfnamefont {I.}~\bibnamefont {Agresti}}, \bibinfo {author} {\bibfnamefont
  {A.}~\bibnamefont {Suprano}}, \bibinfo {author} {\bibfnamefont {M.~R.}\
  \bibnamefont {Barros}}, \bibinfo {author} {\bibfnamefont {G.}~\bibnamefont
  {Carvacho}}, \bibinfo {author} {\bibfnamefont {E.}~\bibnamefont {Wolfe}},
  \bibinfo {author} {\bibfnamefont {A.}~\bibnamefont {Canabarro}}, \emph
  {et~al.},\ }\bibfield  {title} {\enquote {\bibinfo {title} {Causal networks
  and freedom of choice in bell’s theorem},}\ }\href
  {https://doi.org/https://doi.org/10.1103/PRXQuantum.2.040323} {\bibfield
  {journal} {\bibinfo  {journal} {PRX Quantum}\ }\textbf {\bibinfo {volume}
  {2}},\ \bibinfo {pages} {040323} (\bibinfo {year} {2021})}\BibitemShut
  {NoStop}%
\bibitem [{\citenamefont {Boreiri}\ \emph {et~al.}(2022)\citenamefont
  {Boreiri}, \citenamefont {Girardin}, \citenamefont {Ulu}, \citenamefont
  {Lypka-Bartosik}, \citenamefont {Brunner},\ and\ \citenamefont
  {Sekatski}}]{Boreiri2022towards}%
  \BibitemOpen
  \bibfield  {author} {\bibinfo {author} {\bibfnamefont {S.}~\bibnamefont
  {Boreiri}}, \bibinfo {author} {\bibfnamefont {A.}~\bibnamefont {Girardin}},
  \bibinfo {author} {\bibfnamefont {B.}~\bibnamefont {Ulu}}, \bibinfo {author}
  {\bibfnamefont {P.}~\bibnamefont {Lypka-Bartosik}}, \bibinfo {author}
  {\bibfnamefont {N.}~\bibnamefont {Brunner}},\ and\ \bibinfo {author}
  {\bibfnamefont {P.}~\bibnamefont {Sekatski}},\ }\bibfield  {title} {\enquote
  {\bibinfo {title} {{Towards a minimal example of quantum nonlocality without
  inputs}},}\ }\href {https://arxiv.org/abs/2207.08532} {\bibfield  {journal}
  {\bibinfo  {journal} {arXiv:2009.03297}\ } (\bibinfo {year}
  {2022})}\BibitemShut {NoStop}%
\bibitem [{\citenamefont {Chaves}\ \emph
  {et~al.}(2014{\natexlab{a}})\citenamefont {Chaves}, \citenamefont {Luft},\
  and\ \citenamefont {Gross}}]{chaves2014causal}%
  \BibitemOpen
  \bibfield  {author} {\bibinfo {author} {\bibfnamefont {R.}~\bibnamefont
  {Chaves}}, \bibinfo {author} {\bibfnamefont {L.}~\bibnamefont {Luft}},\ and\
  \bibinfo {author} {\bibfnamefont {D.}~\bibnamefont {Gross}},\ }\bibfield
  {title} {\enquote {\bibinfo {title} {{Causal structures from entropic
  information geometry and novel scenarios}},}\ }\href
  {https://doi.org/10.1088/1367-2630/16/4/043001} {\bibfield  {journal}
  {\bibinfo  {journal} {New J. Phys.}\ }\textbf {\bibinfo {volume} {16}},\
  \bibinfo {pages} {043001} (\bibinfo {year} {2014}{\natexlab{a}})}\BibitemShut
  {NoStop}%
\bibitem [{\citenamefont {Steudel}\ and\ \citenamefont
  {Ay}(2015)}]{steudel2015information}%
  \BibitemOpen
  \bibfield  {author} {\bibinfo {author} {\bibfnamefont {B.}~\bibnamefont
  {Steudel}}\ and\ \bibinfo {author} {\bibfnamefont {N.}~\bibnamefont {Ay}},\
  }\bibfield  {title} {\enquote {\bibinfo {title} {{Information-theoretic
  inference of common ancestors}},}\ }\href {https://doi.org/10.3390/e17042304}
  {\bibfield  {journal} {\bibinfo  {journal} {Entropy}\ }\textbf {\bibinfo
  {volume} {17}},\ \bibinfo {pages} {2304} (\bibinfo {year}
  {2015})}\BibitemShut {NoStop}%
\bibitem [{\citenamefont {Fraser}\ and\ \citenamefont
  {Wolfe}(2018)}]{fraser2018causal}%
  \BibitemOpen
  \bibfield  {author} {\bibinfo {author} {\bibfnamefont {T.~C.}\ \bibnamefont
  {Fraser}}\ and\ \bibinfo {author} {\bibfnamefont {E.}~\bibnamefont {Wolfe}},\
  }\bibfield  {title} {\enquote {\bibinfo {title} {{Causal compatibility
  inequalities admitting quantum violations in the triangle structure}},}\
  }\href {https://doi.org/10.1103/PhysRevA.98.022113} {\bibfield  {journal}
  {\bibinfo  {journal} {Phys. Rev. A}\ }\textbf {\bibinfo {volume} {98}},\
  \bibinfo {pages} {022113} (\bibinfo {year} {2018})}\BibitemShut {NoStop}%
\bibitem [{\citenamefont {Pusey}(2019)}]{pusey2019quantum}%
  \BibitemOpen
  \bibfield  {author} {\bibinfo {author} {\bibfnamefont {M.~F.}\ \bibnamefont
  {Pusey}},\ }\bibfield  {title} {\enquote {\bibinfo {title} {{Quantum
  correlations take a new shape}},}\ }\href
  {https://doi.org/10.1103/Physics.12.106} {\bibfield  {journal} {\bibinfo
  {journal} {Physics}\ }\textbf {\bibinfo {volume} {12}},\ \bibinfo {pages}
  {113043} (\bibinfo {year} {2019})}\BibitemShut {NoStop}%
\bibitem [{\citenamefont {Kraft}\ \emph {et~al.}(2021)\citenamefont {Kraft},
  \citenamefont {Designolle}, \citenamefont {Ritz}, \citenamefont {Brunner},
  \citenamefont {G{\"u}hne},\ and\ \citenamefont {Huber}}]{kraft2021quantum}%
  \BibitemOpen
  \bibfield  {author} {\bibinfo {author} {\bibfnamefont {T.}~\bibnamefont
  {Kraft}}, \bibinfo {author} {\bibfnamefont {S.}~\bibnamefont {Designolle}},
  \bibinfo {author} {\bibfnamefont {C.}~\bibnamefont {Ritz}}, \bibinfo {author}
  {\bibfnamefont {N.}~\bibnamefont {Brunner}}, \bibinfo {author} {\bibfnamefont
  {O.}~\bibnamefont {G{\"u}hne}},\ and\ \bibinfo {author} {\bibfnamefont
  {M.}~\bibnamefont {Huber}},\ }\bibfield  {title} {\enquote {\bibinfo {title}
  {Quantum entanglement in the triangle network},}\ }\href@noop {} {\bibfield
  {journal} {\bibinfo  {journal} {Physical Review A}\ }\textbf {\bibinfo
  {volume} {103}},\ \bibinfo {pages} {L060401} (\bibinfo {year}
  {2021})}\BibitemShut {NoStop}%
\bibitem [{\citenamefont {\ifmmode \check{S}\else
  \v{S}\fi{}upi\ifmmode~\acute{c}\else \'{c}\fi{}}\ \emph
  {et~al.}(2020)\citenamefont {\ifmmode \check{S}\else
  \v{S}\fi{}upi\ifmmode~\acute{c}\else \'{c}\fi{}}, \citenamefont {Bancal},\
  and\ \citenamefont {Brunner}}]{vsupic2020quantum}%
  \BibitemOpen
  \bibfield  {author} {\bibinfo {author} {\bibfnamefont {I.}~\bibnamefont
  {\ifmmode \check{S}\else \v{S}\fi{}upi\ifmmode~\acute{c}\else \'{c}\fi{}}},
  \bibinfo {author} {\bibfnamefont {J.-D.}\ \bibnamefont {Bancal}},\ and\
  \bibinfo {author} {\bibfnamefont {N.}~\bibnamefont {Brunner}},\ }\bibfield
  {title} {\enquote {\bibinfo {title} {{Quantum Nonlocality in Networks Can Be
  Demonstrated with an Arbitrarily Small Level of Independence between the
  Sources}},}\ }\href {https://doi.org/10.1103/PhysRevLett.125.240403}
  {\bibfield  {journal} {\bibinfo  {journal} {Phys. Rev. Lett.}\ }\textbf
  {\bibinfo {volume} {125}},\ \bibinfo {pages} {240403} (\bibinfo {year}
  {2020})}\BibitemShut {NoStop}%
\bibitem [{\citenamefont {Kriv\'{a}chy}\ \emph {et~al.}(2020)\citenamefont
  {Kriv\'{a}chy}, \citenamefont {Cai}, \citenamefont {Cavalcanti},
  \citenamefont {Tavakoli}, \citenamefont {Gisin},\ and\ \citenamefont
  {Brunner}}]{krivachy2019neural}%
  \BibitemOpen
  \bibfield  {author} {\bibinfo {author} {\bibfnamefont {T.}~\bibnamefont
  {Kriv\'{a}chy}}, \bibinfo {author} {\bibfnamefont {Y.}~\bibnamefont {Cai}},
  \bibinfo {author} {\bibfnamefont {D.}~\bibnamefont {Cavalcanti}}, \bibinfo
  {author} {\bibfnamefont {A.}~\bibnamefont {Tavakoli}}, \bibinfo {author}
  {\bibfnamefont {N.}~\bibnamefont {Gisin}},\ and\ \bibinfo {author}
  {\bibfnamefont {N.}~\bibnamefont {Brunner}},\ }\bibfield  {title} {\enquote
  {\bibinfo {title} {{A neural network oracle for quantum nonlocality problems
  in networks}},}\ }\href {https://dx.doi.org/10.1038/s41534-020-00305-x}
  {\bibfield  {journal} {\bibinfo  {journal} {npj Quant. Inf.}\ }\textbf
  {\bibinfo {volume} {6}} (\bibinfo {year} {2020})}\BibitemShut {NoStop}%
\bibitem [{\citenamefont {Renou}\ \emph
  {et~al.}(2019{\natexlab{b}})\citenamefont {Renou}, \citenamefont {Wang},
  \citenamefont {Boreiri}, \citenamefont {Beigi}, \citenamefont {Gisin},\ and\
  \citenamefont {Brunner}}]{renou2019limits}%
  \BibitemOpen
  \bibfield  {author} {\bibinfo {author} {\bibfnamefont {M.-O.}\ \bibnamefont
  {Renou}}, \bibinfo {author} {\bibfnamefont {Y.}~\bibnamefont {Wang}},
  \bibinfo {author} {\bibfnamefont {S.}~\bibnamefont {Boreiri}}, \bibinfo
  {author} {\bibfnamefont {S.}~\bibnamefont {Beigi}}, \bibinfo {author}
  {\bibfnamefont {N.}~\bibnamefont {Gisin}},\ and\ \bibinfo {author}
  {\bibfnamefont {N.}~\bibnamefont {Brunner}},\ }\bibfield  {title} {\enquote
  {\bibinfo {title} {{Limits on correlations in networks for quantum and
  no-signaling resources}},}\ }\href
  {https://doi.org/10.1103/PhysRevLett.123.070403} {\bibfield  {journal}
  {\bibinfo  {journal} {Phys. Rev. Lett.}\ }\textbf {\bibinfo {volume} {123}},\
  \bibinfo {pages} {070403} (\bibinfo {year} {2019}{\natexlab{b}})}\BibitemShut
  {NoStop}%
\bibitem [{\citenamefont {B{\"a}umer}\ \emph {et~al.}(2021)\citenamefont
  {B{\"a}umer}, \citenamefont {Gisin},\ and\ \citenamefont
  {Tavakoli}}]{baumer2021demonstrating}%
  \BibitemOpen
  \bibfield  {author} {\bibinfo {author} {\bibfnamefont {E.}~\bibnamefont
  {B{\"a}umer}}, \bibinfo {author} {\bibfnamefont {N.}~\bibnamefont {Gisin}},\
  and\ \bibinfo {author} {\bibfnamefont {A.}~\bibnamefont {Tavakoli}},\
  }\bibfield  {title} {\enquote {\bibinfo {title} {Demonstrating the power of
  quantum computers, certification of highly entangled measurements and
  scalable quantum nonlocality},}\ }\href@noop {} {\bibfield  {journal}
  {\bibinfo  {journal} {npj Quantum Information}\ }\textbf {\bibinfo {volume}
  {7}},\ \bibinfo {pages} {117} (\bibinfo {year} {2021})}\BibitemShut {NoStop}%
\bibitem [{\citenamefont {Sekatski}\ \emph {et~al.}(2022)\citenamefont
  {Sekatski}, \citenamefont {Boreiri},\ and\ \citenamefont
  {Brunner}}]{Sekatski2022partial}%
  \BibitemOpen
  \bibfield  {author} {\bibinfo {author} {\bibfnamefont {P.}~\bibnamefont
  {Sekatski}}, \bibinfo {author} {\bibfnamefont {S.}~\bibnamefont {Boreiri}},\
  and\ \bibinfo {author} {\bibfnamefont {N.}~\bibnamefont {Brunner}},\ }\href
  {https://doi.org/10.48550/ARXIV.2209.09921} {\enquote {\bibinfo {title}
  {Partial self-testing and randomness certification in the triangle
  network},}\ } (\bibinfo {year} {2022})\BibitemShut {NoStop}%
\bibitem [{\citenamefont {Greenberger}\ \emph {et~al.}(1989)\citenamefont
  {Greenberger}, \citenamefont {Horne},\ and\ \citenamefont
  {Zeilinger}}]{greenberger1989going}%
  \BibitemOpen
  \bibfield  {author} {\bibinfo {author} {\bibfnamefont {D.~M.}\ \bibnamefont
  {Greenberger}}, \bibinfo {author} {\bibfnamefont {M.~A.}\ \bibnamefont
  {Horne}},\ and\ \bibinfo {author} {\bibfnamefont {A.}~\bibnamefont
  {Zeilinger}},\ }\bibfield  {title} {\enquote {\bibinfo {title} {{Going beyond
  Bell's theorem}},}\ }in\ \href
  {https://doi.org/10.1007/978-94-017-0849-4\_10} {\emph {\bibinfo {booktitle}
  {Bell's theorem, quantum theory and conceptions of the universe}}}\ (\bibinfo
   {publisher} {Springer},\ \bibinfo {year} {1989})\ pp.\ \bibinfo {pages}
  {69--72}\BibitemShut {NoStop}%
\bibitem [{\citenamefont {Clauser}\ \emph {et~al.}(1969)\citenamefont
  {Clauser}, \citenamefont {Horne}, \citenamefont {Shimony},\ and\
  \citenamefont {Holt}}]{clauser1969proposed}%
  \BibitemOpen
  \bibfield  {author} {\bibinfo {author} {\bibfnamefont {J.~F.}\ \bibnamefont
  {Clauser}}, \bibinfo {author} {\bibfnamefont {M.~A.}\ \bibnamefont {Horne}},
  \bibinfo {author} {\bibfnamefont {A.}~\bibnamefont {Shimony}},\ and\ \bibinfo
  {author} {\bibfnamefont {R.~A.}\ \bibnamefont {Holt}},\ }\bibfield  {title}
  {\enquote {\bibinfo {title} {{Proposed experiment to test local
  hidden-variable theories}},}\ }\href
  {https://doi.org/10.1103/PhysRevLett.23.880} {\bibfield  {journal} {\bibinfo
  {journal} {Phys. Rev. Lett.}\ }\textbf {\bibinfo {volume} {23}},\ \bibinfo
  {pages} {880} (\bibinfo {year} {1969})}\BibitemShut {NoStop}%
\bibitem [{\citenamefont {Mermin}(1990)}]{mermin1990}%
  \BibitemOpen
  \bibfield  {author} {\bibinfo {author} {\bibfnamefont {N.~D.}\ \bibnamefont
  {Mermin}},\ }\bibfield  {title} {\enquote {\bibinfo {title} {{Extreme quantum
  entanglement in a superposition of macroscopically distinct states}},}\
  }\href {https://doi.org/10.1103PhysRevLett.65.1838} {\bibfield  {journal}
  {\bibinfo  {journal} {Phys. Rev. Lett.}\ }\textbf {\bibinfo {volume} {65}},\
  \bibinfo {pages} {1838} (\bibinfo {year} {1990})}\BibitemShut {NoStop}%
\bibitem [{\citenamefont {Geiger}\ and\ \citenamefont
  {Meek}(1999)}]{geiger1999quantifier}%
  \BibitemOpen
  \bibfield  {author} {\bibinfo {author} {\bibfnamefont {D.}~\bibnamefont
  {Geiger}}\ and\ \bibinfo {author} {\bibfnamefont {C.}~\bibnamefont {Meek}},\
  }\bibfield  {title} {\enquote {\bibinfo {title} {{Quantifier elimination for
  statistical problems}},}\ }in\ \href {https://arxiv.org/abs/1301.6698} {\emph
  {\bibinfo {booktitle} {Proc. 15th Conf. on Uncertainty in Artificial
  Intelligence}}}\ (\bibinfo {organization} {Morgan Kaufmann Publishers Inc.},\
  \bibinfo {year} {1999})\ pp.\ \bibinfo {pages} {226--235}\BibitemShut
  {NoStop}%
\bibitem [{\citenamefont {Hall}(2011)}]{hall2011relaxed}%
  \BibitemOpen
  \bibfield  {author} {\bibinfo {author} {\bibfnamefont {M.~J.}\ \bibnamefont
  {Hall}},\ }\bibfield  {title} {\enquote {\bibinfo {title} {Relaxed bell
  inequalities and kochen-specker theorems},}\ }\href
  {https://doi.org/https://doi.org/10.1103/PhysRevA.84.022102} {\bibfield
  {journal} {\bibinfo  {journal} {Physical Review A}\ }\textbf {\bibinfo
  {volume} {84}},\ \bibinfo {pages} {022102} (\bibinfo {year}
  {2011})}\BibitemShut {NoStop}%
\bibitem [{\citenamefont {Chaves}\ \emph
  {et~al.}(2015{\natexlab{b}})\citenamefont {Chaves}, \citenamefont {Kueng},
  \citenamefont {Brask},\ and\ \citenamefont {Gross}}]{chaves2015unifying}%
  \BibitemOpen
  \bibfield  {author} {\bibinfo {author} {\bibfnamefont {R.}~\bibnamefont
  {Chaves}}, \bibinfo {author} {\bibfnamefont {R.}~\bibnamefont {Kueng}},
  \bibinfo {author} {\bibfnamefont {J.~B.}\ \bibnamefont {Brask}},\ and\
  \bibinfo {author} {\bibfnamefont {D.}~\bibnamefont {Gross}},\ }\bibfield
  {title} {\enquote {\bibinfo {title} {{Unifying framework for relaxations of
  the causal assumptions in Bell's theorem}},}\ }\href
  {https://doi.org/10.1103/PhysRevLett.114.140403} {\bibfield  {journal}
  {\bibinfo  {journal} {Phys. Rev. Lett.}\ }\textbf {\bibinfo {volume} {114}},\
  \bibinfo {pages} {140403} (\bibinfo {year} {2015}{\natexlab{b}})}\BibitemShut
  {NoStop}%
\bibitem [{\citenamefont {Pearl}(2009)}]{pearl2009causality}%
  \BibitemOpen
  \bibfield  {author} {\bibinfo {author} {\bibfnamefont {J.}~\bibnamefont
  {Pearl}},\ }\href {https://books.google.com/books?id=LLkhAwAAQBAJ} {\emph
  {\bibinfo {title} {{Causality}}}}\ (\bibinfo  {publisher} {Cambridge
  University Press},\ \bibinfo {year} {2009})\BibitemShut {NoStop}%
\bibitem [{\citenamefont {Suprano}\ \emph {et~al.}(2022)\citenamefont
  {Suprano}, \citenamefont {Poderini}, \citenamefont {Polino}, \citenamefont
  {Agresti}, \citenamefont {Carvacho}, \citenamefont {Canabarro}, \citenamefont
  {Wolfe}, \citenamefont {Chaves},\ and\ \citenamefont
  {Sciarrino}}]{suprano2022experimental}%
  \BibitemOpen
  \bibfield  {author} {\bibinfo {author} {\bibfnamefont {A.}~\bibnamefont
  {Suprano}}, \bibinfo {author} {\bibfnamefont {D.}~\bibnamefont {Poderini}},
  \bibinfo {author} {\bibfnamefont {E.}~\bibnamefont {Polino}}, \bibinfo
  {author} {\bibfnamefont {I.}~\bibnamefont {Agresti}}, \bibinfo {author}
  {\bibfnamefont {G.}~\bibnamefont {Carvacho}}, \bibinfo {author}
  {\bibfnamefont {A.}~\bibnamefont {Canabarro}}, \bibinfo {author}
  {\bibfnamefont {E.}~\bibnamefont {Wolfe}}, \bibinfo {author} {\bibfnamefont
  {R.}~\bibnamefont {Chaves}},\ and\ \bibinfo {author} {\bibfnamefont
  {F.}~\bibnamefont {Sciarrino}},\ }\bibfield  {title} {\enquote {\bibinfo
  {title} {Experimental genuine tripartite nonlocality in a quantum triangle
  network},}\ }\href {https://doi.org/10.1103/PRXQuantum.3.030342} {\bibfield
  {journal} {\bibinfo  {journal} {PRX Quantum}\ }\textbf {\bibinfo {volume}
  {3}},\ \bibinfo {pages} {030342} (\bibinfo {year} {2022})}\BibitemShut
  {NoStop}%
\bibitem [{\citenamefont {Fritz}(2012)}]{fritz2012beyond}%
  \BibitemOpen
  \bibfield  {author} {\bibinfo {author} {\bibfnamefont {T.}~\bibnamefont
  {Fritz}},\ }\bibfield  {title} {\enquote {\bibinfo {title} {{Beyond Bell's
  theorem: correlation scenarios}},}\ }\href
  {https://doi.org/10.1088/1367-2630/14/10/103001} {\bibfield  {journal}
  {\bibinfo  {journal} {New J. Phys.}\ }\textbf {\bibinfo {volume} {14}},\
  \bibinfo {pages} {103001} (\bibinfo {year} {2012})}\BibitemShut {NoStop}%
\bibitem [{\citenamefont {Evans}(2016)}]{evans2016margins}%
  \BibitemOpen
  \bibfield  {author} {\bibinfo {author} {\bibfnamefont {R.~J.}\ \bibnamefont
  {Evans}},\ }\bibfield  {title} {\enquote {\bibinfo {title} {{Graphs for
  margins of Bayesian networks}},}\ }\href {https://doi.org/10.1111/sjos.12194}
  {\bibfield  {journal} {\bibinfo  {journal} {Scandinavian J. Stat.}\ }\textbf
  {\bibinfo {volume} {43}},\ \bibinfo {pages} {625} (\bibinfo {year}
  {2016})}\BibitemShut {NoStop}%
\bibitem [{\citenamefont {Sun}\ \emph {et~al.}(2019)\citenamefont {Sun},
  \citenamefont {Jiang}, \citenamefont {Bai}, \citenamefont {Zhang},
  \citenamefont {Li}, \citenamefont {Jiang}, \citenamefont {Zhang},
  \citenamefont {You}, \citenamefont {Chen}, \citenamefont {Wang} \emph
  {et~al.}}]{sun2019experimental}%
  \BibitemOpen
  \bibfield  {author} {\bibinfo {author} {\bibfnamefont {Q.-C.}\ \bibnamefont
  {Sun}}, \bibinfo {author} {\bibfnamefont {Y.-F.}\ \bibnamefont {Jiang}},
  \bibinfo {author} {\bibfnamefont {B.}~\bibnamefont {Bai}}, \bibinfo {author}
  {\bibfnamefont {W.}~\bibnamefont {Zhang}}, \bibinfo {author} {\bibfnamefont
  {H.}~\bibnamefont {Li}}, \bibinfo {author} {\bibfnamefont {X.}~\bibnamefont
  {Jiang}}, \bibinfo {author} {\bibfnamefont {J.}~\bibnamefont {Zhang}},
  \bibinfo {author} {\bibfnamefont {L.}~\bibnamefont {You}}, \bibinfo {author}
  {\bibfnamefont {X.}~\bibnamefont {Chen}}, \bibinfo {author} {\bibfnamefont
  {Z.}~\bibnamefont {Wang}}, \emph {et~al.},\ }\bibfield  {title} {\enquote
  {\bibinfo {title} {{Experimental demonstration of non-bilocality with truly
  independent sources and strict locality constraints}},}\ }\href
  {https://doi.org/10.1038/s41566-019-0502-7} {\bibfield  {journal} {\bibinfo
  {journal} {Nature Photonics}\ }\textbf {\bibinfo {volume} {13}},\ \bibinfo
  {pages} {687} (\bibinfo {year} {2019})}\BibitemShut {NoStop}%
\bibitem [{\citenamefont {Poderini}\ \emph {et~al.}(2020)\citenamefont
  {Poderini}, \citenamefont {Agresti}, \citenamefont {Marchese}, \citenamefont
  {Polino}, \citenamefont {Giordani}, \citenamefont {Suprano}, \citenamefont
  {Valeri}, \citenamefont {Milani}, \citenamefont {Spagnolo}, \citenamefont
  {Carvacho} \emph {et~al.}}]{poderini2020experimental}%
  \BibitemOpen
  \bibfield  {author} {\bibinfo {author} {\bibfnamefont {D.}~\bibnamefont
  {Poderini}}, \bibinfo {author} {\bibfnamefont {I.}~\bibnamefont {Agresti}},
  \bibinfo {author} {\bibfnamefont {G.}~\bibnamefont {Marchese}}, \bibinfo
  {author} {\bibfnamefont {E.}~\bibnamefont {Polino}}, \bibinfo {author}
  {\bibfnamefont {T.}~\bibnamefont {Giordani}}, \bibinfo {author}
  {\bibfnamefont {A.}~\bibnamefont {Suprano}}, \bibinfo {author} {\bibfnamefont
  {M.}~\bibnamefont {Valeri}}, \bibinfo {author} {\bibfnamefont
  {G.}~\bibnamefont {Milani}}, \bibinfo {author} {\bibfnamefont
  {N.}~\bibnamefont {Spagnolo}}, \bibinfo {author} {\bibfnamefont
  {G.}~\bibnamefont {Carvacho}}, \emph {et~al.},\ }\bibfield  {title} {\enquote
  {\bibinfo {title} {{Experimental violation of n-locality in a star quantum
  network}},}\ }\href {https://doi.org/10.1038/s41467-020-16189-6} {\bibfield
  {journal} {\bibinfo  {journal} {Nature comm.}\ }\textbf {\bibinfo {volume}
  {11}},\ \bibinfo {pages} {1} (\bibinfo {year} {2020})}\BibitemShut {NoStop}%
\bibitem [{\citenamefont {Carvacho}\ \emph {et~al.}(2022)\citenamefont
  {Carvacho}, \citenamefont {Roccia}, \citenamefont {Valeri}, \citenamefont
  {Basset}, \citenamefont {Poderini}, \citenamefont {Pardo}, \citenamefont
  {Polino}, \citenamefont {Carosini}, \citenamefont {Rota}, \citenamefont
  {Neuwirth} \emph {et~al.}}]{carvacho2022quantum}%
  \BibitemOpen
  \bibfield  {author} {\bibinfo {author} {\bibfnamefont {G.}~\bibnamefont
  {Carvacho}}, \bibinfo {author} {\bibfnamefont {E.}~\bibnamefont {Roccia}},
  \bibinfo {author} {\bibfnamefont {M.}~\bibnamefont {Valeri}}, \bibinfo
  {author} {\bibfnamefont {F.~B.}\ \bibnamefont {Basset}}, \bibinfo {author}
  {\bibfnamefont {D.}~\bibnamefont {Poderini}}, \bibinfo {author}
  {\bibfnamefont {C.}~\bibnamefont {Pardo}}, \bibinfo {author} {\bibfnamefont
  {E.}~\bibnamefont {Polino}}, \bibinfo {author} {\bibfnamefont
  {L.}~\bibnamefont {Carosini}}, \bibinfo {author} {\bibfnamefont {M.~B.}\
  \bibnamefont {Rota}}, \bibinfo {author} {\bibfnamefont {J.}~\bibnamefont
  {Neuwirth}}, \emph {et~al.},\ }\bibfield  {title} {\enquote {\bibinfo {title}
  {{Quantum violation of local causality in an urban network using hybrid
  photonic technologies}},}\ }\href {https://doi.org/10.1364/OPTICA.451523}
  {\bibfield  {journal} {\bibinfo  {journal} {Optica}\ }\textbf {\bibinfo
  {volume} {9}},\ \bibinfo {pages} {572} (\bibinfo {year} {2022})}\BibitemShut
  {NoStop}%
\bibitem [{\citenamefont {Kim}\ \emph {et~al.}(2006)\citenamefont {Kim},
  \citenamefont {Fiorentino},\ and\ \citenamefont {Wong}}]{kim2006phase}%
  \BibitemOpen
  \bibfield  {author} {\bibinfo {author} {\bibfnamefont {T.}~\bibnamefont
  {Kim}}, \bibinfo {author} {\bibfnamefont {M.}~\bibnamefont {Fiorentino}},\
  and\ \bibinfo {author} {\bibfnamefont {F.~N.}\ \bibnamefont {Wong}},\
  }\bibfield  {title} {\enquote {\bibinfo {title} {{Phase-stable source of
  polarization-entangled photons using a polarization Sagnac
  interferometer}},}\ }\href
  {https://link.aps.org/doi/10.1103/PhysRevA.73.012316} {\bibfield  {journal}
  {\bibinfo  {journal} {Phys. Rev. A}\ }\textbf {\bibinfo {volume} {73}},\
  \bibinfo {pages} {012316} (\bibinfo {year} {2006})}\BibitemShut {NoStop}%
\bibitem [{\citenamefont {Fedrizzi}\ \emph {et~al.}(2007)\citenamefont
  {Fedrizzi}, \citenamefont {Herbst}, \citenamefont {Poppe}, \citenamefont
  {Jennewein},\ and\ \citenamefont {Zeilinger}}]{fedrizzi2007wavelength}%
  \BibitemOpen
  \bibfield  {author} {\bibinfo {author} {\bibfnamefont {A.}~\bibnamefont
  {Fedrizzi}}, \bibinfo {author} {\bibfnamefont {T.}~\bibnamefont {Herbst}},
  \bibinfo {author} {\bibfnamefont {A.}~\bibnamefont {Poppe}}, \bibinfo
  {author} {\bibfnamefont {T.}~\bibnamefont {Jennewein}},\ and\ \bibinfo
  {author} {\bibfnamefont {A.}~\bibnamefont {Zeilinger}},\ }\bibfield  {title}
  {\enquote {\bibinfo {title} {{A wavelength-tunable fiber-coupled source of
  narrowband entangled photons}},}\ }\href
  {https://doi.org/10.1364/OE.15.015377} {\bibfield  {journal} {\bibinfo
  {journal} {Optics Express}\ }\textbf {\bibinfo {volume} {15}},\ \bibinfo
  {pages} {15377} (\bibinfo {year} {2007})}\BibitemShut {NoStop}%
\bibitem [{\citenamefont {Carvacho}\ \emph {et~al.}(2017)\citenamefont
  {Carvacho}, \citenamefont {Andreoli}, \citenamefont {Santodonato},
  \citenamefont {Bentivegna}, \citenamefont {Chaves},\ and\ \citenamefont
  {Sciarrino}}]{carvacho2017experimental}%
  \BibitemOpen
  \bibfield  {author} {\bibinfo {author} {\bibfnamefont {G.}~\bibnamefont
  {Carvacho}}, \bibinfo {author} {\bibfnamefont {F.}~\bibnamefont {Andreoli}},
  \bibinfo {author} {\bibfnamefont {L.}~\bibnamefont {Santodonato}}, \bibinfo
  {author} {\bibfnamefont {M.}~\bibnamefont {Bentivegna}}, \bibinfo {author}
  {\bibfnamefont {R.}~\bibnamefont {Chaves}},\ and\ \bibinfo {author}
  {\bibfnamefont {F.}~\bibnamefont {Sciarrino}},\ }\bibfield  {title} {\enquote
  {\bibinfo {title} {{Experimental violation of local causality in a quantum
  network}},}\ }\href {https://doi.org/10.1038/ncomms14775} {\bibfield
  {journal} {\bibinfo  {journal} {Nature comm.}\ }\textbf {\bibinfo {volume}
  {8}},\ \bibinfo {pages} {1} (\bibinfo {year} {2017})}\BibitemShut {NoStop}%
\bibitem [{\citenamefont {Saunders}\ \emph {et~al.}(2017)\citenamefont
  {Saunders}, \citenamefont {Bennet}, \citenamefont {Branciard},\ and\
  \citenamefont {Pryde}}]{saunders2017experimental}%
  \BibitemOpen
  \bibfield  {author} {\bibinfo {author} {\bibfnamefont {D.~J.}\ \bibnamefont
  {Saunders}}, \bibinfo {author} {\bibfnamefont {A.~J.}\ \bibnamefont
  {Bennet}}, \bibinfo {author} {\bibfnamefont {C.}~\bibnamefont {Branciard}},\
  and\ \bibinfo {author} {\bibfnamefont {G.~J.}\ \bibnamefont {Pryde}},\
  }\bibfield  {title} {\enquote {\bibinfo {title} {{Experimental demonstration
  of nonbilocal quantum correlations}},}\ }\href
  {https://doi.org/10.1126/sciadv.1602743} {\bibfield  {journal} {\bibinfo
  {journal} {Science Advances}\ }\textbf {\bibinfo {volume} {3}},\ \bibinfo
  {pages} {e1602743} (\bibinfo {year} {2017})}\BibitemShut {NoStop}%
\bibitem [{\citenamefont {Hossenfelder}\ and\ \citenamefont
  {Palmer}(2020)}]{hossenfelder2020rethinking}%
  \BibitemOpen
  \bibfield  {author} {\bibinfo {author} {\bibfnamefont {S.}~\bibnamefont
  {Hossenfelder}}\ and\ \bibinfo {author} {\bibfnamefont {T.}~\bibnamefont
  {Palmer}},\ }\bibfield  {title} {\enquote {\bibinfo {title} {{Rethinking
  superdeterminism}},}\ }\href {https://doi.org/10.3389/fphy.2020.00139}
  {\bibfield  {journal} {\bibinfo  {journal} {Frontiers in Physics}\ }\textbf
  {\bibinfo {volume} {8}},\ \bibinfo {pages} {139} (\bibinfo {year}
  {2020})}\BibitemShut {NoStop}%
\bibitem [{\citenamefont {Tavakoli}\ \emph {et~al.}(2014)\citenamefont
  {Tavakoli}, \citenamefont {Skrzypczyk}, \citenamefont {Cavalcanti},\ and\
  \citenamefont {Ac{\'i}n}}]{tavakoli2014nonlocal}%
  \BibitemOpen
  \bibfield  {author} {\bibinfo {author} {\bibfnamefont {A.}~\bibnamefont
  {Tavakoli}}, \bibinfo {author} {\bibfnamefont {P.}~\bibnamefont
  {Skrzypczyk}}, \bibinfo {author} {\bibfnamefont {D.}~\bibnamefont
  {Cavalcanti}},\ and\ \bibinfo {author} {\bibfnamefont {A.}~\bibnamefont
  {Ac{\'i}n}},\ }\bibfield  {title} {\enquote {\bibinfo {title} {{Nonlocal
  correlations in the star-network configuration}},}\ }\href
  {https://doi.org/10.1103/PhysRevA.90.062109} {\bibfield  {journal} {\bibinfo
  {journal} {Phys. Rev. A}\ }\textbf {\bibinfo {volume} {90}},\ \bibinfo
  {pages} {062109} (\bibinfo {year} {2014})}\BibitemShut {NoStop}%
\bibitem [{\citenamefont {Bharti}\ \emph {et~al.}(2020)\citenamefont {Bharti},
  \citenamefont {Haug}, \citenamefont {Vedral},\ and\ \citenamefont
  {Kwek}}]{bharti2020machine}%
  \BibitemOpen
  \bibfield  {author} {\bibinfo {author} {\bibfnamefont {K.}~\bibnamefont
  {Bharti}}, \bibinfo {author} {\bibfnamefont {T.}~\bibnamefont {Haug}},
  \bibinfo {author} {\bibfnamefont {V.}~\bibnamefont {Vedral}},\ and\ \bibinfo
  {author} {\bibfnamefont {L.-C.}\ \bibnamefont {Kwek}},\ }\bibfield  {title}
  {\enquote {\bibinfo {title} {{Machine learning meets quantum foundations: A
  brief survey}},}\ }\href {https://doi.org/10.1116/5.0007529} {\bibfield
  {journal} {\bibinfo  {journal} {AVS Quantum Science}\ }\textbf {\bibinfo
  {volume} {2}},\ \bibinfo {pages} {034101} (\bibinfo {year}
  {2020})}\BibitemShut {NoStop}%
\bibitem [{\citenamefont {Canabarro}\ \emph {et~al.}(2019)\citenamefont
  {Canabarro}, \citenamefont {Brito},\ and\ \citenamefont
  {Chaves}}]{canabarro2019machine}%
  \BibitemOpen
  \bibfield  {author} {\bibinfo {author} {\bibfnamefont {A.}~\bibnamefont
  {Canabarro}}, \bibinfo {author} {\bibfnamefont {S.}~\bibnamefont {Brito}},\
  and\ \bibinfo {author} {\bibfnamefont {R.}~\bibnamefont {Chaves}},\
  }\bibfield  {title} {\enquote {\bibinfo {title} {{Machine learning nonlocal
  correlations}},}\ }\href {https://doi.org/10.1103/PhysRevLett.122.200401}
  {\bibfield  {journal} {\bibinfo  {journal} {Phys. Rev. Lett.}\ }\textbf
  {\bibinfo {volume} {122}},\ \bibinfo {pages} {200401} (\bibinfo {year}
  {2019})}\BibitemShut {NoStop}%
\bibitem [{\citenamefont {Navascu{\'e}s}\ and\ \citenamefont
  {Wolfe}(2020)}]{navascues2020inflation}%
  \BibitemOpen
  \bibfield  {author} {\bibinfo {author} {\bibfnamefont {M.}~\bibnamefont
  {Navascu{\'e}s}}\ and\ \bibinfo {author} {\bibfnamefont {E.}~\bibnamefont
  {Wolfe}},\ }\bibfield  {title} {\enquote {\bibinfo {title} {{The Inflation
  Technique Completely Solves the Causal Compatibility Problem}},}\ }\href
  {https://doi.org/10.1515/jci-2018-0008} {\bibfield  {journal} {\bibinfo
  {journal} {J. Causal Inference}\ }\textbf {\bibinfo {volume} {8}},\ \bibinfo
  {pages} {70} (\bibinfo {year} {2020})}\BibitemShut {NoStop}%
\bibitem [{\citenamefont {Barrett}\ and\ \citenamefont
  {Gisin}(2011)}]{barrett2011much}%
  \BibitemOpen
  \bibfield  {author} {\bibinfo {author} {\bibfnamefont {J.}~\bibnamefont
  {Barrett}}\ and\ \bibinfo {author} {\bibfnamefont {N.}~\bibnamefont
  {Gisin}},\ }\bibfield  {title} {\enquote {\bibinfo {title} {How much
  measurement independence is needed to demonstrate nonlocality?}}\ }\href
  {https://doi.org/https://doi.org/10.1103/PhysRevLett.106.100406} {\bibfield
  {journal} {\bibinfo  {journal} {Physical review letters}\ }\textbf {\bibinfo
  {volume} {106}},\ \bibinfo {pages} {100406} (\bibinfo {year}
  {2011})}\BibitemShut {NoStop}%
\bibitem [{\citenamefont {P{u}tz}\ \emph {et~al.}(2014)\citenamefont {P{u}tz},
  \citenamefont {Rosset}, \citenamefont {Barnea}, \citenamefont {Liang},\ and\
  \citenamefont {Gisin}}]{putz2014arbitrarily}%
  \BibitemOpen
  \bibfield  {author} {\bibinfo {author} {\bibfnamefont {G.}~\bibnamefont
  {P{u}tz}}, \bibinfo {author} {\bibfnamefont {D.}~\bibnamefont {Rosset}},
  \bibinfo {author} {\bibfnamefont {T.~J.}\ \bibnamefont {Barnea}}, \bibinfo
  {author} {\bibfnamefont {Y.-C.}\ \bibnamefont {Liang}},\ and\ \bibinfo
  {author} {\bibfnamefont {N.}~\bibnamefont {Gisin}},\ }\bibfield  {title}
  {\enquote {\bibinfo {title} {{Arbitrarily small amount of measurement
  independence is sufficient to manifest quantum nonlocality}},}\ }\href
  {https://doi.org/10.1103/PhysRevLett.113.190402} {\bibfield  {journal}
  {\bibinfo  {journal} {Phys. Rev. Lett.}\ }\textbf {\bibinfo {volume} {113}},\
  \bibinfo {pages} {190402} (\bibinfo {year} {2014})}\BibitemShut {NoStop}%
\bibitem [{\citenamefont {Brans}(1988)}]{brans1988bell}%
  \BibitemOpen
  \bibfield  {author} {\bibinfo {author} {\bibfnamefont {C.~H.}\ \bibnamefont
  {Brans}},\ }\bibfield  {title} {\enquote {\bibinfo {title} {Bell's theorem
  does not eliminate fully causal hidden variables},}\ }\href
  {https://doi.org/https://doi.org/10.1007/BF00670750} {\bibfield  {journal}
  {\bibinfo  {journal} {International Journal of Theoretical Physics}\ }\textbf
  {\bibinfo {volume} {27}},\ \bibinfo {pages} {219} (\bibinfo {year}
  {1988})}\BibitemShut {NoStop}%
\bibitem [{\citenamefont {Hall}(2010)}]{hall2010local}%
  \BibitemOpen
  \bibfield  {author} {\bibinfo {author} {\bibfnamefont {M.~J.}\ \bibnamefont
  {Hall}},\ }\bibfield  {title} {\enquote {\bibinfo {title} {{Local
  deterministic model of singlet state correlations based on relaxing
  measurement independence}},}\ }\href
  {https://doi.org/10.1103/PhysRevLett.105.250404} {\bibfield  {journal}
  {\bibinfo  {journal} {Phys. Rev. Lett.}\ }\textbf {\bibinfo {volume} {105}},\
  \bibinfo {pages} {250404} (\bibinfo {year} {2010})}\BibitemShut {NoStop}%
\bibitem [{\citenamefont {Hall}\ and\ \citenamefont
  {Branciard}(2020)}]{hall2020measurement}%
  \BibitemOpen
  \bibfield  {author} {\bibinfo {author} {\bibfnamefont {M.~J.}\ \bibnamefont
  {Hall}}\ and\ \bibinfo {author} {\bibfnamefont {C.}~\bibnamefont
  {Branciard}},\ }\bibfield  {title} {\enquote {\bibinfo {title}
  {{Measurement-dependence cost for Bell nonlocality Causal versus retrocausal
  models}},}\ }\href {https://doi.org/10.1103/PhysRevA.102.052228} {\bibfield
  {journal} {\bibinfo  {journal} {Phys. Rev. A}\ }\textbf {\bibinfo {volume}
  {102}},\ \bibinfo {pages} {052228} (\bibinfo {year} {2020})}\BibitemShut
  {NoStop}%
\bibitem [{\citenamefont {Chaves}\ \emph
  {et~al.}(2014{\natexlab{b}})\citenamefont {Chaves}, \citenamefont {Luft},
  \citenamefont {Maciel}, \citenamefont {Gross}, \citenamefont {Janzing},\ and\
  \citenamefont {Sch{\"o}lkopf}}]{chaves2014inferring}%
  \BibitemOpen
  \bibfield  {author} {\bibinfo {author} {\bibfnamefont {R.}~\bibnamefont
  {Chaves}}, \bibinfo {author} {\bibfnamefont {L.}~\bibnamefont {Luft}},
  \bibinfo {author} {\bibfnamefont {T.~O.}\ \bibnamefont {Maciel}}, \bibinfo
  {author} {\bibfnamefont {D.}~\bibnamefont {Gross}}, \bibinfo {author}
  {\bibfnamefont {D.}~\bibnamefont {Janzing}},\ and\ \bibinfo {author}
  {\bibfnamefont {B.}~\bibnamefont {Sch{\"o}lkopf}},\ }\bibfield  {title}
  {\enquote {\bibinfo {title} {{Inferring latent structures via information
  inequalities}},}\ }\href@noop {} {\bibfield  {journal} {\bibinfo  {journal}
  {arXiv:1407.2256}\ } (\bibinfo {year} {2014}{\natexlab{b}})}\BibitemShut
  {NoStop}%
\bibitem [{\citenamefont {Fritz}\ and\ \citenamefont
  {Chaves}(2012)}]{fritz2012entropic}%
  \BibitemOpen
  \bibfield  {author} {\bibinfo {author} {\bibfnamefont {T.}~\bibnamefont
  {Fritz}}\ and\ \bibinfo {author} {\bibfnamefont {R.}~\bibnamefont {Chaves}},\
  }\bibfield  {title} {\enquote {\bibinfo {title} {{Entropic inequalities and
  marginal problems}},}\ }\href {https://doi.org/10.1109/TIT.2012.2222863}
  {\bibfield  {journal} {\bibinfo  {journal} {IEEE Trans. Info. Theo.}\
  }\textbf {\bibinfo {volume} {59}},\ \bibinfo {pages} {803} (\bibinfo {year}
  {2012})}\BibitemShut {NoStop}%
\bibitem [{\citenamefont {Li}\ \emph {et~al.}(2022)\citenamefont {Li},
  \citenamefont {Mao}, \citenamefont {Weilenmann}, \citenamefont {Tavakoli},
  \citenamefont {Chen}, \citenamefont {Feng}, \citenamefont {Yang},
  \citenamefont {Renou}, \citenamefont {Trillo}, \citenamefont {Le} \emph
  {et~al.}}]{li2022testing}%
  \BibitemOpen
  \bibfield  {author} {\bibinfo {author} {\bibfnamefont {Z.-D.}\ \bibnamefont
  {Li}}, \bibinfo {author} {\bibfnamefont {Y.-L.}\ \bibnamefont {Mao}},
  \bibinfo {author} {\bibfnamefont {M.}~\bibnamefont {Weilenmann}}, \bibinfo
  {author} {\bibfnamefont {A.}~\bibnamefont {Tavakoli}}, \bibinfo {author}
  {\bibfnamefont {H.}~\bibnamefont {Chen}}, \bibinfo {author} {\bibfnamefont
  {L.}~\bibnamefont {Feng}}, \bibinfo {author} {\bibfnamefont {S.-J.}\
  \bibnamefont {Yang}}, \bibinfo {author} {\bibfnamefont {M.-O.}\ \bibnamefont
  {Renou}}, \bibinfo {author} {\bibfnamefont {D.}~\bibnamefont {Trillo}},
  \bibinfo {author} {\bibfnamefont {T.~P.}\ \bibnamefont {Le}}, \emph
  {et~al.},\ }\bibfield  {title} {\enquote {\bibinfo {title} {{Testing real
  quantum theory in an optical quantum network}},}\ }\href
  {https://link.aps.org/doi/10.1103/PhysRevLett.128.040402} {\bibfield
  {journal} {\bibinfo  {journal} {Phys. Rev. Lett.}\ }\textbf {\bibinfo
  {volume} {128}},\ \bibinfo {pages} {040402} (\bibinfo {year}
  {2022})}\BibitemShut {NoStop}%
\bibitem [{\citenamefont {Wu}\ \emph {et~al.}(2022)\citenamefont {Wu},
  \citenamefont {Jiang}, \citenamefont {Gu}, \citenamefont {Huang},
  \citenamefont {Bai}, \citenamefont {Sun}, \citenamefont {Zhang},
  \citenamefont {Gong}, \citenamefont {Mao}, \citenamefont {Zhong} \emph
  {et~al.}}]{wu2022experimental}%
  \BibitemOpen
  \bibfield  {author} {\bibinfo {author} {\bibfnamefont {D.}~\bibnamefont
  {Wu}}, \bibinfo {author} {\bibfnamefont {Y.-F.}\ \bibnamefont {Jiang}},
  \bibinfo {author} {\bibfnamefont {X.-M.}\ \bibnamefont {Gu}}, \bibinfo
  {author} {\bibfnamefont {L.}~\bibnamefont {Huang}}, \bibinfo {author}
  {\bibfnamefont {B.}~\bibnamefont {Bai}}, \bibinfo {author} {\bibfnamefont
  {Q.-C.}\ \bibnamefont {Sun}}, \bibinfo {author} {\bibfnamefont
  {X.}~\bibnamefont {Zhang}}, \bibinfo {author} {\bibfnamefont {S.-Q.}\
  \bibnamefont {Gong}}, \bibinfo {author} {\bibfnamefont {Y.}~\bibnamefont
  {Mao}}, \bibinfo {author} {\bibfnamefont {H.-S.}\ \bibnamefont {Zhong}},
  \emph {et~al.},\ }\bibfield  {title} {\enquote {\bibinfo {title}
  {Experimental refutation of real-valued quantum mechanics under strict
  locality conditions},}\ }\href@noop {} {\bibfield  {journal} {\bibinfo
  {journal} {Physical Review Letters}\ }\textbf {\bibinfo {volume} {129}},\
  \bibinfo {pages} {140401} (\bibinfo {year} {2022})}\BibitemShut {NoStop}%
\bibitem [{\citenamefont {Van~Himbeeck}\ \emph {et~al.}(2019)\citenamefont
  {Van~Himbeeck}, \citenamefont {Brask}, \citenamefont {Pironio}, \citenamefont
  {Ramanathan}, \citenamefont {Sainz},\ and\ \citenamefont
  {Wolfe}}]{van2019quantum}%
  \BibitemOpen
  \bibfield  {author} {\bibinfo {author} {\bibfnamefont {T.}~\bibnamefont
  {Van~Himbeeck}}, \bibinfo {author} {\bibfnamefont {J.~B.}\ \bibnamefont
  {Brask}}, \bibinfo {author} {\bibfnamefont {S.}~\bibnamefont {Pironio}},
  \bibinfo {author} {\bibfnamefont {R.}~\bibnamefont {Ramanathan}}, \bibinfo
  {author} {\bibfnamefont {A.~B.}\ \bibnamefont {Sainz}},\ and\ \bibinfo
  {author} {\bibfnamefont {E.}~\bibnamefont {Wolfe}},\ }\bibfield  {title}
  {\enquote {\bibinfo {title} {{Quantum violations in the Instrumental scenario
  and their relations to the Bell scenario}},}\ }\href
  {https://doi.org/10.22331/q-2019-09-16-186} {\bibfield  {journal} {\bibinfo
  {journal} {Quantum}\ }\textbf {\bibinfo {volume} {3}},\ \bibinfo {pages}
  {186} (\bibinfo {year} {2019})}\BibitemShut {NoStop}%
\bibitem [{\citenamefont {Elliott}(2009)}]{elliott2009linear}%
  \BibitemOpen
  \bibfield  {author} {\bibinfo {author} {\bibfnamefont {M.~B.}\ \bibnamefont
  {Elliott}},\ }\bibfield  {title} {\enquote {\bibinfo {title} {A linear
  program for testing local realism},}\ }\href
  {https://doi.org/10.48550/arXiv.0905.2950} {\bibfield  {journal} {\bibinfo
  {journal} {arXiv preprint arXiv:0905.2950}\ } (\bibinfo {year}
  {2009})}\BibitemShut {NoStop}%
\bibitem [{\citenamefont {Zhang}\ \emph {et~al.}(2011)\citenamefont {Zhang},
  \citenamefont {Glancy},\ and\ \citenamefont
  {Knill}}]{zhang2011asymptotically}%
  \BibitemOpen
  \bibfield  {author} {\bibinfo {author} {\bibfnamefont {Y.}~\bibnamefont
  {Zhang}}, \bibinfo {author} {\bibfnamefont {S.}~\bibnamefont {Glancy}},\ and\
  \bibinfo {author} {\bibfnamefont {E.}~\bibnamefont {Knill}},\ }\bibfield
  {title} {\enquote {\bibinfo {title} {Asymptotically optimal data analysis for
  rejecting local realism},}\ }\href
  {https://doi.org/10.1103/PhysRevA.84.062118} {\bibfield  {journal} {\bibinfo
  {journal} {Physical Review A}\ }\textbf {\bibinfo {volume} {84}},\ \bibinfo
  {pages} {062118} (\bibinfo {year} {2011})}\BibitemShut {NoStop}%
\bibitem [{\citenamefont {Gisin}\ \emph {et~al.}(2020)\citenamefont {Gisin},
  \citenamefont {Bancal}, \citenamefont {Cai}, \citenamefont {Remy},
  \citenamefont {Tavakoli}, \citenamefont {Cruzeiro}, \citenamefont {Popescu},\
  and\ \citenamefont {Brunner}}]{gisin2020constraints}%
  \BibitemOpen
  \bibfield  {author} {\bibinfo {author} {\bibfnamefont {N.}~\bibnamefont
  {Gisin}}, \bibinfo {author} {\bibfnamefont {J.-D.}\ \bibnamefont {Bancal}},
  \bibinfo {author} {\bibfnamefont {Y.}~\bibnamefont {Cai}}, \bibinfo {author}
  {\bibfnamefont {P.}~\bibnamefont {Remy}}, \bibinfo {author} {\bibfnamefont
  {A.}~\bibnamefont {Tavakoli}}, \bibinfo {author} {\bibfnamefont {E.~Z.}\
  \bibnamefont {Cruzeiro}}, \bibinfo {author} {\bibfnamefont {S.}~\bibnamefont
  {Popescu}},\ and\ \bibinfo {author} {\bibfnamefont {N.}~\bibnamefont
  {Brunner}},\ }\bibfield  {title} {\enquote {\bibinfo {title} {{Constraints on
  nonlocality in networks from no-signaling and independence}},}\ }\href
  {https://doi.org/10.1038/s41467-020-16137-4} {\bibfield  {journal} {\bibinfo
  {journal} {Nature comm.}\ }\textbf {\bibinfo {volume} {11}},\ \bibinfo
  {pages} {2378} (\bibinfo {year} {2020})}\BibitemShut {NoStop}%
\bibitem [{\citenamefont {Andersen}(2001)}]{Andersen2001}%
  \BibitemOpen
  \bibfield  {author} {\bibinfo {author} {\bibfnamefont {E.~D.}\ \bibnamefont
  {Andersen}},\ }\bibfield  {title} {\enquote {\bibinfo {title} {{Certificates
  of Primal or Dual Infeasibility in Linear Programming}},}\ }\href
  {https://doi.org/10.1023/a:1011259103627} {\bibfield  {journal} {\bibinfo
  {journal} {Comp. Optim. Appl.}\ }\textbf {\bibinfo {volume} {20}},\ \bibinfo
  {pages} {171} (\bibinfo {year} {2001})}\BibitemShut {NoStop}%
\bibitem [{\citenamefont {Dinh}\ and\ \citenamefont
  {Jeyakumar}(2014)}]{Dinh2014}%
  \BibitemOpen
  \bibfield  {author} {\bibinfo {author} {\bibfnamefont {N.}~\bibnamefont
  {Dinh}}\ and\ \bibinfo {author} {\bibfnamefont {V.}~\bibnamefont
  {Jeyakumar}},\ }\bibfield  {title} {\enquote {\bibinfo {title} {{Farkas'
  lemma: three decades of generalizations for mathematical optimization}},}\
  }\href {https://doi.org/10.1007/s11750-014-0319-y} {\bibfield  {journal}
  {\bibinfo  {journal} {{TOP}}\ }\textbf {\bibinfo {volume} {22}},\ \bibinfo
  {pages} {1} (\bibinfo {year} {2014})}\BibitemShut {NoStop}%
\bibitem [{\citenamefont {Cao}\ \emph {et~al.}(2022)\citenamefont {Cao},
  \citenamefont {Renou}, \citenamefont {Zhang}, \citenamefont {Mass{\'e}},
  \citenamefont {Coiteux-Roy}, \citenamefont {Liu}, \citenamefont {Huang},
  \citenamefont {Li}, \citenamefont {Guo},\ and\ \citenamefont
  {Wolfe}}]{cao2022experimental}%
  \BibitemOpen
  \bibfield  {author} {\bibinfo {author} {\bibfnamefont {H.}~\bibnamefont
  {Cao}}, \bibinfo {author} {\bibfnamefont {M.-O.}\ \bibnamefont {Renou}},
  \bibinfo {author} {\bibfnamefont {C.}~\bibnamefont {Zhang}}, \bibinfo
  {author} {\bibfnamefont {G.}~\bibnamefont {Mass{\'e}}}, \bibinfo {author}
  {\bibfnamefont {X.}~\bibnamefont {Coiteux-Roy}}, \bibinfo {author}
  {\bibfnamefont {B.-H.}\ \bibnamefont {Liu}}, \bibinfo {author} {\bibfnamefont
  {Y.-F.}\ \bibnamefont {Huang}}, \bibinfo {author} {\bibfnamefont {C.-F.}\
  \bibnamefont {Li}}, \bibinfo {author} {\bibfnamefont {G.-C.}\ \bibnamefont
  {Guo}},\ and\ \bibinfo {author} {\bibfnamefont {E.}~\bibnamefont {Wolfe}},\
  }\bibfield  {title} {\enquote {\bibinfo {title} {Experimental demonstration
  that no tripartite-nonlocal causal theory explains nature’s
  correlations},}\ }\href@noop {} {\bibfield  {journal} {\bibinfo  {journal}
  {Physical Review Letters}\ }\textbf {\bibinfo {volume} {129}},\ \bibinfo
  {pages} {150402} (\bibinfo {year} {2022})}\BibitemShut {NoStop}%
\bibitem [{\citenamefont {Bancal}\ \emph {et~al.}(2010)\citenamefont {Bancal},
  \citenamefont {Gisin},\ and\ \citenamefont {Pironio}}]{BancalSymmetricBell}%
  \BibitemOpen
  \bibfield  {author} {\bibinfo {author} {\bibfnamefont {J.-D.}\ \bibnamefont
  {Bancal}}, \bibinfo {author} {\bibfnamefont {N.}~\bibnamefont {Gisin}},\ and\
  \bibinfo {author} {\bibfnamefont {S.}~\bibnamefont {Pironio}},\ }\bibfield
  {title} {\enquote {\bibinfo {title} {Looking for symmetric bell
  inequalities},}\ }\href {https://doi.org/10.1088/1751-8113/43/38/385303}
  {\bibfield  {journal} {\bibinfo  {journal} {J. Phys. A.}\ }\textbf {\bibinfo
  {volume} {43}},\ \bibinfo {pages} {385303} (\bibinfo {year}
  {2010})}\BibitemShut {NoStop}%
\bibitem [{\citenamefont {Bremner}\ \emph {et~al.}(2009)\citenamefont
  {Bremner}, \citenamefont {Sikiric},\ and\ \citenamefont
  {Sch{\"u}rmann}}]{bremner2009polyhedral}%
  \BibitemOpen
  \bibfield  {author} {\bibinfo {author} {\bibfnamefont {D.}~\bibnamefont
  {Bremner}}, \bibinfo {author} {\bibfnamefont {M.~D.}\ \bibnamefont
  {Sikiric}},\ and\ \bibinfo {author} {\bibfnamefont {A.}~\bibnamefont
  {Sch{\"u}rmann}},\ }\bibfield  {title} {\enquote {\bibinfo {title}
  {Polyhedral representation conversion up to symmetries},}\ }in\ \href@noop {}
  {\emph {\bibinfo {booktitle} {CRM proceedings}}},\ Vol.~\bibinfo {volume}
  {48}\ (\bibinfo {year} {2009})\ pp.\ \bibinfo {pages} {45--72}\BibitemShut
  {NoStop}%
\bibitem [{\citenamefont {L\~{A}\textparagraph{}rwald}\ and\ \citenamefont
  {Reinelt}(2015)}]{PANDA}%
  \BibitemOpen
  \bibfield  {author} {\bibinfo {author} {\bibfnamefont {S.}~\bibnamefont
  {L\~{A}\textparagraph{}rwald}}\ and\ \bibinfo {author} {\bibfnamefont
  {G.}~\bibnamefont {Reinelt}},\ }\bibfield  {title} {{\selectlanguage
  {English}\enquote {\bibinfo {title} {Panda: a software for polyhedral
  transformations},}\ }}\href {https://doi.org/10.1007/s13675-015-0040-0}
  {\bibfield  {journal} {\bibinfo  {journal} {EURO Journal on Computational
  Optimization}\ ,\ \bibinfo {pages} {1}} (\bibinfo {year} {2015})}\BibitemShut
  {NoStop}%
\bibitem [{\citenamefont {Ioannou}\ and\ \citenamefont
  {Rosset}(2021)}]{SDPSymmetry}%
  \BibitemOpen
  \bibfield  {author} {\bibinfo {author} {\bibfnamefont {M.}~\bibnamefont
  {Ioannou}}\ and\ \bibinfo {author} {\bibfnamefont {D.}~\bibnamefont
  {Rosset}},\ }\bibfield  {title} {\enquote {\bibinfo {title} {Noncommutative
  polynomial optimization under symmetry},}\ }\href
  {https://arxiv.org/abs/2112.10803} {\bibfield  {journal} {\bibinfo  {journal}
  {arXiv:2112.10803}\ } (\bibinfo {year} {2021})}\BibitemShut {NoStop}%
\end{thebibliography}

%

\section*{Acknowledgements} 
The authors thank Tamás Kriváchy for discussions about the ML implementation. This work was supported by The John Templeton Foundation via the grant Q-CAUSAL No 61084, via The Quantum Information Structure of Spacetime (QISS) Project (qiss.fr) (the opinions expressed in this publication are those of the author(s) and do not necessarily reflect the views of the John Templeton Foundation)  Grant Agreement No.  61466 and via QISS2  Grant Agreement No. 62312, by MIUR via PRIN 2017 (Progetto di Ricerca di Interesse Nazionale): project QUSHIP (2017SRNBRK), by the Regione Lazio programme “Progetti di Gruppi di ricerca” legge Regionale n. 13/2008 (SINFONIA project, prot. n. 85-2017-15200) via LazioInnova spa and by the ERC Advanced Grant QU-BOSS
(Grant agreement no. 884676). RC and AC acknowledge the Serrapilheira Institute (Grant No. Serra-1708-15763), the Brazilian National Council for Scientific and Technological Development (CNPq) via the National Institute for Science and Technology on Quantum Information (INCT-IQ) and Grants 307295/2020-6 and No. 311375/2020-0, the Brazilian agencies MCTIC and MEC. Research at Perimeter Institute is supported in part by the Government of Canada through the Department of Innovation, Science and Industry Canada and by the Province of Ontario through the Ministry of Colleges and Universities.

 \section*{Author contributions}
 
 E.P., D.P., G.R., I.A., G.C., F.S., E.W., R.S., A.C., and R.C. developed the project, E.P., D.P., I.A., A.S., G.Mi., G.C. and F.S. devised the experiment; E.P., D.P., G.R., I.A., A.S., G.Mi., G.C. and F.S. performed the experiment; E.P., D.P., G.R., I.A., A.S., G.C., F.S., E.W., R.S., A.C., G.Mo. and R.C. performed data analysis and modeling; E.W. and R.S. developed the theoretical tools of the inflation technique; A.C., G.Mo. and R.C. developed the theoretical tools of machine learning; all authors discussed the results and contributed to the writing of the manuscript.

\section*{Competing interests}

The authors declare no competing interest.

\end{document}


\title{Supplementary Information for \\
Experimental nonclassicality in a causal network without assuming freedom of choice}

\author{Emanuele Polino}
\author{Davide Poderini}
\author{Giovanni Rodari}
\author{Iris Agresti}
\author{Alessia Suprano}
\affiliation{Dipartimento di Fisica - Sapienza Universit\`{a} di Roma, P.le Aldo Moro 5, I-00185 Roma, Italy}

\author{Gonzalo Carvacho}

\affiliation{Dipartimento di Fisica - Sapienza Universit\`{a} di Roma, P.le Aldo Moro 5, I-00185 Roma, Italy}

\author{Elie Wolfe}
\email{ewolfe@perimeterinstitute.ca}

\affiliation{Perimeter Institute for Theoretical Physics, 31 Caroline St. N, Waterloo, Ontario, N2L 2Y5, Canada}

\author{Askery Canabarro}
\author{George Moreno}
\affiliation{International Institute of Physics, Federal University of Rio Grande do Norte, 59078-970, Natal, RN, Brazil}

\affiliation{Grupo de F\'isica da Mat\'eria Condensada, N\'ucleo de Ci\^encias Exatas - NCEx, Campus Arapiraca, Universidade Federal de Alagoas, 57309-005, Arapiraca, AL, Brazil}

\author{Giorgio Milani}
\affiliation{Dipartimento di Fisica - Sapienza Universit\`{a} di Roma, P.le Aldo Moro 5, I-00185 Roma, Italy}

\author{Robert W. Spekkens}
\affiliation{Perimeter Institute for Theoretical Physics, 31 Caroline St. N, Waterloo, Ontario, N2L 2Y5, Canada}

\author{Rafael Chaves}
\email{rchaves@iip.ufrn.br}
\affiliation{International Institute of Physics $\&$ School of Science and Technology, Federal University of Rio Grande do Norte, 59078-970, P. O. Box 1613, Natal, Brazil}

\author{Fabio Sciarrino} 
\email{fabio.sciarrino@uniroma1.it}

\affiliation{Dipartimento di Fisica - Sapienza Universit\`{a} di Roma, P.le Aldo Moro 5, I-00185 Roma, Italy}

\maketitle
\clearpage
\section{Data analysis}

The data acquisition was performed using three independent time-to-digital converters (TDC) with a resolution of $\approx 81$ ps.
The TDCs, one for each measurement station, were synchronized using a shared random signal, acting as time reference.
The synchronization was performed in real time, via dedicated software running in a separated machine which received data from the three nodes on the
local network.
After the synchronization, the data coming from the TDCs are filtered to retain only the events corresponding to a two-fold coincidence detection between events coming from the same source. Such events are then saved to disk for further analysis and post processing.
The time window in which two events are considered coincident is $\approx 4.1$ ns, this allows to filter out most noise sources affecting the measurement.
The array of two-fold coincidences obtained is further analyzed to count coincidence events among all the three parties. 
A six-fold event extracted in this way is a set of three two-fold events that can be considered coincident in a given window, larger than the one used to extract two-fold coincidences.
When more than three two-fold coincidences are found in the same time window, the additional events are discarded. 
More specifically, when more than one coincidence from a same source are found within the window interval, we select only the first detected. 
Selecting the first coincidence represents an arbitrary and unbiased choice, that is a  useful convention for resolving possible ambiguities in the definition of six-fold events.

\clearpage
\section{Experimental Fritz distribution}

\begin{figure*}[htp!]
\includegraphics[width=.95\textwidth]{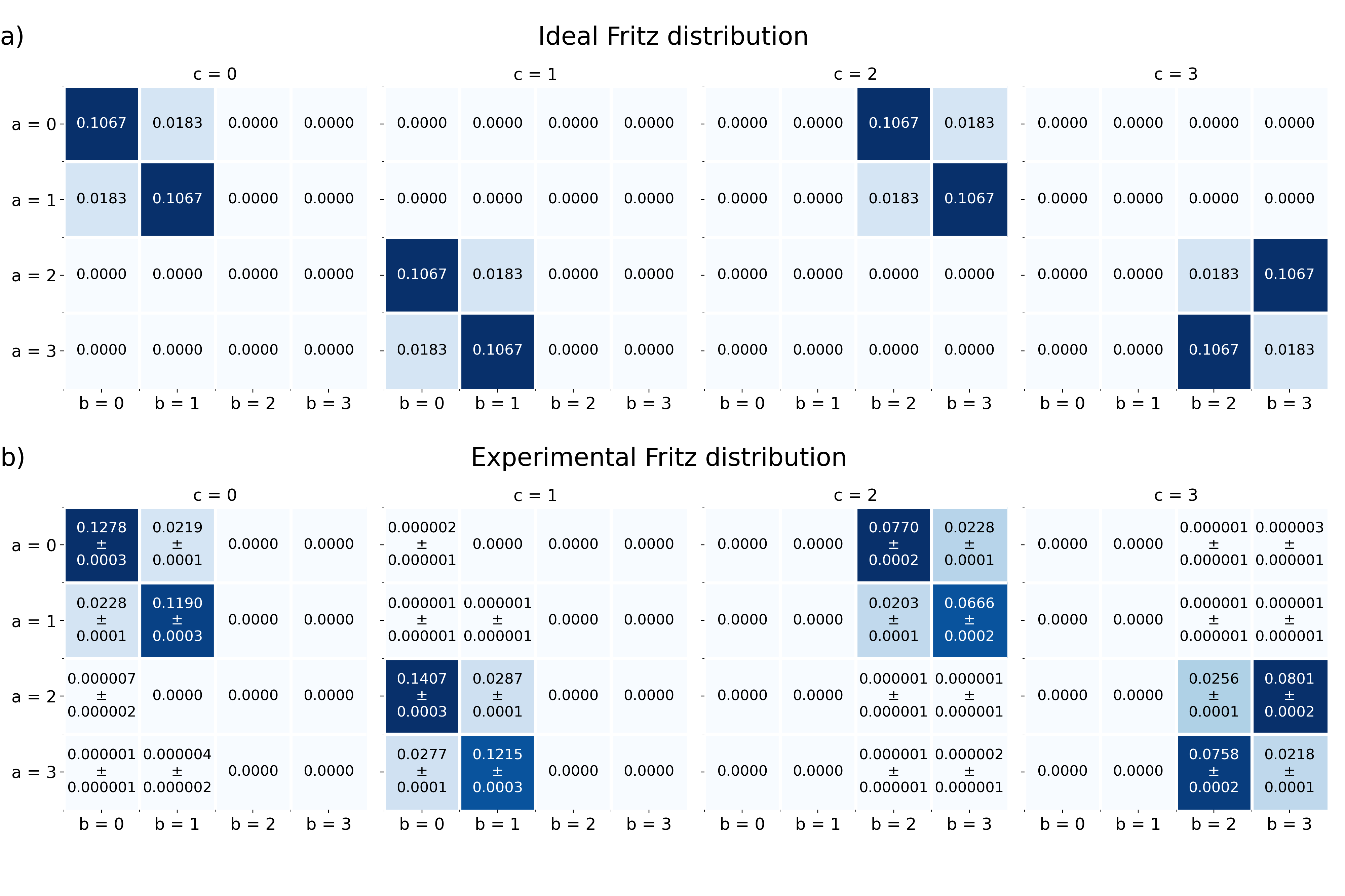}x
\caption{\textbf{Experimental Fritz distribution.}  \textbf{a)} Theoretical Fritz distribution considering ideal noiseless singlet states. \textbf{b)} Experimental distribution measured in an experimental run.
}
\label{fig:fritz_}
\end{figure*}

We show the experimental frequencies corresponding to the terms of Fritz distribution in Supp.~Fig.~\ref{fig:fritz_}-a, in comparison with the ideal values (Supp.~Fig.~\ref{fig:fritz_}-b). The overall statistics is composed of $\sim 10^6$ events.

\FloatBarrier

\section{Neural Network oracles}


Given a causal structure and a distribution over the observed outputs, we used an ensemble of neural networks to determine whether the distribution could have been reproduced by using exclusively local resources. The general idea is to encode the causal structure into neural networks and request them to reproduce the target distribution. Both the assumed causal structures and a feedforward neural networks can have their information flow determined by a directed acyclic graph (DAG). Therefore, we train diverse neural networks taking into consideration the causal structure to reproduce the target distribution. 

Consider the scenario in which three sources ($\lambda_{AB}, \lambda_{AC}, \lambda_{BC}$), send information either via a classical or quantum channel to three parties, Alice (A), Bob (B), and Charlie (C) with the following constrained flow of information: each source only sends information to two parties of the three, constituting a triangle network with quaternary (four possible flag numbers) outputs and no inputs, as described in the main text. In this manner, Alice, Bob and Charlie receives, respectively, the pairs $\lambda_{A}= \{ \lambda_{AB}, \lambda_{AC} \}$, $\lambda_{B}= \{ \lambda_{AB}, \lambda_{BC} \}$ and $\lambda_{C}= \{ \lambda_{AC}, \lambda_{BC} \}$. Therefore, individual inputs $\in \mathbb{R}^2$ (they have length 2). For training, we provide batches of ($N_{\text{batch}},2$) dimension for the corresponding MLP for each party.

The parties process their inputs by means of arbitrary local response functions, characterized by the conditional probabilities $p(a \vert \lambda_{AB},\lambda_{AC})$, $p(b \vert \lambda_{AB},\lambda_{BC})$ and $p(c \vert \lambda_{AC},\lambda_{BC})$, where $a, b, c \in \{0, 1, 2, 3\}$ are the flag numbers by the parties $A,B,C$, respectively. Although any other distribution can be reabsorbed by the established parties’ response functions $p_X(x\vert{} \lambda_{xy} ,\lambda_{xz})$, for the classical setup one can assume that the sources send a random variable taken from an uniform distribution in the unit interval, i.e. $\lambda_{AB},\lambda_{BC}, \lambda_{AC} \in [0,1]$. 

Therefore, given these constraints and the assumption that each source is independent, such a scenario is well characterized by the probability distribution $p(a,b,c)$ over the random variables of the outputs \cite{tamas2020}, which be written as:


\begin{equation}
\label{eq:p_abc}
p(a,b,c)= \iiint_0^1  p(a \vert \lambda_{AB},\lambda_{AC}) p(b \vert \lambda_{AB},\lambda_{BC}) p(c \vert \lambda_{AC},\lambda_{BC})d\lambda_{AB} d\lambda_{AC} d\lambda_{BC} \;.
\end{equation}

The aim in the machine learning (ML) part of this work is to construct neural networks capable of approximating distributions given by Eq. \ref{eq:p_abc}. Amid the myriad of ML algorithms that exist nowadays, feedforward neural networks are the one to emulate a directed acyclic graph (DAG) corresponding to a causal structure. This symmetry between causal structure and neural network  is a powerful trait making the method applicable, \textit{a priori}, to any causal structure \cite{tamas2020}.  

For each party (A,B,C), the corresponding response function will be incorporated by means of a fully connected multilayer perceptron (MLP). However, in principle, more advanced neural networks architectures could be used as well, such as convolutional neural networks (CNNs),  recurrent neural networks (RNNs) and so on, but this would require a dedicated investigation that would be far beyond the scope of this paper, although the implementation of an ensemble of MLPs is already a natural advance of the inaugural method shown in Ref. \cite{tamas2020}. 

From the machine learning perspective, the input layers to the MLPs are composed of the independent uniformly distributed random numbers in the unit interval, i.e. $\lambda_{AB},\lambda_{BC}, \lambda_{AC} \in [0,1]$, with the restriction in the flow of information imposed by the triangle network: each source only sends information to two parties of the three. In this manner, Alice receives ($\lambda_{AB}, \lambda_{AC}$), Bob receives ($\lambda_{AB}, \lambda_{BC}$) and Charlie receives ($\lambda_{BC}, \lambda_{AC}$). Therefore, individual inputs $\in \mathbb{R}^2$ (they have length 2). For training, we provide batches of ($N_{\text{batch}},2$) dimension for the corresponding MLP for each party. The output layers retrieve the corresponding probabilities conditioned on the respective inputs: $p(a \vert \lambda_{AB},\lambda_{AC})$, $p(b \vert \lambda_{AB},\lambda_{BC})$ and $p(c \vert \lambda_{AC},\lambda_{BC})$. In this step, we use a softmax activation function so that the outputs are three normalized vectors $\in \mathbb{R}^4$ (length 4). 


We then evaluate the neural network for $N_{batch}$ sample values of random variables ($\lambda_{AB}, \lambda_{AC}, \lambda_{BC}$) in order to approximate the joint probability distribution Eq. \ref{eq:p_abc}, averaging the Cartesian product of the conditional probabilities~($\in \mathbb{R}^4$),

\begin{equation}
\label{eq:p_abc_ml}
\tilde{p}(a,b,c) = \frac{1}{ N_{\text{batch}} } \sum_{i = 1}^{N_{\text{batch}}} p(a \vert {\lambda_{AB}}_i,{\lambda_{AC}}_i) p(b \vert {\lambda_{AB}}_i,{\lambda_{BC}}_i)  p(c \vert {\lambda_{AC}}_i,{\lambda_{BC}}_i),
\end{equation}
\sloppy yielding the approximated joint probability distribution ${\tilde{p}(a,b,c)\in \mathbb{R}^{64}}$ with the following index ordering: ${[0,1,2,3] \times [0,1,2,3] \times [0,1,2,3] \rightarrow [000, 001, 002, ... ,333]}$. In this manner, the method is order sensitive. This step formally differs from the Monte Carlo approximation in Ref. \cite{tamas2020}, however, in practice, they yield the same result for the examples we tested.

The approximation of $p(a,b,c)$ is achieved by setting a loss function quantifying the discrepancy between the target distribution $p(a,b,c)$ and the neural network’s constructed output $\tilde{p}(a,b,c)$. A natural candidate is therefore the relative Kullback–Leibler divergence, 

\begin{equation}
\label{eq:KL}
KL(p, \tilde{p})  = \sum_{abc} p(a,b,c) \log \left( \frac{p(a,b,c)}{\tilde{p}(a,b,c)} \right).
\end{equation}
In fact, any differentiable discrepancy measure between $p$ and $\tilde{p}$ works (for instance, the MSE (l2-norm), the MAE (l1-norm), Euclidean distance, and so on). However, the KL seems to be the best on the practical side (faster and better convergence), a point also registered in Ref. \cite{tamas2020}.

\begin{figure}[H]
\centering
\includegraphics[scale=0.6]{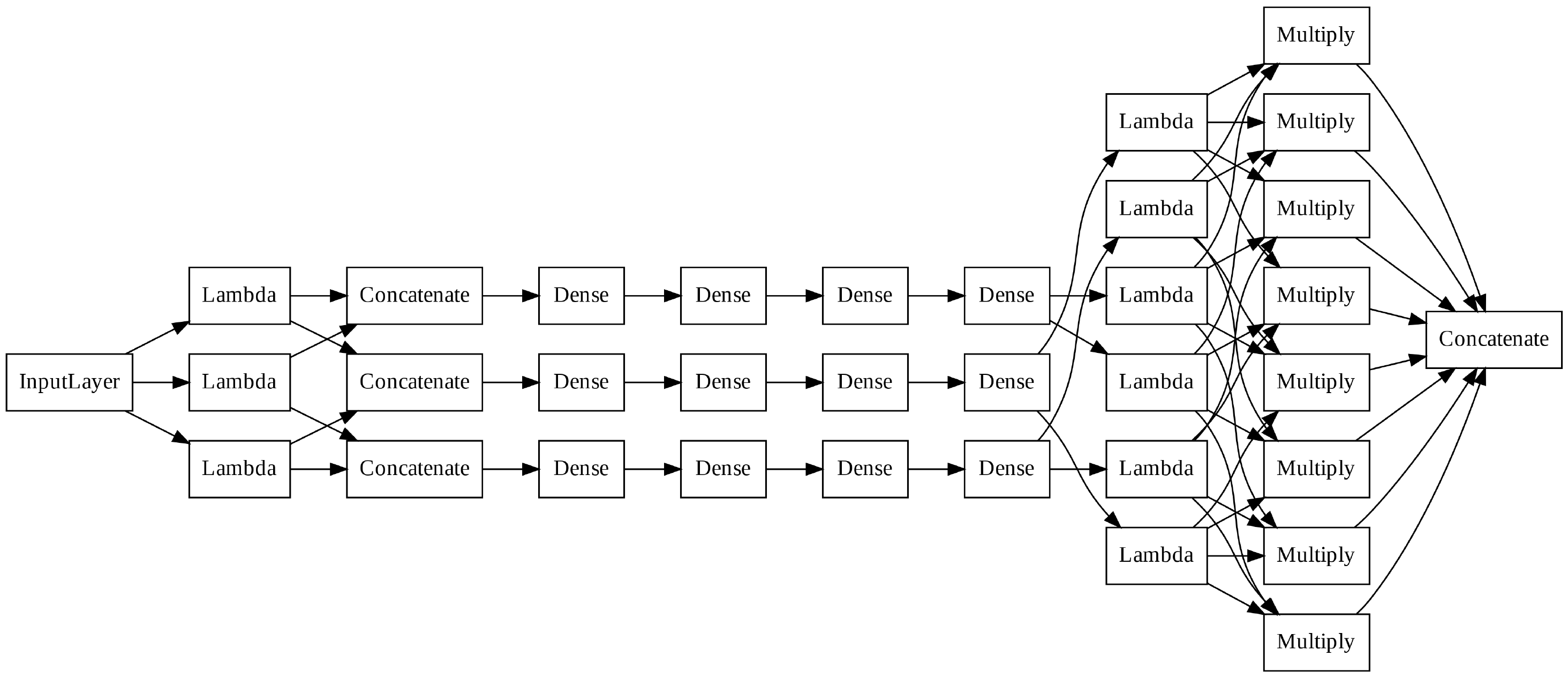}
\caption[]{ML diagram}
\end{figure}

As already mentioned, for an individual neural network forming the committee, we used a similar approach as in \cite{tamas2020}. The ensemble of neural networks, which can be imagined as an assembly of oracles, provides a certificate that a distribution is local once it is learned. In other words, if a target distribution is inside the local set, then a sufficiently expressive neural network should be able to learn the appropriate response functions and reproduce it. As the "right" expressivity is not known \textit{a priori}, we decided to use the ensemble approach, in which we train diverse neural networks with distinct number of layers and neurons. For distributions outside the local set, we should see that the machine can not approximate well the given target, no matter the expressitivy power we employ for the neural network. So, the ensemble is a good indication (not a definite proof) of this impossibility. Altogether, this gives us a criterion for deciding whether a target distribution is inside the local set or not.

Heuristically, it means that, by construction, the neural network can learn the local responses of the parties to their inputs. Notwithstanding, if a given target is surely outside the local set (quantum channels are permitted from the sources to the parties), then by adding noise according to a convex sum we should see a clear transition in function of visibility in the learner’s behavior when entering the set of local correlations, as demonstrated in the main text. The distance between the target and learned distributions can be computed in a number of ways. However, the element-wise mean square error (MSE) also known as l2-norm error between $p(a,b,c)$ and $\tilde{p}(a,b,c)$, is a better guide to the eyes for identifying the transition.

The advantage of using multiple neural networks architectures is well perceived by examining Supp.~Tab.~\ref{table:v}. Note that for distinct visibility values, a different neural network architecture yields the best MSE distance, strengthening the advantage of using an assembly of oracles.   

\begin{table}[htpb]
\centering

\begin{tabular}{ccccc}
\hline \hline
 Visibility (v) & Optimum Architecture \\ 
 \hline
 0.00 & 5 layers, 16 neurons \\
 0.05 & 4 layers, 16 neurons \\
 0.10 & 5 layers, 16 neurons \\
 0.15 & 4 layers, 16 neurons \\
 0.20 & 5 layers, 32 neurons \\
 0.25 & 4 layers, 32 neurons \\
 0.30 & 6 layers, 32 neurons \\
 0.35 & 4 layers, 32 neurons \\
 0.40 & 3 layers, 32 neurons \\
 0.45 & 3 layers, 32 neurons \\
 0.50 & 3 layers, 32 neurons \\
 0.55 & 6 layers, 32 neurons \\
 0.60 & 3 layers, 16 neurons \\
 0.65 & 5 layers, 16 neurons \\
 0.70 & 5 layers, 32 neurons \\
 0.75 & 6 layers, 16 neurons \\
 0.80 & 3 layers, 32 neurons \\
 0.85 & 3 layers, 32 neurons \\
 0.90 & 5 layers, 16 neurons \\
 0.95 & 3 layers, 32 neurons \\
 1.00 & 5 layers, 16 neurons \\
 \hline \hline
\end{tabular}
\caption{The corresponding best architecture yielding the minimum MSE distance for each visibility value $v$. \label{table:v}}
\end{table}

\FloatBarrier
\clearpage
\section{Causal compatibility inequality from the inflation technique}
\begin{figure}[H]
\centering
\includegraphics[scale=0.8]{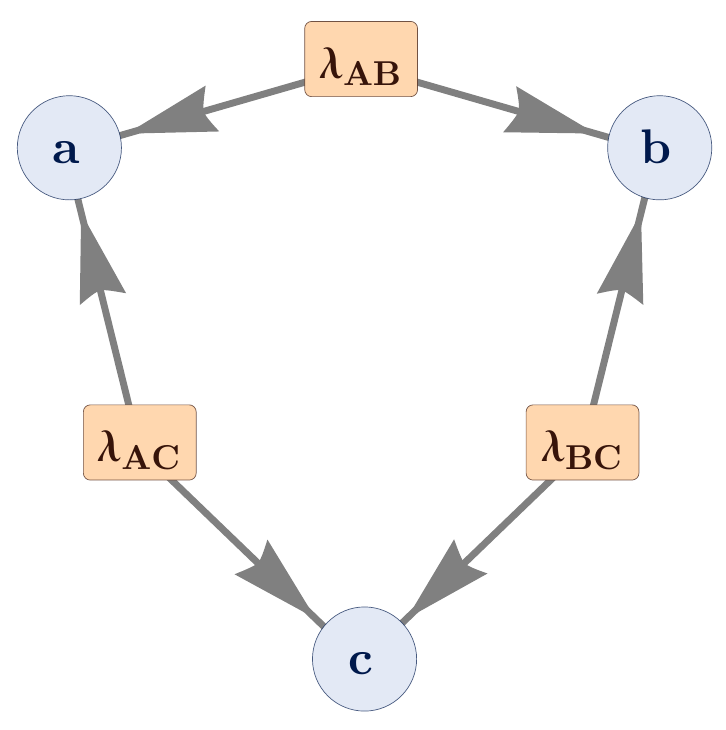}
\caption[]{The causal structure of the classical triangle network, a falsified hypothesis for explaining our experimental statistics.
\label{fig:reprodtri}}
\end{figure}

\begin{figure}[t]
\centering
\includegraphics[scale=0.8]{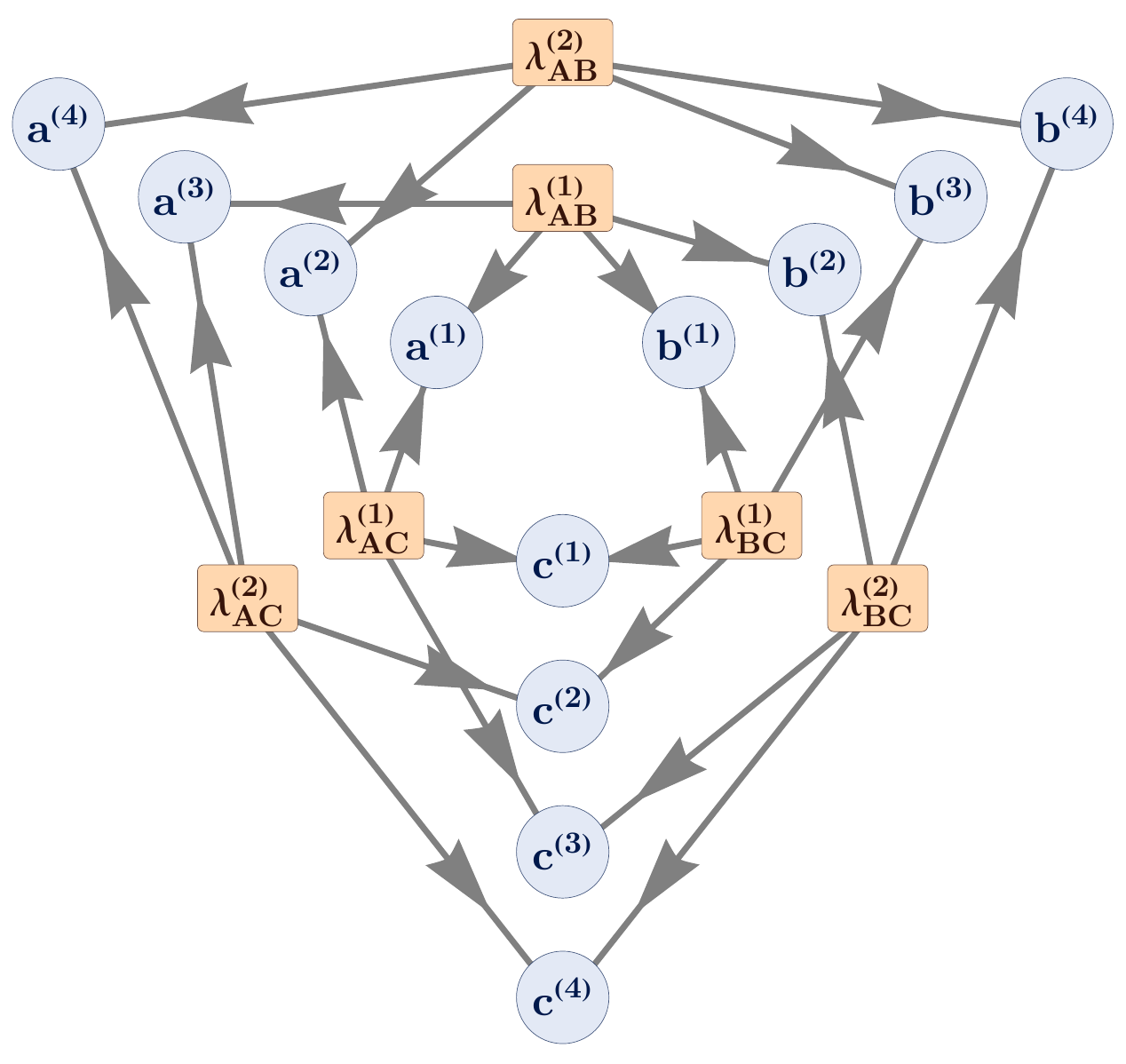}
\caption[]{The second order inflation graph of the triangle network.
\label{fig:inflation}}
\end{figure}

Here's how the second order inflation test works for the triangle network. Suppose that a distribution $p(a,b,c)$ is compatible with the triangle causal structure in Fig.~\ref{fig:reprodtri}, in the sense of
\begin{align}\label{eq:trianglecompat}
p(abc)={\sum_{{\lambda_{AB}\lambda_{AC}\lambda_{BC}}}}
p(a|\lambda_{AB},\lambda_{AC})
p(b|\lambda_{AB},\lambda_{BC})
p(c|\lambda_{AC},\lambda_{BC})
p(\lambda_{AB}) 
p(\lambda_{AC})
p(\lambda_{BC})
\,.
\end{align}
Now, imagine (gedankenexperiment) recycling the same causal components to create some hypothetically observable distribution ${p'(a^{(1)},b^{(1)},c^{(1)},a^{(2)},b^{(2)},c^{(2)},a^{(3)},b^{(3)},c^{(3)},a^{(4)},b^{(4)},c^{(4)})}$ compatible with the second order inflation graph of the triangle scenario, depicted in Fig.~\ref{fig:inflation}. Relative to the inflation graph, compatibility means
\begin{multline}
p'({a^{(1)}},{b^{(1)}},{c^{(1)}},
{a^{(2)}},{b^{(2)}},{c^{(2)}},
{a^{(3)}},{b^{(3)}},{c^{(3)}},
{a^{(4)}},{b^{(4)}},{c^{(4)}})
=\sum_{\substack{
\lambda_{AB}^{(1)},\lambda_{AB}^{(2)},\\
\lambda_{AC}^{(1)},\lambda_{AC}^{(2)},\\
\lambda_{BC}^{(1)},\lambda_{BC}^{(2)}
}}
\begin{psmallmatrix*}[r]
p'({a^{(1)}|\lambda_{AB}^{(1)},\lambda_{AC}^{(1)}})
p'({b^{(1)}|\lambda_{AB}^{(1)},\lambda_{BC}^{(1)}})
p'({c^{(1)}|\lambda_{AC}^{(1)},\lambda_{BC}^{(1)}})
\\\times p'({a^{(2)}|\lambda_{AB}^{(2)},\lambda_{AC}^{(1)}})
p'({b^{(2)}|\lambda_{AB}^{(1)},\lambda_{BC}^{(2)}})
p'({c^{(2)}|\lambda_{AC}^{(2)},\lambda_{BC}^{(1)}})
\\\times p'({a^{(3)}|\lambda_{AB}^{(1)},\lambda_{AC}^{(2)}})
p'({b^{(3)}|\lambda_{AB}^{(2)},\lambda_{BC}^{(1)}})
p'({c^{(3)}|\lambda_{AC}^{(1)},\lambda_{BC}^{(2)}})
\\\times p'({a^{(4)}|\lambda_{AB}^{(2)},\lambda_{AC}^{(2)}})
p'({b^{(4)}|\lambda_{AB}^{(2)},\lambda_{BC}^{(2)}})
p'({c^{(4)}|\lambda_{AC}^{(2)},\lambda_{BC}^{(2)}})
\\\times p'({\lambda_{AB}^{(1)}})
p'({\lambda_{AB}^{(2)}})
p'({\lambda_{AC}^{(1)}})p'({\lambda_{AC}^{(2)}})
p'({\lambda_{BC}^{(1)}})p'({\lambda_{BC}^{(2)}})\hphantom{\times}
\end{psmallmatrix*}
\end{multline}
By \enquote{recycling causal components} we mean that the \emph{functional dependence} of every variable on its parents in Fig.~\ref{fig:inflation} is presumed to be the same functional dependence as the analogous variable's functional dependence on its parents in Fig.~\ref{fig:reprodtri}. The  Fig.~\ref{fig:reprodtri} analog of each individual variable in Fig.~\ref{fig:inflation} is obtained by dropping the superscript \emph{copy index}. Hereafter, when describing random variables being assigned particular values we place the name of the variable in underscript and the value it takes in overscript.

 That is,
\begin{align}\begin{split}\label{eq:inflationrawlemma}
&p'(\underset{a^{(1)}|\lambda_{AB}^{(1)},\lambda_{AC}^{(1)}}{v|\mu_1,\mu_2})
=p'(\underset{a^{(2)}|\lambda_{AB}^{(2)},\lambda_{AC}^{(1)}}{v|\mu_1,\mu_2})
=p'(\underset{a^{(3)}|\lambda_{AB}^{(1)},\lambda_{AC}^{(2)}}{v|\mu_1,\mu_2})
=p'(\underset{a^{(4)}|\lambda_{AB}^{(2)},\lambda_{AC}^{(2)}}{v|\mu_1,\mu_2})
=p(\underset{a|\lambda_{AB},\lambda_{AC}}{v|\mu_1,\mu_2})\,,
\\\text{and}\quad &p'(\underset{b^{(1)}|\lambda_{AB}^{(1)},\lambda_{BC}^{(1)}}{v|\mu_1,\mu_2})
=p'(\underset{b^{(2)}|\lambda_{AB}^{(1)},\lambda_{BC}^{(2)}}{v|\mu_1,\mu_2})
=p'(\underset{b^{(3)}|\lambda_{AB}^{(2)},\lambda_{BC}^{(1)}}{v|\mu_1,\mu_2})
=p'(\underset{b^{(4)}|\lambda_{AB}^{(2)},\lambda_{BC}^{(2)}}{v|\mu_1,\mu_2})
=p(\underset{b|\lambda_{AB},\lambda_{BC}}{v|\mu_1,\mu_2})\,,
\\\text{and}\quad &p'(\underset{c^{(1)}|\lambda_{AC}^{(1)},\lambda_{BC}^{(1)}}{v|\mu_1,\mu_2})
=p'(\underset{c^{(2)}|\lambda_{AC}^{(2)},\lambda_{BC}^{(1)}}{v|\mu_1,\mu_2})
=p'(\underset{c^{(3)}|\lambda_{AC}^{(1)},\lambda_{BC}^{(2)}}{v|\mu_1,\mu_2})
=p'(\underset{c^{(4)}|\lambda_{AC}^{(2)},\lambda_{BC}^{(2)}}{v|\mu_1,\mu_2})
=p(\underset{c|\lambda_{AC},\lambda_{BC}}{v|\mu_1,\mu_2})\,,
\\\text{and}\quad &p'(\underset{\lambda_{AB}^{(1)}}{\mu})=p'(\underset{\lambda_{AB}^{(2)}}{\mu})=p(\underset{\lambda_{AB}}{\mu})\,,
\quad\text{and}\quad p'(\underset{\lambda_{AC}^{(1)}}{\mu})=p'(\underset{\lambda_{AC}^{(2)}}{\mu})=p(\underset{\lambda_{AC}}{\mu})\,,
\quad\text{and}\quad p'(\underset{\lambda_{BC}^{(1)}}{\mu})=p'(\underset{\lambda_{BC}^{(2)}}{\mu})=p(\underset{\lambda_{BC}}{\mu})\,.
\end{split}\end{align}

Now, Eq.~\eqref{eq:inflationrawlemma} holds whenever $p({abc})$ is compatible with the triangle causal structure. Naturally, in our case, Eq.~\eqref{eq:inflationrawlemma}  by itself is not operationally testable. We only have access to the observed distribution $p({abc})$; if we had access to the underlying functional dependencies such as $p({a|\lambda_{AB},\lambda_{AC}})$ we wouldn't be bother with inflation, as we would then be able to access the (in)compatibility of $p({abc})$ by means of Eq.~\eqref{eq:trianglecompat} alone. The key insight is to extract \emph{easily testable implications} of Eq.~\eqref{eq:inflationrawlemma} which follow merely from the \emph{(non)existence} of suitable functional dependencies. The two implications of Eq.~\eqref{eq:inflationrawlemma} which we elect to incorporate into our linear program feasibility test are  \textbf{marginalization} and \textbf{symmetry}.

The marginalization condition is that
\begin{align}\label{eq:marginalization}\begin{split}
p'(\underset{a^{(1)}}{v_1},\underset{b^{(1)}}{v_2},\underset{c^{(1)}}{v_3},
\underset{a^{(4)}}{v_{10}},\underset{b^{(4)}}{v_{11}},\underset{c^{(4)}}{v_{12}})
&\coloneqq\smashoperator{\sum_{\substack{v_4, v_5, v_6, \\v_7, v_8, v_9}}}
p'(\underset{a^{(1)}}{v_1},\underset{b^{(1)}}{v_2},\underset{c^{(1)}}{v_3},
\underset{a^{(2)}}{v_4},\underset{b^{(2)}}{v_5},\underset{c^{(2)}}{v_6},
\underset{a^{(3)}}{v_7},\underset{b^{(3)}}{v_8},\underset{c^{(3)}}{v_9},
\underset{a^{(4)}}{v_{10}},\underset{b^{(4)}}{v_{11}},\underset{c^{(4)}}{v_{12}})
\\&=p(\underset{a}{v_1},\underset{b}{v_2},\underset{c}{v_3}) p(\underset{a}{v_{10}}, \underset{b}{v_{11}}, \underset{c}{v_{12}})\,,
\\\text{which we express as}\quad&{\boldsymbol{M}}\cdot p'= p^{\otimes 2}\,,
\end{split}\end{align}
where ${\boldsymbol{M}}$ is a zero/one valued marginalization matrix; it encodes the appropriate linear combinations of probabilities of $p'$ to make up the marginal probabilities. Accordingly, Eq.~\eqref{eq:marginalization} is a matrix equality constraint, or $4^6$ different equality constraints corresponding to the $4^6$ rows of ${\boldsymbol{M}}$, as the cardinality of all observable variables in our triangle network is  4. 

The symmetry condition follows from the fact that the copy indices are dummy indices. That is, since ${p'(\underset{\lambda_{AB}^{(1)}}{\mu})=p'(\underset{\lambda_{AB}^{(2)}}{\mu})}$ and so on, it follows that $p'$ must be invariant under the relabellings
\begin{align}
\begin{array}{c|c|c}
\pi_1 & \pi_2 & \pi_3 \\\hline
\lambda_{AB}^{(1)} \leftrightarrow \lambda_{AB}^{(2)} & \lambda_{AC}^{(1)} \leftrightarrow \lambda_{AC}^{(2)} & \lambda_{BC}^{(1)} \leftrightarrow \lambda_{BC}^{(2)} \\\hline
 a^{(1)}\leftrightarrow a^{(2)} & a^{(1)}\leftrightarrow a^{(3)} &  b^{(1)}\leftrightarrow b^{(2)} \\
 a^{(3)}\leftrightarrow a^{(4)} & a^{(2)}\leftrightarrow a^{(4)} &  b^{(3)}\leftrightarrow b^{(4)} \\
 b^{(1)}\leftrightarrow b^{(3)} & c^{(1)}\leftrightarrow c^{(2)} &  c^{(1)}\leftrightarrow c^{(3)} \\
 b^{(3)}\leftrightarrow b^{(4)} & c^{(3)}\leftrightarrow c^{(4)} &  c^{(2)}\leftrightarrow c^{(4)}
\end{array}
\end{align}
which generates an order-8 symmetry group. 
For maximum clarity, the cumulative effect of the symmetry implications of Eq.~\eqref{eq:inflationrawlemma}  are
\begin{align}\begin{split}
&p'(\underset{a^{(1)}}{v_{1}},\underset{b^{(1)}}{v_{2}},\underset{c^{(1)}}{v_{3}},
\underset{a^{(2)}}{v_{4}},\underset{b^{(2)}}{v_{5}},\underset{c^{(2)}}{v_{6}},
\underset{a^{(3)}}{v_{7}},\underset{b^{(3)}}{v_{8}},\underset{c^{(3)}}{v_{9}},
\underset{a^{(4)}}{v_{10}},\underset{b^{(4)}}{v_{11}},\underset{c^{(4)}}{v_{12}})\\=&
p'(\underset{a^{(1)}}{v_{1}},\underset{b^{(1)}}{v_{2}},\underset{c^{(1)}}{v_{3}},
\underset{a^{(2)}}{v_{4}},\underset{b^{(2)}}{v_{6}},\underset{c^{(2)}}{v_{5}},
\underset{a^{(3)}}{v_{8}},\underset{b^{(3)}}{v_{7}},\underset{c^{(3)}}{v_{11}},
\underset{a^{(4)}}{v_{12}},\underset{b^{(4)}}{v_{9}},\underset{c^{(4)}}{v_{10}})\\=&
p'(\underset{a^{(1)}}{v_{2}},\underset{b^{(1)}}{v_{1}},\underset{c^{(1)}}{v_{4}},
\underset{a^{(2)}}{v_{3}},\underset{b^{(2)}}{v_{7}},\underset{c^{(2)}}{v_{8}},
\underset{a^{(3)}}{v_{5}},\underset{b^{(3)}}{v_{6}},\underset{c^{(3)}}{v_{9}},
\underset{a^{(4)}}{v_{10}},\underset{b^{(4)}}{v_{11}},\underset{c^{(4)}}{v_{12}})\\=&
p'(\underset{a^{(1)}}{v_{2}},\underset{b^{(1)}}{v_{1}},\underset{c^{(1)}}{v_{4}},
\underset{a^{(2)}}{v_{3}},\underset{b^{(2)}}{v_{8}},\underset{c^{(2)}}{v_{7}},
\underset{a^{(3)}}{v_{6}},\underset{b^{(3)}}{v_{5}},\underset{c^{(3)}}{v_{11}},
\underset{a^{(4)}}{v_{12}},\underset{b^{(4)}}{v_{9}},\underset{c^{(4)}}{v_{10}})\\=&
p'(\underset{a^{(1)}}{v_{3}},\underset{b^{(1)}}{v_{4}},\underset{c^{(1)}}{v_{1}},
\underset{a^{(2)}}{v_{2}},\underset{b^{(2)}}{v_{5}},\underset{c^{(2)}}{v_{6}},
\underset{a^{(3)}}{v_{7}},\underset{b^{(3)}}{v_{8}},\underset{c^{(3)}}{v_{10}},
\underset{a^{(4)}}{v_{9}},\underset{b^{(4)}}{v_{12}},\underset{c^{(4)}}{v_{11}})\\=&
p'(\underset{a^{(1)}}{v_{3}},\underset{b^{(1)}}{v_{4}},\underset{c^{(1)}}{v_{1}},
\underset{a^{(2)}}{v_{2}},\underset{b^{(2)}}{v_{6}},\underset{c^{(2)}}{v_{5}},
\underset{a^{(3)}}{v_{8}},\underset{b^{(3)}}{v_{7}},\underset{c^{(3)}}{v_{12}},
\underset{a^{(4)}}{v_{11}},\underset{b^{(4)}}{v_{10}},\underset{c^{(4)}}{v_{9}})\\=&
p'(\underset{a^{(1)}}{v_{4}},\underset{b^{(1)}}{v_{3}},\underset{c^{(1)}}{v_{2}},
\underset{a^{(2)}}{v_{1}},\underset{b^{(2)}}{v_{7}},\underset{c^{(2)}}{v_{8}},
\underset{a^{(3)}}{v_{5}},\underset{b^{(3)}}{v_{6}},\underset{c^{(3)}}{v_{10}},
\underset{a^{(4)}}{v_{9}},\underset{b^{(4)}}{v_{12}},\underset{c^{(4)}}{v_{11}})\\=&
p'(\underset{a^{(1)}}{v_{4}},\underset{b^{(1)}}{v_{3}},\underset{c^{(1)}}{v_{2}},
\underset{a^{(2)}}{v_{1}},\underset{b^{(2)}}{v_{8}},\underset{c^{(2)}}{v_{7}},
\underset{a^{(3)}}{v_{6}},\underset{b^{(3)}}{v_{5}},\underset{c^{(3)}}{v_{12}},
\underset{a^{(4)}}{v_{11}},\underset{b^{(4)}}{v_{10}},\underset{c^{(4)}}{v_{9}})\,.
\end{split}\end{align}
As a linear program, we essentially have
\begin{align}\label{eq:primalLP1}
\exists_{p'|p'\geq {\boldsymbol{0}}}\,:\quad {\boldsymbol{M}}\cdot p'=p^{\otimes 2} \quad\text{and}\quad G_{\boldsymbol{\pi}}\circ p' =p',
\end{align}
where $G_{\boldsymbol{\pi}}\circ$ is the group \emph{twirling} operation, i.e. the projection onto the symmetric subspace of the group.

We can simplify the linear program~\eqref{eq:primalLP1} by replacing the invariance of $p'$ under symmetry with the statement that $p^{\otimes 2}$ is recovered by marginalizing the twirled version of $p'$, that is, ${\boldsymbol{M}}\cdot\left(G_{\boldsymbol{\pi}}\circ p'\right)=p^{\otimes 2}$. For that matter, we can just as well apply the left action of the twirling operator on the marginalization matrix, i.e.
\begin{align}
\left(G^{-1}_{\boldsymbol{\pi}}\circ {\boldsymbol{M}}\right)\cdot p' \coloneqq \left(G_{\boldsymbol{\pi}}\circ {\boldsymbol{M}}^T\right)^T \cdot p' = {\boldsymbol{M}}\cdot\left(G_{\boldsymbol{\pi}}\circ p'\right)
\end{align}
and hence we can equivalently express the linear program~\eqref{eq:primalLP1} as 
\begin{align}\label{eq:primalLP2}
\exists_{p'|p'\geq {\boldsymbol{0}}}\,:\quad \left(G^{-1}_{\boldsymbol{\pi}}\circ {\boldsymbol{M}}\right) \cdot p'=p^{\otimes 2}
\end{align}
which has the advantage of being easily dualized, i.e. the dual of LP~\eqref{eq:primalLP2} is
\begin{align}
\nexists_{{\boldsymbol{y}}|{\boldsymbol{y}}\cdot\left(G^{-1}_{\boldsymbol{\pi}}\circ {\boldsymbol{M}}\right)\geq {\boldsymbol{0}}}\,:\quad {\boldsymbol{y}}\cdot p^{\otimes 2}<0\,.
\end{align}
Indeed, the existence of a ${\boldsymbol{y}}$ vector which yields both ${\left(G^{-1}_{\boldsymbol{\pi}}\circ {\boldsymbol{M}}\right)\cdot {\boldsymbol{y}}\geq  {\boldsymbol{0}}}$ and ${{\boldsymbol{y}}\cdot p^{\otimes 2}<0}$ constitutes a certificate of infeasibility for LP~\eqref{eq:primalLP2} by Farkas' duality lemma~\cite{Andersen2001}. Accordingly, we discover the polynomial inequality which witnesses the incompatibility of $p$ with the classical triangle network by explicitly minimizing ${{\boldsymbol{y}}\cdot p^{\otimes 2}}$ subject to the constraints  ${\left(G^{-1}_{\boldsymbol{\pi}}\circ {\boldsymbol{M}}\right)\cdot {\boldsymbol{y}}\geq  {\boldsymbol{0}}}$.
\color{black}

In Supp.~Tab.~\ref{tab:ineq} we show the coefficients ${\boldsymbol{y}}$ pertaining to the products of probabilities generating the causal compatibility inequality 
. The coefficient is zero unless explicitly specified otherwise in Supp.~Tab.~\ref{tab:ineq}. 


\begin{table}
\centering 
\caption{\textbf{Tabular representation of the causal inequality.} This table specified the coefficients relative to the pairs of multiplied probabilities that give rise to the polynomial causal inequality. All unspecified coefficients are taken to be zero.}
\(
\begin{array}{rl}
 -1\to & \fbox{$
\begin{array}{llllllll}
 202 021 & 202 131 & 233 000 & 233 110 & 312 021 & 312 131 & 323 000 & 323 110
\end{array}$} \\
 2 \to &\fbox{$
\begin{array}{lllllllllllll}
023 010 & 031 002 & 031 012 & 033 010 & 100 023 & 100 033 & 102 031 & 112 031 & 121 002 & 121 012 & 121 102 & 121 112 & 123 010 \\
123 100 & 133 010 & 133 100 & 203 010 & 203 100 & 212 001 & 212 011 & 212 101 & 212 111 & 213 010 & 213 100 & 221 212 & 222 031 \\ 222 121 & 223 020 & 223 030 & 223 120 & 223 130 & 223 200 & 223 210 & 231 212 & 232 031 & 232 121 & 300 223 & 302 001 & 302 011 \\ 302 101 & 302 111 & 302 221 & 302 231 & 303 010 & 303 100 & 310 223 & 313 010 & 313 100 & 321 212 & 321 302 & 322 031 & 322 121 \\ 331 212 & 331 302 & 332 031 & 332 121 & 333 020 & 333 030 & 333 120 & 333 130 & 333 200 & 333 210 & 333 300 & 333 310 
\end{array}$} \\
 3\to &\fbox{$
\begin{array}{llllllll}
 212 031 & 212 121 & 223 010 & 223 100 & 302 031 & 302 121 & 333 010 & 333 100
\end{array}$}\\
 1\to &\fbox{$
\begin{array}{lllllllllllll}
002 001 & 003 000 & 010 003 & 011 002 & 012 001 & 012 011 & 013 000 & 013 010 & 020 003 & 020 013 & 022 001 & 022 011 & 022 021 \\
023 020 & 030 003 & 030 013 & 030 023 & 031 022 & 032 001 & 032 011 & 032 021 & 032 031 & 033 020 & 033 030 & 100 003 & 100 013 \\ 101 002 & 101 012 & 101 022 & 101 032 & 102 001 & 102 011 & 102 101 & 103 000 & 103 010 & 103 020 & 103 030 & 103 100 & 110 003 \\ 110 013 & 110 103 & 111 002 & 111 012 & 111 022 & 111 032 & 111 102 & 112 001 & 112 011 & 112 101 & 112 111 & 113 000 & 113 010 \\ 113 020 & 113 030 & 113 100 & 113 110 & 120 003 & 120 013 & 120 023 & 120 033 & 120 103 & 120 113 & 121 022 & 121 032 & 122 001 \\ 122 011 & 122 021 & 122 031 & 122 101 & 122 111 & 122 121 & 123 020 & 123 030 & 123 120 & 130 003 & 130 013 & 130 023 & 130 033 \\ 130 103 & 130 113 & 130 123 & 131 022 & 131 032 & 131 122 & 132 001 & 132 011 & 132 021 & 132 031 & 132 101 & 132 111 & 132 121 \\ 132 131 & 133 020 & 133 030 & 133 120 & 133 130 & 200 003 & 200 013 & 200 020 & 200 021 & 200 022 & 200 023 & 200 030 & 200 031 \\ 200 032 & 200 033 & 200 103 & 200 113 & 200 120 & 200 121 & 200 122 & 200 123 & 200 130 & 200 131 & 200 132 & 200 133 & 201 002 \\ 201 012 & 201 020 & 201 021 & 201 022 & 201 023 & 201 030 & 201 031 & 201 032 & 201 033 & 201 102 & 201 112 & 201 120 & 201 121 \\ 201 122 & 201 123 & 201 130 & 201 131 & 201 132 & 201 133 & 202 020 & 202 022 & 202 023 & 202 030 & 202 031 & 202 032 & 202 033 \\ 202 120 & 202 121 & 202 122 & 202 123 & 202 130 & 202 132 & 202 133 & 202 201 & 203 020 & 203 021 & 203 022 & 203 023 & 203 030 \\ 203 031 & 203 032 & 203 033 & 203 120 & 203 121 & 203 122 & 203 123 & 203 130 & 203 131 & 203 132 & 203 133 & 203 200 & 210 003 \\ 210 013 & 210 020 & 210 021 & 210 022 & 210 023 & 210 030 & 210 031 & 210 032 & 210 033 & 210 103 & 210 113 & 210 120 & 210 121 \\ 210 122 & 210 123 & 210 130 & 210 131 & 210 132 & 210 133 & 210 203 & 211 002 & 211 012 & 211 020 & 211 021 & 211 022 & 211 023 \\ 211 030 & 211 031 & 211 032 & 211 033 & 211 102 & 211 112 & 211 120 & 211 121 & 211 122 & 211 123 & 211 130 & 211 131 & 211 132 \\ 211 133 & 211 202 & 212 020 & 212 021 & 212 022 & 212 023 & 212 030 & 212 032 & 212 033 & 212 120 & 212 122 & 212 123 & 212 130 \\ 212 131 & 212 132 & 212 133 & 212 201 & 212 211 & 213 020 & 213 021 & 213 022 & 213 023 & 213 030 & 213 031 & 213 032 & 213 033 \\ 213 120 & 213 121 & 213 122 & 213 123 & 213 130 & 213 131 & 213 132 & 213 133 & 213 200 & 213 210 & 220 000 & 220 001 & 220 002 \\ 220 003 & 220 010 & 220 011 & 220 012 & 220 013 & 220 023 & 220 033 & 220 100 & 220 101 & 220 102 & 220 103 & 220 110 & 220 111 \\ 220 112 & 220 113 & 220 123 & 220 133 & 220 203 & 220 213 & 221 000 & 221 001 & 221 002 & 221 003 & 221 010 & 221 011 & 221 012 \\ 221 013 & 221 022 & 221 032 & 221 100 & 221 101 & 221 102 & 221 103 & 221 110 & 221 111 & 221 112 & 221 113 & 221 122 & 221 132 \\ 222 000 & 222 001 & 222 002 & 222 003 & 222 010 & 222 011 & 222 012 & 222 013 & 222 100 & 222 101 & 222 102 & 222 103 & 222 110 \\ 222 111 & 222 112 & 222 113 & 222 201 & 222 211 & 222 221 & 223 000 & 223 001 & 223 002 & 223 003 & 223 011 & 223 012 & 223 013 \\ 223 101 & 223 102 & 223 103 & 223 110 & 223 111 & 223 112 & 223 113 & 223 220 & 230 000 & 230 001 & 230 002 & 230 003 & 230 010 \\ 230 011 & 230 012 & 230 013 & 230 023 & 230 033 & 230 100 & 230 101 & 230 102 & 230 103 & 230 110 & 230 111 & 230 112 & 230 113 \\ 230 123 & 230 133 & 230 203 & 230 213 & 230 223 & 231 000 & 231 001 & 231 002 & 231 003 & 231 010 & 231 011 & 231 012 & 231 013 \\ 231 022 & 231 032 & 231 100 & 231 101 & 231 102 & 231 103 & 231 110 & 231 111 & 231 112 & 231 113 & 231 122 & 231 132 & 231 222 \\ 232 000 & 232 001 & 232 002 & 232 003 & 232 010 & 232 011 & 232 012 & 232 013 & 232 100 & 232 101 & 232 102 & 232 103 & 232 110 \\ 232 111 & 232 112 & 232 113 & 232 201 & 232 211 & 232 221 & 232 231 & 233 001 & 233 002 & 233 003 & 233 010 & 233 011 & 233 012 \\ 233 013 & 233 100 & 233 101 & 233 102 & 233 103 & 233 111 & 233 112 & 233 113 & 233 220 & 233 230 & 300 003 & 300 013 & 300 020 \\ 300 021 & 300 022 & 300 023 & 300 030 & 300 031 & 300 032 & 300 033 & 300 103 & 300 113 & 300 120 & 300 121 & 300 122 & 300 123 \\ 300 130 & 300 131 & 300 132 & 300 133 & 300 203 & 300 213 & 301 002 & 301 012 & 301 020 & 301 021 & 301 022 & 301 023 & 301 030 \\ 301 031 & 301 032 & 301 033 & 301 102 & 301 112 & 301 120 & 301 121 & 301 122 & 301 123 & 301 130 & 301 131 & 301 132 & 301 133 \\ 301 202 & 301 212 & 301 222 & 301 232 & 302 020 & 302 021 & 302 022 & 302 023 & 302 030 & 302 032 & 302 033 & 302 120 & 302 122 \\ 302 123 & 302 130 & 302 131 & 302 132 & 302 133 & 302 201 & 302 211 & 302 301 & 303 020 & 303 021 & 303 022 & 303 023 & 303 030 \\ 303 031 & 303 032 & 303 033 & 303 120 & 303 121 & 303 122 & 303 123 & 303 130 & 303 131 & 303 132 & 303 133 & 303 200 & 303 210 \\ 303 220 & 303 230 & 303 300 & 310 003 & 310 013 & 310 020 & 310 021 & 310 022 & 310 023 & 310 030 & 310 031 & 310 032 & 310 033 \\ 310 103 & 310 113 & 310 120 & 310 121 & 310 122 & 310 123 & 310 130 & 310 131 & 310 132 & 310 133 & 310 203 & 310 213 & 310 303 \\ 311 002 & 311 012 & 311 020 & 311 021 & 311 022 & 311 023 & 311 030 & 311 031 & 311 032 & 311 033 & 311 102 & 311 112 & 311 120 \\ 311 121 & 311 122 & 311 123 & 311 130 & 311 131 & 311 132 & 311 133 & 311 202 & 311 212 & 311 222 & 311 232 & 311 302 & 312 020 \\ 312 022 & 312 023 & 312 030 & 312 031 & 312 032 & 312 033 & 312 120 & 312 121  \\
\end{array}$}\\
\end{array}
\)
\label{tab:ineq}
\end{table}

\begin{table*}
\centering 
\(
\begin{array}{rl}
 1\to &\fbox{$
\begin{array}{lllllllllllll}
 312 122 & 312 123 & 312 130 & 312 132 & 312 133 & 312 201 & 312 211 & 312 301 & 312 311 & 313 020 & 313 021 & 313 022 & 313 023 \\
 313 030 & 313 031 & 313 032 & 313 033 & 313 120 & 313 121 & 313 122 & 313 123 & 313 130 & 313 131 & 313 132 & 313 133 & 313 200 \\ 313 210 & 313 220 & 313 230 & 313 300 & 313 310 & 320 000 & 320 001 & 320 003 & 320 010 & 320 011 & 320 012 & 320 013 & 320 023 \\ 320 033 & 320 100 & 320 101 & 320 102 & 320 103 & 320 110 & 320 111 & 320 112 & 320 113 & 320 123 & 320 133 & 320 203 & 320 213 \\ 320 223 & 320 233 & 320 303 & 320 313 & 321 001 & 321 002 & 321 003 & 321 010 & 321 011 & 321 012 & 321 013 & 321 022 & 321 032 \\ 321 100 & 321 101 & 321 102 & 321 103 & 321 110 & 321 111 & 321 112 & 321 113 & 321 122 & 321 132 & 321 222 & 321 232 & 322 000 \\ 322 001 & 322 002 & 322 003 & 322 010 & 322 011 & 322 012 & 322 013 & 322 100 & 322 101 & 322 102 & 322 103 & 322 110 & 322 111 \\ 322 112 & 322 113 & 322 201 & 322 211 & 322 221 & 322 231 & 322 301 & 322 311 & 322 321 & 323 001 & 323 002 & 323 003 & 323 010 \\ 323 011 & 323 012 & 323 013 & 323 100 & 323 101 & 323 102 & 323 103 & 323 111 & 323 112 & 323 113 & 323 220 & 323 230 & 323 320 \\ 330 000 & 330 001 & 330 002 & 330 003 & 330 010 & 330 011 & 330 012 & 330 013 & 330 023 & 330 033 & 330 100 & 330 101 & 330 102 \\ 330 103 & 330 110 & 330 111 & 330 112 & 330 113 & 330 123 & 330 133 & 330 203 & 330 213 & 330 223 & 330 233 & 330 303 & 330 313 \\ 330 323 & 331 000 & 331 001 & 331 002 & 331 003 & 331 010 & 331 011 & 331 012 & 331 013 & 331 022 & 331 032 & 331 100 & 331 101 \\ 331 102 & 331 103 & 331 110 & 331 111 & 331 112 & 331 113 & 331 122 & 331 132 & 331 222 & 331 232 & 331 322 & 332 000 & 332 001 \\ 332 002 & 332 003 & 332 010 & 332 011 & 332 012 & 332 013 & 332 100 & 332 101 & 332 102 & 332 103 & 332 110 & 332 111 & 332 112 \\ 332 113 & 332 201 & 332 211 & 332 221 & 332 231 & 332 301 & 332 311 & 332 321 & 332 331 & 333 000 & 333 001 & 333 002 & 333 003 \\ 333 011 & 333 012 & 333 013 & 333 101 & 333 102 & 333 103 & 333 110 & 333 111 & 333 112 & 333 113 & 333 220 & 333 230 & 333 320 \\ 333 330  \\
\end{array}$} \\
\end{array}
\)
\end{table*}